\documentclass{ws-spin}
\usepackage{multicol}

\newcommand{\japref}{42}
\newcommand{\srepref}{53}
\newcommand{\jphysdref}{87}
\newcommand{\aplSAW}{138}

\usepackage{subfigure}
\usepackage{graphicx}

\DeclareGraphicsExtensions{.eps}

\begin{document}

\catchline{xx}{xx}{2016}{}{}
\markboth{Kuntal Roy}{Ultra-low-energy Electric field-induced Magnetization Switching in Multiferroic Heterostructures}

\title{Ultra-low-energy Electric field-induced Magnetization Switching in Multiferroic Heterostructures}

\author{Kuntal Roy}

\address{School of Electrical and Computer Engineering\\
Purdue University, West Lafayette, Indiana 47907, USA\\
\email{royk@purdue.edu}
}

\maketitle

\begin{history}
\received{05 Jun 2016}
\accepted{Day Month Year}
\end{history}

\begin{abstract}
Electric field-induced magnetization switching in multiferroics is intriguing for both fundamental studies and potential technological applications. Here, we review the recent developments on electric field-induced magnetization switching in multiferroic heterostructures. Particularly, we study the dynamics of magnetization switching between the two stable states in a shape-anisotropic single-domain nanomagnet using stochastic Landau-Lifshitz-Gilbert (LLG) equation in the presence of thermal fluctuations. For magnetostrictive nanomagnets in strain-coupled multiferroic composites, such study of magnetization dynamics, contrary to steady-state scenario, revealed intriguing new phenomena on binary switching mechanism. While the traditional method of binary switching requires to tilt the potential profile to the desired state of switching, we show that no such tilting is necessary to switch successfully since the magnetization's excursion out of magnet's plane can generate a built-in asymmetry during switching. We also study the switching dynamics in multiferroic heterostructures having magnetoelectric coupling at the interface and magnetic exchange coupling that can facilitate to maintain the direction of switching with the polarity of the applied electric field. We calculate the performance metrics like switching delay and energy dissipation during switching while simulating LLG dynamics. The performance metrics turn out to be very encouraging for potential technological applications.
\end{abstract}


\begin{multicols}{2}

\section{Introduction}

Multiferroics are intriguing materials in which there is \emph{intrinsic} coupling between one or more ferroic properties, e.g., ferroelectric, ferromagnetic, ferroelastic ordering.\cite{RefWorks:841,RefWorks:664,RefWorks:164,RefWorks:897,RefWorks:898,RefWorks:899,roy13_spin,RefWorks:890} In multiferroic \emph{magnetoelectrics}, application of an electric field can rotate the magnetization via converse magnetoelectric effect, however, such materials in \emph{single-phase} were thought to be rare.\cite{RefWorks:512} Usually such single-phase multiferroics have issues of weak coupling between polarization and magnetization, and also they usually operate only at low temperatures.\cite{RefWorks:558,RefWorks:665} Although, new concepts may come along on switching in single-phase materials\cite{RefWorks:665,RefWorks:707,RefWorks:667,RefWorks:669,RefWorks:699,RefWorks:700,RefWorks:701,RefWorks:846} possibly utilizing  Dzyaloshinsky-Moriya (DM) interaction,\cite{RefWorks:697,RefWorks:698,RefWorks:745,roy14_6} magnetoelectric coupling in \emph{strain-mediated} multiferroic composites consisting of a piezoelectric layer coupled to a magnetostrictive nanomagnet (see Fig.~\ref{fig:multiferroic_composite}) is shown to be very effective.\cite{RefWorks:558,RefWorks:521,RefWorks:164,RefWorks:165,RefWorks:842,RefWorks:843,RefWorks:519} Electric field-induced magnetization switching in multiferroics uses a voltage directly thereby eliminating the need to switch magnetization by a cumbersome magnetic field or by a spin-polarized current,\cite{RefWorks:8,RefWorks:155,RefWorks:7,roy11_3} although new concepts are being investigated e.g., utilizing giant spin-Hall effect.\cite{roy14_3}

The calculation of performance metrics using stochastic Landau-Lifshitz-Gilbert (LLG) equation of magnetization dynamics in strain-mediated multiferroic composites is shown to hold profound promise for computing\cite{roy13_spin,roy11_news,roy13,roy14,roy14_4} in beyond Moore's law era.\cite{RefWorks:553,moore65,RefWorks:126,RefWorks:211,nri} With a suitable choice of materials and dimensions, when a voltage of few millivolts is applied across such heterostructures, the piezoelectric layer gets strained and the strain is elastically transferred to the magnetostrictive nanomagnet, which can rotate the magnetization. Such switching mechanism dissipates a minuscule amount of energy of $\sim$1 attojoule (aJ) in sub-nanosecond switching delay at room-temperature.\cite{roy13_spin,roy11_6} This study has opened up a new field called \emph{straintronics}.\cite{roy13_spin,roy11_news,roy13,roy14} Experimental efforts have been undergoing to investigate such device functionality and the induced stress anisotropy in magnetostrictive nanomagnets is demonstrated.\cite{RefWorks:559,RefWorks:806,RefWorks:836,RefWorks:609,RefWorks:844,RefWorks:790,RefWorks:838,RefWorks:868} The direct experimental demonstration of switching speed (rather than ferromagnetic resonance experiments to get the time-scale of switching) and using low-thickness piezoelectric layers avoiding considerable degradation of the piezoelectric constants [e.g., $<$ 100 nm of lead magnesium niobate-lead titanate (PMN-PT)]\cite{RefWorks:823,RefWorks:820} are still under investigation.

\vspace*{5mm}
\begin{figurehere}
\centerline{\includegraphics[width=80mm]{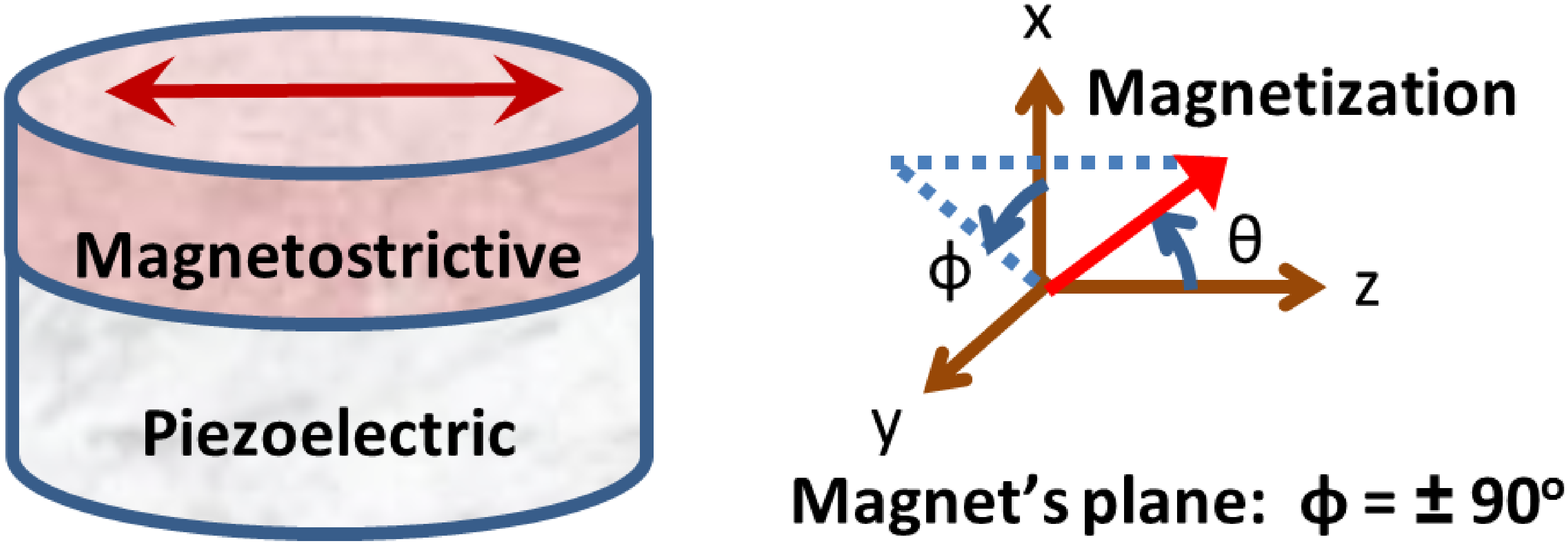}}
\caption{\label{fig:multiferroic_composite}
Schematic diagram of a strain-mediated multiferroic composite (piezoelectric-magnetostrictive heterostructure), and axis assignment. The magnetostrictive nanomagnet is shaped like an elliptical cylinder and it has a single ferromagnetic domain. The mutually anti-parallel magnetization states along the $z$-axis can store a binary bit of information (0 or 1). In standard spherical coordinate system, $\theta$ is the polar angle and $\phi$ is the azimuthal angle. According to the dimensions of the elliptical cylinder in the respective directions, we term the $z$-axis the easy axis, the $y$-axis the in-plane hard axis, and the $x$-axis the out-of-plane hard axis for the nanomagnet.}
\end{figurehere}

The binary switching of magnetization between two stable states of a shape-anisotropic single-domain magnetostrictive nanomagnet in a strain-mediated multiferroic composite is very intriguing.\cite{roy13_2,roy11_5} The understanding behind binary switching from one stable state to another enables us to design a better switch to address our ever-increasing demand to store, process, and communicate information. From around mid-nineties, the methodology of binary switching for information processing as conceived by famous scientist Landauer and others says that an \emph{externally} introduced tilt or \emph{asymmetry} in potential landscape of a bistable element in the \emph{desired direction of switching} is \emph{necessary}.\cite{RefWorks:148,RefWorks:149,RefWorks:144,RefWorks:557} This \emph{asymmetry} in potential landscape can be achieved by utilizing an \emph{external} magnetic field in a single-domain nanomagnet with two stable states acting as a binary switch.\cite{RefWorks:557} The tilt generates a motion along the direction of switching and the degree of tilt should be sufficient enough to dissuade thermal fluctuations with a tolerable error probability. Such tilt or asymmetry in potential landscape is deemed to be \emph{necessary} for switching to take place successfully. Fig.~\ref{fig:switching_asymm_symm3D}(a) depicts such traditional methodology of binary switching.

\vspace*{2mm}
\begin{figurehere}
\centerline{\includegraphics[width=80mm]{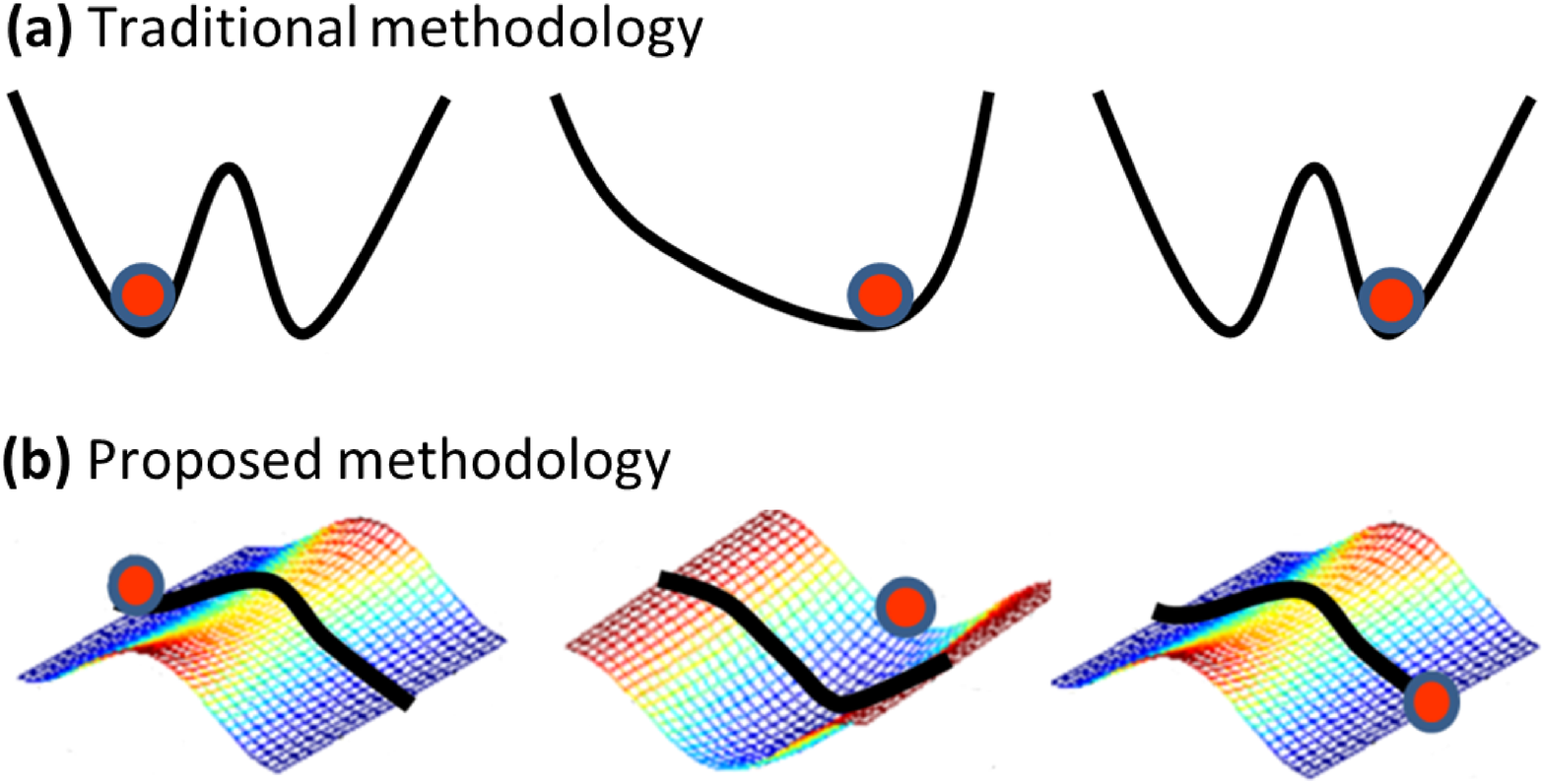}}
\caption{Traditional methodology versus the proposed methodology of binary switching.
(a) In traditional methodology of binary switching, the potential landscape is tilted towards the direction of switching alongwith the lowering of energy barrier separating the two stable states. Note that there are two external agents involved here, one makes the potential landscape monostable and the other one that tilts the potential landscape. At the end, the potential profile is restored back to that of the initial stage to complete the switching process.
(b) In the proposed methodology of binary switching, the potential profile remains always `symmetric', i.e., energy barrier is lowered but the potential landscape is \emph{not} tilted to favor the final state. Switching occurs due to \emph{internal} dynamics considering complete three-dimensional potential landscape and full three-dimensional motion. Note that in this case only one external agent is involved since it does not require tilting the potential landscape. For a nanomagnet acting as a binary switch, the bold line in its potential landscape corresponds to when magnetization resides on magnet's plane. Deflection of magnetization from magnet's plane corresponds to out-of-plane excursion of magnetization, which is the physical mechanism of switching in this case. (Reprinted from Ref.~\srepref.)}
\label{fig:switching_asymm_symm3D}
\end{figurehere}


Note that the magnetization switching mechanism using spin-transfer-torque,\cite{RefWorks:8,RefWorks:155,RefWorks:7} in which a spin-polarized current is passed through a nanomagnet to switch its magnetization, is analogous to the \emph{traditional} methodology of binary switching. It is well known that the Slonczewski-like spin-transfer-torque that acts in-plane of the nanomagnet cannot be treated as an effective potential,\cite{RefWorks:8} however, the \emph{direction} of \emph{externally} applied spin-polarized current induces an \emph{equivalent} tilt or asymmetry and causes the magnetization switching in the desired direction.

In Ref.~\refcite{roy13_2}, it is shown that it is not necessary to tilt the potential landscape by \emph{external} means to switch successfully, even in the presence of thermal fluctuations. Fig.~\ref{fig:switching_asymm_symm3D}(b) illustrates the basic concept underlying such methodology of binary switching. To understand such switching, the complete three-dimensional potential landscape needs to be considered and thereby it is demonstrated in Ref.~\refcite{roy13_2} that the \emph{intrinsic} dynamics can provide an equivalent asymmetry without any requirement of making the potential landscape asymmetric. It needs mention here that for both traditional and proposed methodologies, it is necessary to lower the energy barrier and make the monostable well deep enough to dissuade thermal fluctuations. For a magnetostrictive nanomagnet, stress is the external agent modulating and eventually inverting the potential landscape of the nanomagnet. The voltage-induced stress can be generated on a magnetostrictive nanomagnet by elastically coupling a piezoelectric layer, i.e., using 2-phase multiferroic composites.\cite{RefWorks:558,RefWorks:164,RefWorks:165,RefWorks:562,RefWorks:328,RefWorks:167,roy11_news} Such magnetization switching in multiferroic composites can potentially be the basis of ultra-low-energy computing in our future information processing systems.\cite{roy13_spin,roy11_news,roy13_2,roy11_6}


If we consider only the \emph{steady-state} scenario, we will come to a conclusion that in a strain-mediated multiferroic composite, the strain transferred by the piezoelectric layer to the magnetostrictive nanomagnet, can only rotate the magnetization $90^\circ$ and not the complete $180^\circ$, which is the usual perception. However, a complete $180^\circ$ switching facilitates us to achieve a higher magnetoresistance while reading the magnetization state using magnetic tunnel junctions (MTJs).\cite{RefWorks:577,RefWorks:555,RefWorks:572,RefWorks:76,RefWorks:74,RefWorks:33,RefWorks:300} Although there are proposals of $90^\circ$ switching mechanism,\cite{RefWorks:520,RefWorks:551,RefWorks:559} as shown in Ref.~\refcite{roy13_2} and described above, a complete $180^\circ$ switching is possible if we consider the \emph{dynamics} of magnetization into account rather than \emph{assuming} steady-state scenario. Basically magnetization's excursion out of magnet's plane provides an equivalent asymmetry to cause a complete $180^\circ$ switching.\cite{roy13_2}

\vspace*{5mm}
\begin{figurehere}
\centerline{\includegraphics[width=80mm]{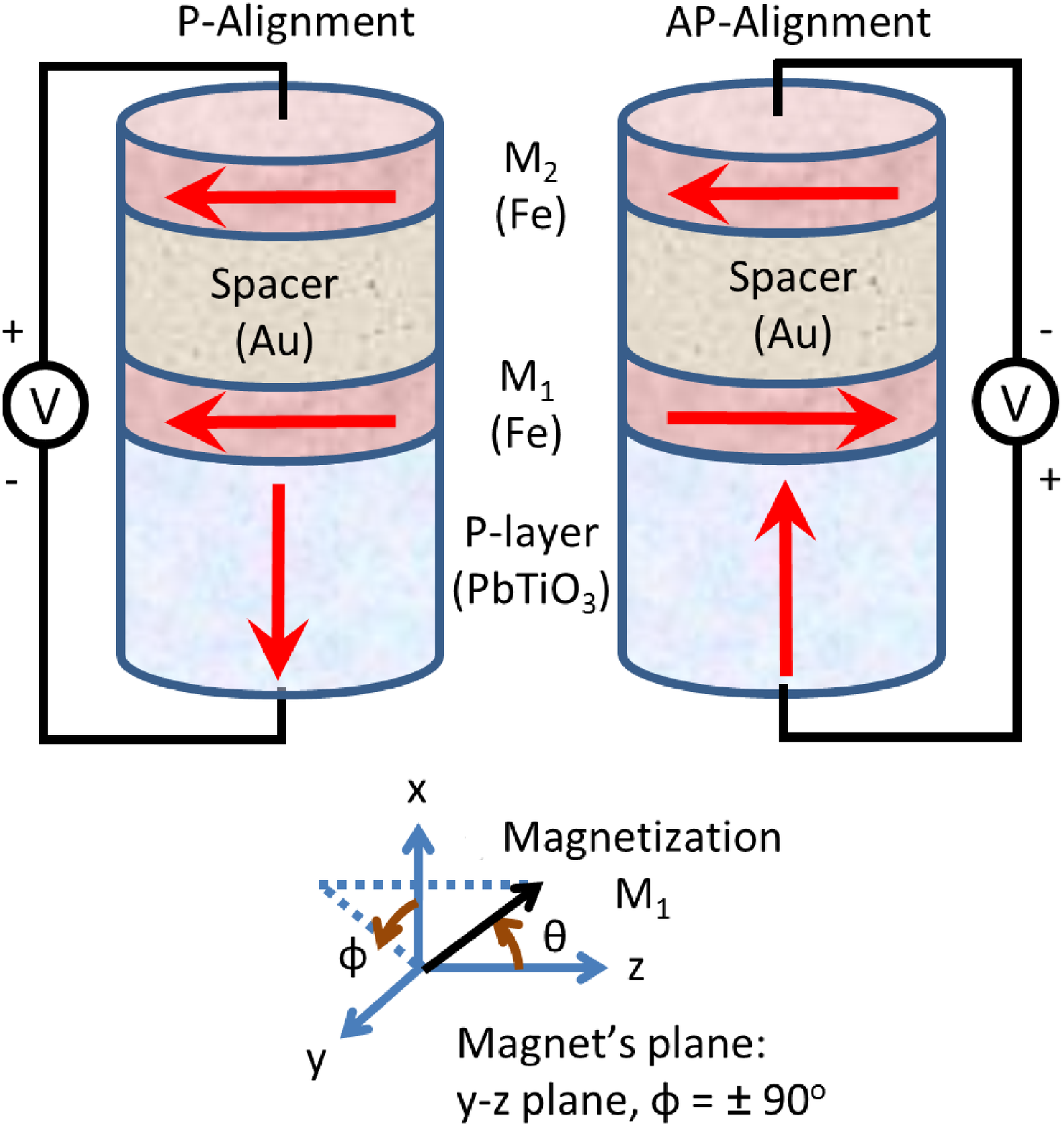}}
\caption{Schematics of the interface and exchange coupled multiferroic heterostructures. The unique coupling between the polarization in the P-layer and the trilayer $M_1$/spacer/$M_2$ allows the polarization direction to dictate the magnetic ground state in the trilayer. If the polarization points downward ($P_\downarrow$), P-alignment in the trilayer is preferred while an upward polarization ($P_\uparrow$) prefers the AP-alignment. Application of a voltage with correct polarity can switch the polarization and hence the magnetization $M_1$ gets switched too due to interface and exchange coupling. At the bottom of the figure, the axis assignment for the dynamical motion of magnetization $M_1$ in standard spherical coordinate system is shown. 
(\textcopyright IOP Publishing.  Reproduced by permission of IOP Publishing from Ref.~\jphysdref. All rights reserved.)} 
\label{fig:schematic_interface_coupled}
\end{figurehere}


Although the aforesaid switching mechanism in \emph{strain-mediated} multiferroic composites is intriguing and promising for technological applications, it would be of substantial interest if there exists a strong coupling between polarization and magnetization at the heterostructure interface to harness additional asymmetry during magnetization switching in a specified direction. In Ref.~\refcite{RefWorks:649}, \emph{interface and exchange coupled} multiferroics are proposed based on density functional theory (DFT) of first-principles calculations. Although this specific case needs to be experimentally demonstrated, the proposed concept is tenable. The first-principles calculations have been proved to be very useful in such respect\cite{RefWorks:512} and with the experimental progress on similar front\cite{RefWorks:793,RefWorks:404,RefWorks:795,RefWorks:796,RefWorks:797} (also using ferromagnetic oxides\cite{RefWorks:792,RefWorks:798} rather than ferromagnetic metals), there is a considerable promise on such polarization-magnetization coupling mechanism.\cite{RefWorks:649,RefWorks:676,RefWorks:680,RefWorks:681,RefWorks:682,RefWorks:683} In Figure~\ref{fig:schematic_interface_coupled}, such interface and exchange coupling between polarization and magnetization in a multiferroic heterostructure is depicted.\cite{RefWorks:649} The polarization direction in the P-layer uniquely determines the ground state of the trilayer $M_1$/spacer/$M_2$, i.e., for downward polarization P$_\downarrow$, parallel alignment (P-alignment) in the trilayer is preferred, while for the upward polarization P$_\uparrow$, antiparallel alignment (AP-alignment) in the trilayer is achieved. This polarization can be switched electrically and such polarization-magnetization coupling mechanism makes the switching of magnetization in the $M_1$-layer \emph{non-volatile}. If a voltage with certain polarity is applied and maintained, the state of the system remains unaltered too. This is advantageous over the strain-mediated switching, which just toggles the magnetization states and therefore requires a read-before-write mechanism. Note that there are other exchanged coupled systems with an insulating spacer layer, however, the interlayer exchange coupling energy is small.\cite{RefWorks:725,RefWorks:726} There are device structures with non-magnetic spacer layer to preserve large interlayer exchange coupling too, but a high electric field is required and the switching becomes volatile.\cite{RefWorks:727}

In Ref.~\refcite{roy14_2}, magnetization dynamics in the interface and exchange coupled multiferroic heterostructures is studied by solving stochastic Landau-Lifshitz-Gilbert equation in the presence of room-temperature thermal fluctuations. Such phenomenological study of switching has been very useful to understand the performance metrics of magnetic devices.\cite{roy13_spin,roy11_news,roy13,roy14,roy11_6,roy13_2} First, an interfacial anisotropy in the interface and exchange coupled multiferroic heterostructures is modeled and subsequently the analysis is performed on the \emph{dynamics} of magnetization. The results show that switching in sub-nanosecond delay is possible while expending only $\sim$1 aJ of energy at room-temperature. The key point is that the \emph{strong} interface anisotropy makes the switching error-resilient and fast, allowing to use nanomagnets with very small dimensions (magnetization is stable with $\sim$10 nm lateral dimensions even in the presence of room-temperature thermal fluctuations). Such performance metrics of area, delay, and energy are particularly promising for computing in our future information processing systems.\cite{RefWorks:553,moore65,RefWorks:126,RefWorks:211,nri}

The rest of the paper is organized as follows. In Section~\ref{sec:model}, we describe in subsequent subsections two models for electric field-induced magnetization switching: (1) strain-mediated multiferroic composites, and (2) interface and exchange coupled multiferroic heterostructures. For both the models, the stochastic Landau-Lifshitz-Gilbert (LLG) equation of magnetization dynamics in the presence of thermal fluctuations is analytically solved to get a coupled set of equations, which need to be solved numerically onwards. Section~\ref{sec:results} presents the simulation results by solving the coupled sets of equations numerically. Then we calculate the performance metrics e.g., switching delay, energy dissipation, which show particularly promising results for technological applications. Finally, Section~\ref{sec:conclusions} summarizes this review and provides the outlook on the electric field-induced magnetization switching in multiferroic heterostructures.

\section{\label{sec:model}Model}

Here, we will review the models developed for strain-mediated multiferroic composites, and interface and exchange coupled multiferroic heterostructures in subsequent subsections. The emphasis would be on the dynamical motion by solving the Landau-Lifshitz-Gilbert (LLG) equation of magnetization dynamics.\cite{RefWorks:162,RefWorks:161} We will also consider thermal fluctuations incorporated in the LLG equation making it of \emph{stochastic} nature.\cite{RefWorks:186,roy13_2,roy11_6,roy14_2}

\subsection{\label{sec:model_strain}Strain-mediated multiferroic composites} 
Consider a nanomagnet shaped like an elliptical cylinder with its elliptical cross section lying on the $y$-$z$ plane; the major axis is aligned along the $z$-direction and the minor axis along the $y$-direction. (See Fig.~\ref{fig:multiferroic_composite}.) The dimension of the major axis, the minor axis, and the thickness are $a$, $b$, and $l$, respectively. The volume of the nanomagnet is $\Omega=(\pi/4)a b l$. In standard spherical coordinate system, as shown in the Fig.~\ref{fig:multiferroic_composite}, $\theta$ is the polar angle and $\phi$ is the azimuthal angle. Note that when $\phi =\pm 90^{\circ}$, the magnetization vector lies on the plane of the nanomagnet ($y$-$z$ plane). Any deviation from $\phi = \pm 90^{\circ}$ is termed as out-of-plane excursion of magnetization.

The total energy of the single-domain, magnetostrictive, polycrystalline (i.e., no net magnetocrystalline anisotropy) nanomagnet,  while subjected to uniaxial stress along the easy axis ($z$-axis, the major axis of the ellipse) is the sum of the shape anisotropy energy and the uniaxial stress anisotropy energy:\cite{RefWorks:157}
\begin{equation}
E(\theta,\phi,t) = E_{shape}(\theta,\phi) + E_{stress}(\theta,t),
\end{equation}
where $E_{shape}(\theta,\phi)$ is the shape anisotropy energy and $E_{stress}(\theta,t)$ is the stress anisotropy energy at time $t$. The former is given by\cite{RefWorks:157}
\begin{equation}
E_{shape}(\theta,\phi) = \frac{\mu_0}{2} M_s^2 \Omega N_d(\theta,\phi),
\label{shape-anisotropy}
\end{equation}
where $M_s$ is the saturation magnetization and $N_d(\theta,\phi)$ is the demagnetization factor expressed 
as\cite{RefWorks:157} 
\begin{multline}
N_d(\theta,\phi) = N_{d-zz} cos^2\theta + N_{d-yy} sin^2\theta \ sin^2\phi \\ + N_{d-xx} sin^2\theta \, cos^2\phi
\end{multline}
with $N_{d-zz}$, $N_{d-yy}$, and $N_{d-xx}$ being the components of the demagnetization factor along the $z$-axis, $y$-axis, and $x$-axis, respectively. The parameters $N_{d-zz}$, $N_{d-yy}$, and $N_{d-xx}$ depend on the shape and dimensions of the nanomagnet\cite{RefWorks:157}  and are determined from the prescription in Ref.~\refcite{RefWorks:402}. 

The in-plane ($\phi=\pm 90^\circ$) shape-anisotropic energy-barrier between the two stable states ($\theta = 0^\circ$ and $180^\circ$) can be expressed as
\begin{equation}
E_{barrier} = \frac{\mu_0}{2} M_s^2 \Omega \left(N_{d-yy} - N_{d-zz}\right).
\label{eq:energy-barrier}
\end{equation}
\noindent
With $a=100\,nm$, $a=90\,nm$, and $l=6\,nm$, $E_{barrier} \simeq 44\, kT$ at room-temperature ($T=300\,K$), which means that the static error probability due to spontaneous fluctuation of magnetization is $e^{-E_{barrier}/kT} = e^{-44}$.

The uniaxial shape anisotropy favors lining up the magnetization along the major axis ($z$-axis) by minimizing $E_{shape}$, which is why we call the major axis the ``easy axis'' and the minor axis ($y$-axis) the ``in-plane hard axis'' of the magnet. The $x$-axis will therefore be the ``out-of-plane hard axis'' of the magnet and it is ``harder'' than the in-plane one since the thickness is much smaller than the magnet's lateral dimensions (i.e., $l << a, b$).


We assume that an uniaxial stress is generated along the $z$-axis (easy axis) upon application of an electric field along the $x$-axis. It is possible to constrain the expansion along $y$-axis to generate the uniaxial stress along the $z$-direction for piezoelectrics like lead-zirconate-titanate (PZT). We can also use piezoelectrics like lead magnesium niobate-lead titanate (PMN-PT), which generates \emph{anisotropic} strain in the lateral plane (i.e., the signs of $d_{31}$ and $d_{32}$ coefficients are different) and therefore it can produce more strain for a given voltage or otherwise a lower voltage is required to generate a specified strain. The stress anisotropy energy is given by\cite{RefWorks:157}
\begin{equation}
E_{stress}(\theta,t) = - (3/2) \lambda_s \sigma(t) \Omega \, cos^2\theta,
\label{eq:e_stress}
\end{equation}
where $(3/2) \lambda_s$ is the magnetostriction coefficient of the single-domain nanomagnet and $\sigma(t)$ is the stress at an instant of time $t$. Note that a positive $\lambda_s \sigma$ product will favor alignment of the magnetization along the major axis ($z$-axis), while a negative $\lambda_s \sigma$ product will favor alignment on the $x$-$y$ plane ($\theta=90^\circ$), because that will minimize $E_{stress}$. In our convention, a compressive stress is negative and tensile stress is positive. Therefore, in a material like Terfenol-D that has positive $\lambda_s$, a compressive stress will favor alignment on the $x$-$y$ plane ($\theta=90^\circ$), and tensile along the major axis. The situation will be opposite with common magnetic materials like iron, nickel, and cobalt, which have negative $\lambda_s$.
 
At any instant of time $t$, the total energy of the nanomagnet can be expressed as 
\begin{equation}
 E(\theta,\phi,t) = B(\phi,t) sin^2\theta + C(t),
\end{equation}
where 
\begin{subequations}
\begin{eqnarray}
B(\phi,t) &=& B_{shape}(\phi) + B_{stress}(t), \\
B_{shape}(\phi) &=& \frac{\mu_0}{2} \, M_s \Omega \lbrack H_k + H_d cos^2\phi \rbrack, \\
H_k &=& (N_{d-yy}-N_{d-zz})M_s, \\
H_d &=& (N_{d-xx}-N_{d-yy})M_s, \\
B_{stress}(t) &=& (3/2) \lambda_s \sigma(t) \Omega, \\
C(t) &=& \frac{\mu_0}{2} M_s^2 \Omega N_{d-zz} - (3/2) \lambda_s \sigma(t) \Omega,
\end{eqnarray}
\end{subequations}
$H_k$ is the Stoner-Wohlfarth switching field,\cite{RefWorks:557} and $H_d$ is the out-of-plane demagnetization field.\cite{RefWorks:157}
Note that $B_{stress}$ has the same sign as the $\lambda _s \sigma$ product. It will be negative if we use stress to rotate the magnetization from the easy axis ($z$-direction) to the plane defined by the in-plane hard axis ($y$-direction) and the out-of-plane hard axis ($x$-direction), i.e., the $x$-$y$ plane ($\theta=90^\circ$).

In the macrospin approximation, the magnetization \textbf{M}(t) of the nanomagnet has a constant magnitude but a variable direction, so that we can represent it by the vector of unit norm $\mathbf{n_m}(t) =\mathbf{M}(t)/|\mathbf{M}| = \mathbf{\hat{e}_r}$ where $\mathbf{\hat{e}_r}$ is the unit vector in the radial direction in the standard spherical coordinate system represented by ($r$,$\theta$,$\phi$). The  unit vectors for the $\theta$- and $\phi$-rotations are denoted by $\mathbf{\hat{e}_\theta}$ and $\mathbf{\hat{e}_\phi}$, respectively. 

\begin{figure*}
\centering
\includegraphics[width=\textwidth]{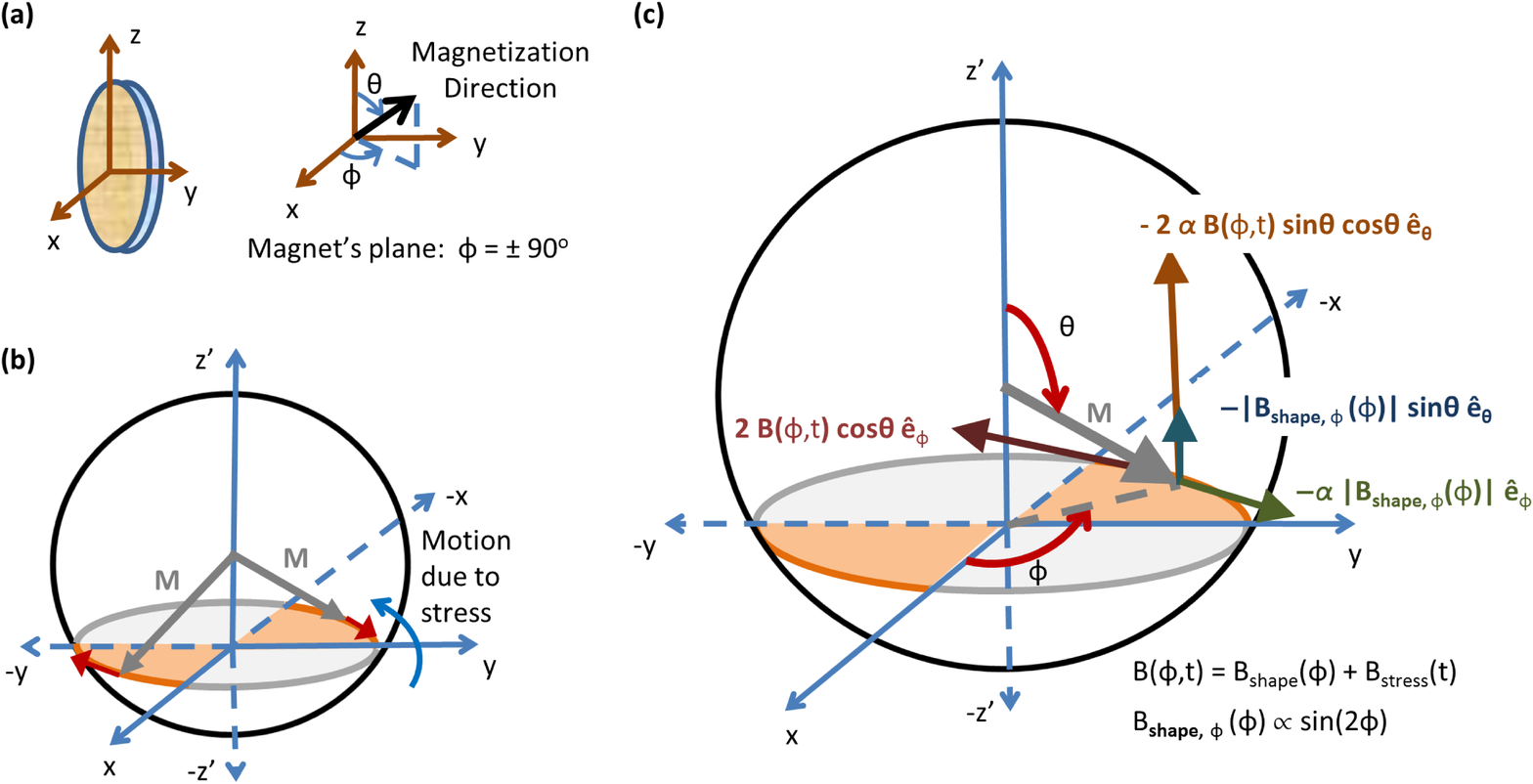}
\caption{\label{fig:dynamics_illustration}
Illustration of magnetization's motion in three-dimensional space.
(a) Cross section of the nanomagnet and axes assignment.
(b) The applied stress tries to lift the magnetization \textbf{M} out of the magnet's plane while the $\hat{\mathbf{e}}_\phi$-component of the shape anisotropy torque due to Gilbert damping tries to bring it back to the plane ($\phi=\pm90^\circ$). This stabilizes the value of $\phi$, but it happens only in the ``good'' quadrants $\phi \in$ ($90^{\circ}$, $180^{\circ}$) and $\phi \in$ ($270^{\circ}$, $360^{\circ}$). For the other two quadrants $\phi \in$ ($0^{\circ}$, $90^{\circ}$) and $\phi \in$ ($180^{\circ}$, $270^{\circ}$), such counteraction does not happen and we term them as ``bad'' quadrants.
(c) Illustration of the motion of magnetization \textbf{M} in three-dimensional space under various torques generated due to shape and stress anisotropy alongwith considering the damping of magnetization ($\alpha$ is the phenomenological damping parameter). Note that the dependence of shape anisotropy energy on $\phi$ has generated two additional motions $-|B_{shape,\phi}(\phi)|sin\,\theta\,\hat{\mathbf{e}}_\theta$ and $-\alpha |B_{shape,\phi}(\phi)|\,\hat{\mathbf{e}}_\phi$. (See text for details.) The quadrant $\phi \in$ ($90^{\circ}$, $180^{\circ}$) is chosen for illustration; choice of the other ``good'' quadrant $\phi \in$ ($270^{\circ}$, $360^{\circ}$) is analogous. (Reprinted from Ref.~\srepref. $B(\phi)$ and $B_{stress}$ are replaced by $B(\phi,t)$ and $B_{stress}(t)$, respectively.)
}
\end{figure*}


The torque due to shape and stress anisotropy is derived from the gradient of potential profile as 
\begin{eqnarray}
\mathbf{T_E}(\theta,\phi,t) &=& - \mathbf{n_m} \times \nabla E(\theta,\phi,t) \nonumber\\
&=& - \mathbf{\hat{e}_r} \times \left( \cfrac{\partial E}{\partial \theta} \, \mathbf{\hat{e}_\theta} + \cfrac{1}{sin\theta} \,\cfrac{\partial E}{\partial \phi} \, \mathbf{\hat{e}_\phi} \right) \nonumber\\
&=& - 2 B(\phi,t) sin\theta \, cos\theta \,\mathbf{\hat{e}_\phi}  \nonumber\\
&& \qquad - B_{shape,\phi}(\phi) sin\theta \,\mathbf{\hat{e}_\theta}, 	 \label{eq:stress_torque}
\end{eqnarray}
where 
\begin{equation}
B_{shape,\phi}(\phi)=\frac{\mu_0}{2} \, M_s^2 \Omega (N_{d-xx}-N_{d-yy}) sin(2\phi). \label{eq:B0e}
\end{equation}

The effect of random thermal fluctuations is incorporated via a random magnetic field $\mathbf{h}(t)= h_x(t)\mathbf{\hat{e}_x} + h_y(t)\mathbf{\hat{e}_y} + h_z(t)\mathbf{\hat{e}_z}$, where $h_i(t)$ ($i=x,y,z$) are the three components of the random thermal field in Cartesian coordinates. We assume the properties of the random field $\mathbf{h}(t)$ as described in Ref.~\refcite{RefWorks:186}. The random thermal field can be written as\cite{RefWorks:186}
\begin{equation}
h_i(t) = \sqrt{\frac{2 \alpha kT}{|\gamma| M_V \Delta t}} \; G_{(0,1)}(t) \quad (i \in x,y,z),
\label{eq:ht}
\end{equation}
\noindent
where $\alpha$ is the dimensionless phenomenological Gilbert damping parameter, $\gamma$ is the gyromagnetic ratio for electrons, $1/\Delta t$ is the attempt frequency of thermal fluctuations, $M_V=\mu_0 M_s \Omega$, $\Omega$ is the volume, $k$ is the Boltzmann constant, $T$ is temperature, and the quantity $G_{(0,1)}(t)$ is a Gaussian distribution with zero mean and unit variance. 

The thermal field and the corresponding torque acting on the magnetization can be written as 
\begin{equation}
\mathbf{H_{TH}}(\theta,\phi,t)=P_\theta(\theta,\phi,t)\,\mathbf{\hat{e}_\theta}+P_\phi(\theta,\phi,t)\,\mathbf{\hat{e}_\phi},
\end{equation}
and
\begin{eqnarray}
\mathbf{T_{TH}}(\theta,\phi,t) &=&\mathbf{n_m} \times \mathbf{H_{TH}}(\theta,\phi,t)\nonumber\\
															 &=&P_\theta(\theta,\phi,t)\,\mathbf{\hat{e}_\phi}-P_\phi(\theta,\phi,t)\,\mathbf{\hat{e}_\theta},\, \label{eq:T_TH}
\end{eqnarray}
respectively, where
\begin{eqnarray}
P_\theta(\theta,\phi,t) &=& M_V  \lbrack h_x(t)\,cos\theta\,cos\phi + h_y(t)\,cos\theta sin\phi \nonumber\\
												&& \quad- h_z(t)\,sin\theta \rbrack, \label{eq:P_theta}\\
P_\phi(\theta,\phi,t) &=& M_V  \lbrack h_y(t)\,cos\phi -h_x(t)\,sin\phi\rbrack \label{eq:P_phi}.
\label{eq:thermal_parts}
\end{eqnarray}

Additionally, there is motion due to Gilbert damping\cite{RefWorks:162,RefWorks:161} (perpendicular to the precessional motion) through which magnetization relaxes towards the minimum energy position on magnet's potential landscape. The magnetization dynamics under the action of these two torques $\mathbf{T_{E}}$ and $\mathbf{T_{TH}}$ is described by the stochastic Landau-Lifshitz-Gilbert (LLG) equation as follows.

\begin{equation}
\frac{d\mathbf{n_m}}{dt} - \alpha \left(\mathbf{n_m} \times \frac{d\mathbf{n_m}}{dt} \right) = -\frac{|\gamma|}{M_V} \, \left\lbrack \mathbf{T_E} +  \mathbf{T_{TH}}\right\rbrack,
\label{LLG}
\end{equation}
where $\alpha$ is the phenomenological Gilbert damping parameter and $\gamma$ is the gyromagnetic ratio of electrons. Solving the above equations, we get the following coupled equations for the dynamics of $\theta$ and $\phi$:
\begin{multline}
\left(1+\alpha^2 \right) \cfrac{d\theta}{dt} = \frac{|\gamma|}{M_V} \lbrack B_{shape,\phi}(\phi) sin\theta \\ - 2\alpha B(\phi,t) sin\theta \, cos\theta  \\ + \left(\alpha P_\theta(\theta,\phi,t) + P_\phi(\theta,\phi,t)\right)\rbrack,
\label{eq:theta_dynamics}
\end{multline}
\begin{multline}
\left(1+\alpha^2 \right) \cfrac{d \phi}{dt} = \frac{|\gamma|}{M_V} \lbrack \alpha B_{shape,\phi}(\phi) + 2 B(\phi,t) cos\theta 
\\ - \{sin\theta\}^{-1} \left(P_\theta(\theta,\phi,t) - \alpha P_\phi(\theta,\phi,t) \right) \rbrack \quad (sin\theta \neq 0).
	\label{eq:phi_dynamics}
\end{multline}

We will ignore the random thermal torque due to room-temperature thermal fluctuations while explaining the magnetization dynamics first, following Ref.~\refcite{roy13_2}, however, we will discuss the key consequences of incorporating thermal fluctuations with simulation results later. We assume that the magnetization starts from $\theta \simeq 180^\circ$ ($-z$-axis) and the applied stress attempts to switch it to $\theta \simeq 0^\circ$ ($+z$-axis). 

\begin{figure*}
\centering
\includegraphics[width=\textwidth]{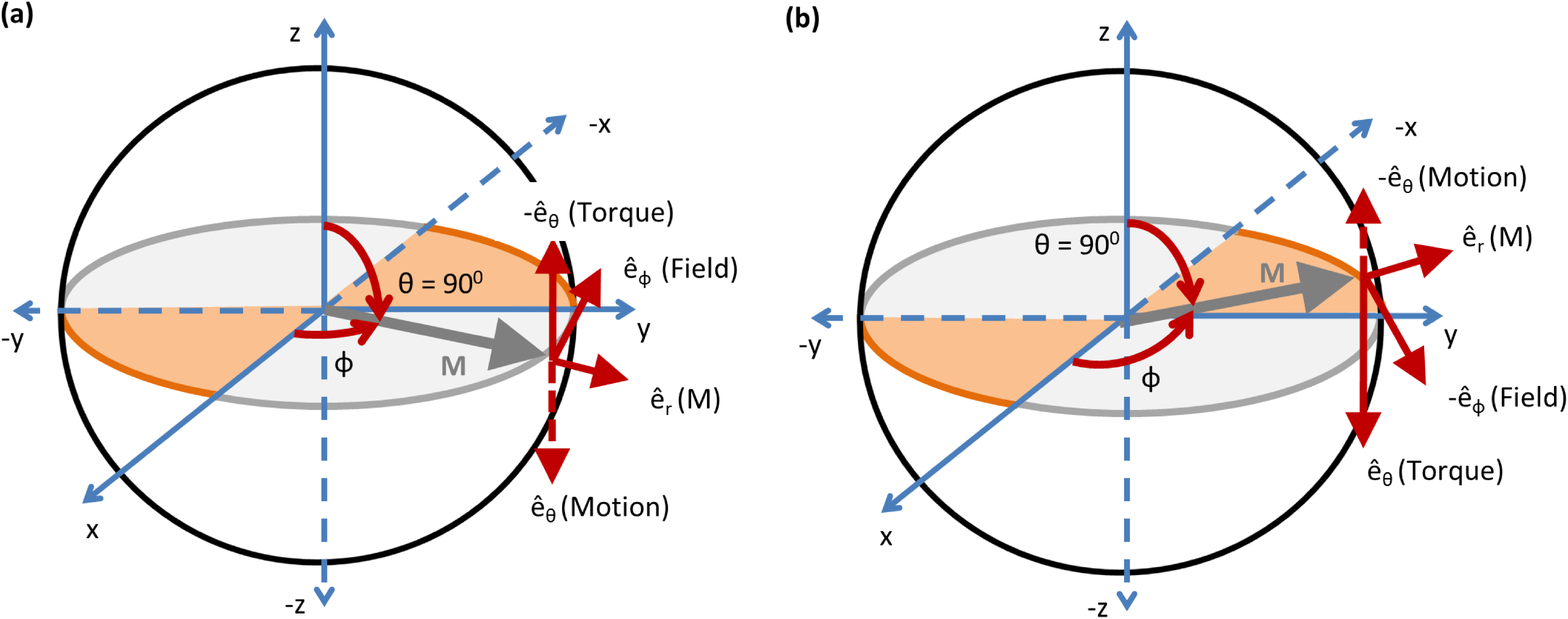}
\caption{\label{fig:motion_illustration_good_bad}
Field and torque acting on the magnetization \textbf{M} when it comes on the $x$-$y$ plane ($\theta = 90^\circ$).
(a) $\phi \in (0^\circ, 90^\circ)$. The field always tries to keep the magnetization on magnet's plane ($\phi=\pm90^\circ$). For this ``bad'' quadrant $\phi \in (0^\circ, 90^\circ)$, magnetization backtracks towards $\theta \simeq 180^\circ$ causing a switching failure. Choice of the other ``bad'' quadrant $\phi \in (180^\circ, 270^\circ)$ is analogous. 
(b) $\phi \in (90^\circ, 180^\circ)$. The field again tries to keep the magnetization on magnet's plane ($\phi=\pm90^\circ$). For this ``good'' quadrant $\phi \in (90^\circ, 180^\circ)$, magnetization can traverse towards its destination $\theta \simeq 0^\circ$. Choice of the other ``good'' quadrant $\phi \in (270^\circ, 360^\circ)$ is analogous. Note that the motion of magnetization is opposite to the direction of torque exerted on it since the Land\'{e} $g$-factor for electrons is negative. If magnetization starts from the other easy axis $\theta \simeq 0^\circ$ and we switch it towards $\theta \simeq 180^\circ$, the roles of the four quadrants of $\phi$ would have been exactly opposite. (Reprinted from Ref.~\srepref.)}
\end{figure*}

\textit{Out-of-plane excursion of magnetization.} We will first intuitively describe how magnetization is deflected from the magnet's plane ($\phi=\pm90^\circ$, i.e., $y$-$z$ plane), and is stabilized out-of-plane as depicted in Fig.~\ref{fig:dynamics_illustration}(b) due to different torques acting on it. The shape anisotropy energy is in general dependent on azimuthal angle $\phi$ (rather than assuming $\phi=\pm90^\circ$) and it generates additional motions of magnetization in $\hat{\mathbf{e}}_\theta$ and $\hat{\mathbf{e}}_\phi$ directions [see the motions containing the term $B_{shape,\phi}(\phi)$ in Fig.~\ref{fig:dynamics_illustration}(c) and equations~(\ref{eq:theta_dynamics}) and~(\ref{eq:phi_dynamics})]. Both of these torques are proportional to $sin(2\phi)$ and vanish when $\phi=\pm90^\circ$. Also, note that the $\phi$-component is proportional to the damping parameter $\alpha$. As shown in the Fig.~\ref{fig:dynamics_illustration}(b), the applied stress generates a torque that attempts to rotate the magnetization anticlockwise and forces the magnetization to deflect from magnet's plane and stay out of magnet's plane. As magnetization is deflected from the plane of the magnet ($\phi=\pm90^\circ$), the $\phi$-component of the torque due to shape anisotropy energy as mentioned earlier [$\propto\,\alpha\,sin(2\phi)$] would attempt to bring the magnetization back to magnet's plane. Because of this \emph{counteraction}, the magnetization is only stable in the second or the fourth quadrant [i.e., ($90^{\circ}$, $180^{\circ}$) or ($270^{\circ}$, $360^{\circ}$)], among the four possible quadrants of $\phi$. Note that $sin(2\phi)$ is a \emph{negative} quantity in these two quadrants and hence the motion due to shape anisotropy energy counteracts the precessional motion due to stress. We would term these two quadrants (second or the fourth) as ``good'' quadrants and the other two (first and third) quadrants as ``bad'' quadrants, the reasoning behind which would be more prominent onwards, i.e., consideration of the torques due to $\phi$-dependence of shape anisotropy energy is crucial to the magnetization dynamics. The key lesson is that $\phi$ becomes stable only in the ``good'' quadrants and it facilitates switching of magnetization in the desired direction.

\textit{Magnetization's motion in three-dimensional space.} We will now describe the motion of magnetization in the full three-dimensional space intuitively in the presence of various torques originating from the shape and stress anisotropy energy as depicted in the Fig.~\ref{fig:dynamics_illustration}(c). Note that the motion of magnetization needs to be along the $-\hat{\mathbf{e}}_\theta$ direction since magnetization is being switched from $\theta \simeq 180^\circ$ towards $\theta \simeq 0^\circ$. The applied stress generates a precessional motion of magnetization in the +$\hat{\mathbf{e}}_\phi$ direction, and the damping of magnetization generates a motion additionally that is perpendicular to both the direction of magnetization ($\hat{\mathbf{e}}_r$) and +$\hat{\mathbf{e}}_\phi$, i.e., in $-\hat{\mathbf{e}}_\theta$ direction. These two motions are shown as $2B(\phi,t)cos\theta\,\hat{\mathbf{e}}_\phi$ and $-2\alpha B(\phi,t)sin\theta cos\theta\,\hat{\mathbf{e}}_\theta$, respectively in Fig.~\ref{fig:dynamics_illustration}(c), where $\alpha$ is the damping parameter and the quantity $B(\phi,t)$ includes terms both due to the shape anisotropy energy $B_{shape}(\phi)$ and the stress anisotropy energy $B_{stress}(t)$. The quantity $B_{stress}(t)$ is negative and it must overcome the shape anisotropy term $B_{shape}(\phi)$ for switching to get started (mathematically, note that both the quantities $B(\phi,t)$ and $cos\theta$ are negative in the interval $180^\circ \geq \theta \geq 90^\circ$). Therefore, magnetization starts switching towards its desired direction due to the applied stress. Note that this damped motion in $-\hat{\mathbf{e}}_\theta$ direction is considerably weak because of the multiplicative damping parameter $\alpha$, which is usually much less than one (e.g., $\alpha$=0.1 for Terfenol-D). 

As described earlier [see Fig.~\ref{fig:dynamics_illustration}(b)], due to applied stress, magnetization rotates out-of-plane and stays in a ``good'' quadrant for $\phi$ [i.e., ($90^{\circ}$, $180^{\circ}$) or ($270^{\circ}$, $360^{\circ}$)], which generates a motion of magnetization in the $-\hat{\mathbf{e}}_\theta$ direction due to $\phi$-dependence of shape anisotropy energy. Thereby a damped motion is generated in the $-\hat{\mathbf{e}}_\phi$ direction. These two motions are shown as $-|B_{shape,\phi}(\phi)|sin\theta\,\hat{\mathbf{e}}_\theta$ and $-\alpha |B_{shape,\phi}(\phi)|\,\hat{\mathbf{e}}_\phi$, respectively in Fig.~\ref{fig:dynamics_illustration}(c), where $B_{shape,\phi}(\phi) \propto sin(2\phi)$. In the ``good'' quadrants for $\phi$, $B_{shape,\phi}(\phi)$ is negative, therefore, if the magnetization stays out of magnet's plane in a ``good'' quadrant, magnetization rotates in its desired direction. Since this motion does not possess any damping factor [note the other damped motion in the $-\hat{\mathbf{e}}_\theta$ direction in Fig.~\ref{fig:dynamics_illustration}(c)], it can eventually increase the magnetization switching speed to a couple of orders of magnitude higher. On the other hand, if the magnetization resides out-of-plane but in a ``bad'' quadrant, the motion of magnetization in its desired direction of switching is hindered. If we apply a higher magnitude of stress, the magnetization is deflected out of magnet's plane \emph{more} inside a ``good'' quadrant (counteracting the random thermal kicks possibly acting in the opposite direction, which will be described later). Note that the damped motion $-\alpha |B_{shape,\phi}(\phi)|\,\hat{\mathbf{e}}_\phi$ attempts to bring magnetization back towards the magnet's plane. As these two motions counteract each other [see Fig.~\ref{fig:dynamics_illustration}(c)], magnetization continues moving in the $-\hat{\mathbf{e}}_\theta$ direction and eventually reaches the $x$-$y$ plane ($\theta=90^\circ$). Note that without damping, such counteraction does not happen and the magnetization just precesses through ``good'' and ``bad'' quadrants consecutively. 

\begin{figure*}
\centering
\includegraphics[width=\textwidth]{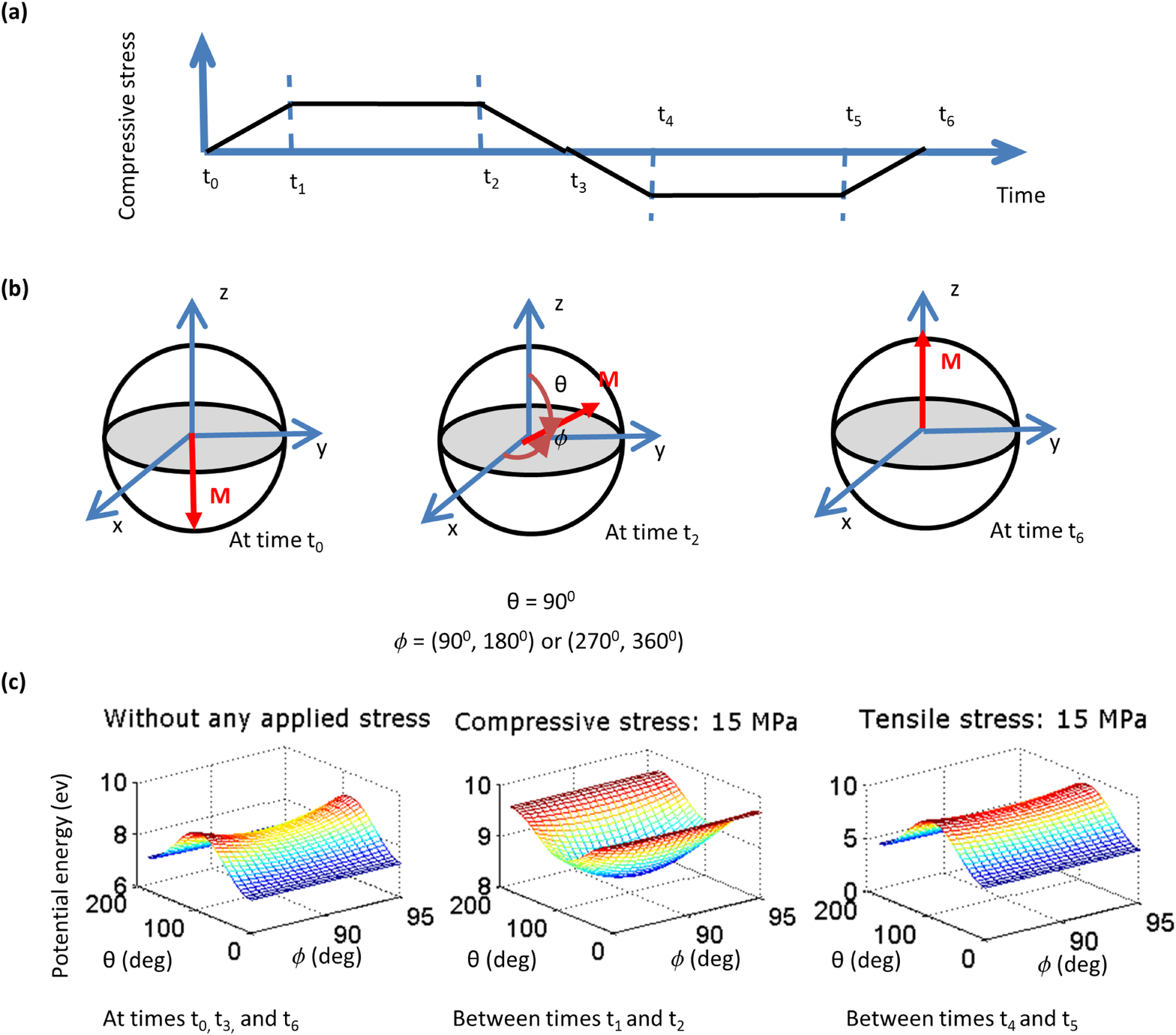}
\caption{\label{fig:stress_cycle_all} Stress cycle, magnetization directions, and potential profiles at different time instants during switching of magnetization. (a) Stress-cycle on the magnetostrictive nanomagnet. (b) Magnetization directions at different instants of time. (c) Potential landscapes of the magnetostrictive nanomagnet in relaxed, compressively stressed, and expansively stressed conditions. Note that the three-dimensional potential landscape has never been made asymmetric to favor the final state during switching. (Reprinted from Ref.~\srepref.)}
\end{figure*}

Upon reaching the $x$-$y$ plane ($\theta=90^\circ$), if magnetization stays in a ``good'' quadrant for $\phi$ [i.e., $(90^\circ,180^\circ)$ or $(270^\circ,360^\circ)$], then the torque on the magnetization will be in the correct direction to facilitate magnetization's traversal towards $\theta \simeq 0^\circ$ [see Figs.~\ref{fig:motion_illustration_good_bad}(a) and~\ref{fig:motion_illustration_good_bad}(b)]. Once again this signifies the merit of terminology (``good'' or ``bad'') used for the four quadrants of $\phi$. At $\theta=90^\circ$ (i.e., $cos\theta=0$), the effect of stress on the magnetization rotation has diminished completely [see Fig.~\ref{fig:dynamics_illustration}(c)]. The only two motions that are active at $\theta=90^\circ$ are $-|B_{shape,\phi}(\phi)|sin\,\theta\,\hat{\mathbf{e}}_\theta$ and $-\alpha |B_{shape,\phi}(\phi)|\,\hat{\mathbf{e}}_\phi$ [Fig.~\ref{fig:dynamics_illustration}(c)]. Since $\alpha \ll 1$, magnetization quickly gets out from $\theta = 90^\circ$ and as the magnetization vector is deflected from $\theta=90^\circ$ towards $\theta=0^\circ$, the effect of stress again comes into play.

\textit{Stress cycle, magnetization directions, and potential profiles.} Fig.~\ref{fig:stress_cycle_all} shows the stress-cycle alongwith the energy profiles and magnetization directions at different instants of time during switching of magnetization. At time $t_0$, the magnetization direction is along the easy axis $\theta \simeq 180^\circ$ with itse potential landscape unperturbed by stress. Note that the potential profile of the magnet is `symmetric' in both $\theta$- and $\phi$-space with two degenerate minima at $\theta = 0^{\circ}$, 180$^{\circ}$ and a maximum at $\theta = 90^{\circ}$ in $\theta$-space, signifying that a binary information can be stored in the nanomagnet corresponding to $\theta = 0^{\circ}$, 180$^{\circ}$. The anisotropy in the barrier is due to shape anisotropy energy of the nanomagnet only, which is $\sim$44 kT at room-temperature using the nanomagnet's dimensions and the material parameters used for the magnetostrictive nanomagnet made of Terfenol-D (see Section~\ref{sec:results} later). Note that the barrier height separating the two stable states ($\theta$ = $0^\circ$ and $180^\circ$) is meant when the magnetization stays in-plane (i.e., $\phi=\pm90^\circ$) of the magnet. The barrier becomes higher when the magnetization is deflected from $\phi=90^\circ$ as shown in the Fig.~\ref{fig:stress_cycle_all}(c) at time $t_0$. The barrier is the highest when the magnetization points along the out-of-plane direction ($\phi$ = $0^\circ$ or $180^\circ$), which is due to the small thickness of the nanomagnet compared to the lateral dimensions. Note that the magnetization can start from any angle $\phi_{initial} \in (0^\circ, 360^\circ)$ in the presence of thermal fluctuations [see Fig.~\ref{fig:thermal_theta_phi_distribution_terfenolD_1000ns}(b) later].

\begin{figure*}
\centering
\includegraphics[width=\textwidth]{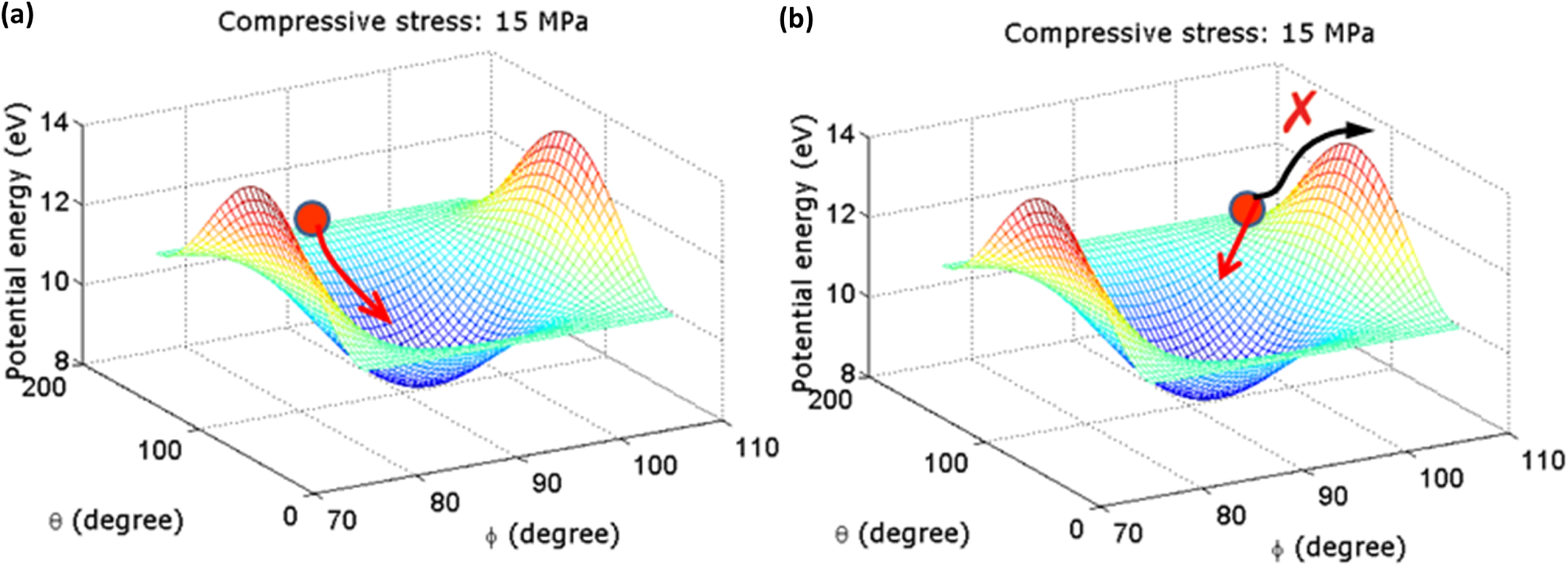}
\caption{\label{fig:motion_phi_init_in_out_of_plane} Illustration of magnetization's motion when magnetization starts switching out of magnet's plane ($\phi \neq 90^\circ$), which can happen due to thermal fluctuations [see Fig.~\ref{fig:thermal_theta_phi_distribution_terfenolD_1000ns}(b)] and the high shape-anisotropy energy barrier therein cannot be overcome by stress anisotropy. 
(a) Magnetization starts in-plane of the magnet ($\phi = 90^\circ$), where the potential landscape is inverted by the stress anisotropy and thus magnetization does not face a potential hill while starting to switch.
(b) Magnetization starts from out-of-plane of the magnet ($\phi \neq 90^\circ$), where the potential landscape cannot be inverted by the stress anisotropy and thus magnetization does face a potential hill at start. However, due to $\phi$-motion of magnetization, it eventually surpasses the potential hill and comes near to magnet's plane, where from it can start switching in $\theta$-space.
(Reprinted from Ref.~\srepref.)}
\end{figure*}

Fig.~\ref{fig:stress_cycle_all}(c) depicts that as a compressive stress is ramped up on the nanomagnet between time instants $t_0$ and $t_1$ and a sufficient stress is applied, the potential landscape in $\theta$-space becomes monostable near $\phi=\pm90^\circ$. Since the barrier height is high near $\phi=0^\circ$ or $180^\circ$, the potential landscape may not become monostable in $\theta$-space therein. However, that is not necessary for switching since application of stress rotates the magnetization in $\phi$-direction and theretofore the magnetization can eventually come near $\phi=\pm90^\circ$, which facilitates switching from $\theta \simeq 180^\circ$ towards $\theta=90^\circ$ (see Fig.~\ref{fig:motion_phi_init_in_out_of_plane}). From Fig.~\ref{fig:stress_cycle_all}(c), we can see that the minimum energy position between time instants $t_1$ and $t_2$ is at ($\theta=90^\circ$, $\phi=\pm90^\circ$) and the potential profile at time instant $t_1$ is still `symmetric'.

Fig.~\ref{fig:stress_cycle_all} shows that stress is held constant between time instants $t_1$ and $t_2$ and the magnetization eventually reaches at $x$-$y$ plane ($\theta=90^\circ$). For a sufficiently fast ramp rate and high stress, magnetization will reside in ``good'' quadrants (see Fig.~\ref{fig:thermal_stress_success_ramp_time_mag} later). This will ensure that magnetization traverses in the correct direction towards $\theta = 0^\circ$ and switches successfully. A sufficiently fast ramp rate ensures that magnetization would not backtrack towards $\theta=180^\circ$ even after crossing $\theta=90^\circ$ towards $\theta=0^\circ$. If stress is held constant for longer time, magnetization will have higher probability to collapse on magnet's plane ($\phi=\pm90^\circ$) following which thermal fluctuations will scuttle the magnetization either in a ``good'' quadrant or in a ``bad'' quadrant with equal probability, so the success rate would be 50\%.


It should be noted from Fig.~\ref{fig:stress_cycle_all} that the stress is reversed (compression to tensile) between time instants $t_2$ and $t_4$ rather than just withdrawn, which makes the potential landscape of the nanomagnet more steep in $\theta$-space. However, it does not necessarily mean that the switching will be completed always faster. The reversal of stress can cause magnetization to traverse into ``bad'' quadrants in $\phi$-space causing magnetization to precess and therefore the switching delay may eventually increase. Particularly for higher stress levels, such increase in switching delay may happen. However, it is noticed that reversing the stress makes the success rate of switching a bit ($<$5\%) higher in the presence of thermal fluctuations. The tensile stress is held constant and when $\theta$ becomes $ \leq 5^\circ$, switching is deemed to have completed. The stress is withdrawn at the end between time instants $t_5$ and $t_6$ to complete the switching cycle.

\begin{figure*}
\centering
\includegraphics[width=\textwidth]{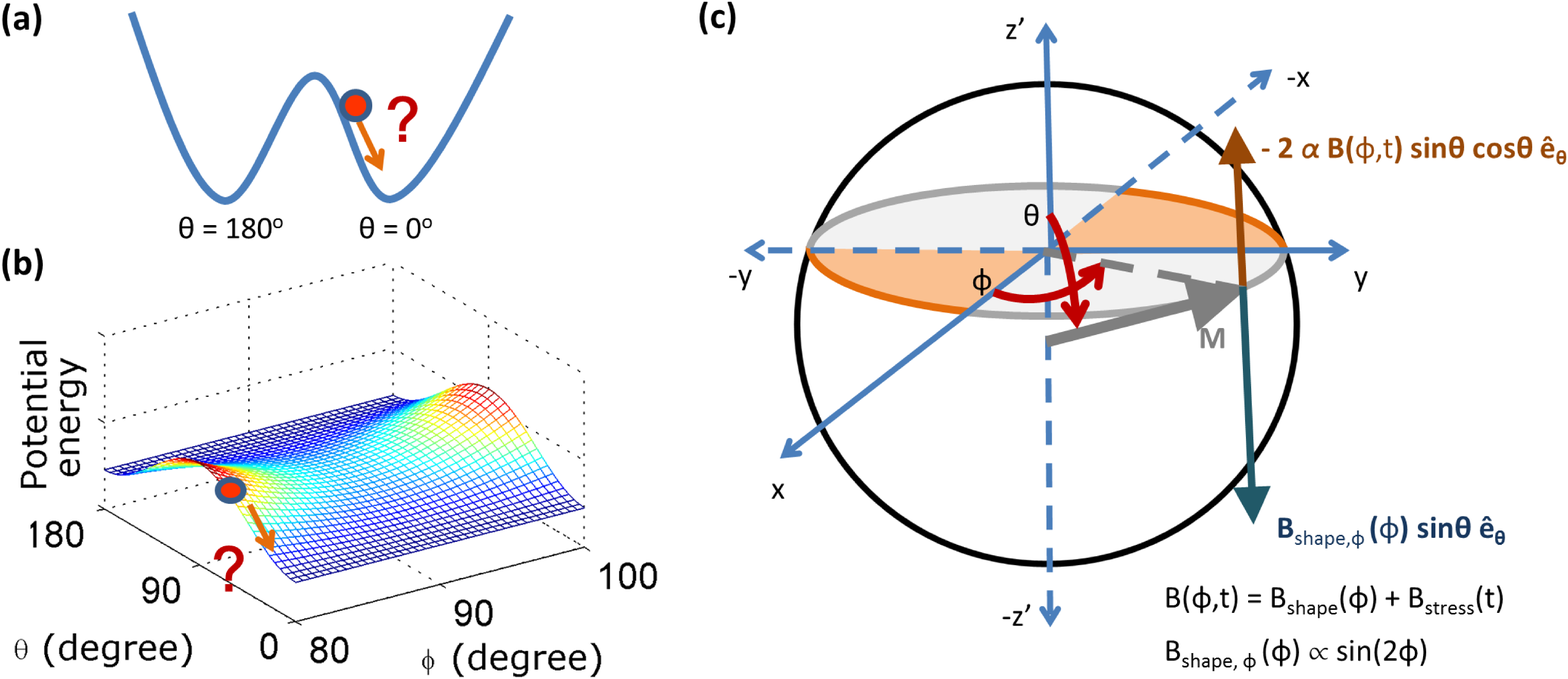}
\caption{\label{fig:theta_phi_motion_bad_quad_fails_illustration} 
Magnetization can backtrack even after it has crossed the hard axis towards its destination.
(a) Magnetization has started from $\theta \simeq 180^\circ$ and crossed the hard axis $\theta = 90^\circ$, but it is well possible that magnetization backtracks towards $\theta \simeq 180^\circ$ even without considering the presence of thermal fluctuations. Looking at the two-dimensional magnet's plane and considering two-dimensional motion of magnetization on magnet's plane, this seems unreasonable in the absence of thermal fluctuations. 
(b) Full three-dimensional potential landscape of magnetization. Considering the complete three-dimensional motion of magnetization in this potential landscape of the nanomagnet, switching failure may be plausible even in the absence of thermal fluctuations since there is consequence of out-of-plane motion of magnetization.
(c) Explanation behind magnetization's backtracking even after it has crossed the hard axis towards its destination. Magnetization may switch to the incorrect direction because it is in a ``bad'' quadrant for $\phi$ and there is a motion of magnetization $B_{shape, \phi}(\phi)\,sin\theta\,\hat{\mathbf{e}}_\theta$ in the unintended direction. The other magnetization's motion $-2\alpha B(\phi,t)sin\theta cos\theta\,\hat{\mathbf{e}}_\theta$ due to damping is in the intended direction but it may be quite small compared to the other motion and thus magnetization may well backtrack. The quadrant $\phi \in$ ($0^{\circ}$, $90^{\circ}$) is chosen for illustration; choice of the other ``bad'' quadrant $\phi \in$ ($180^{\circ}$, $270^{\circ}$) is analogous. (Parts (a) and (c) are reprinted from Ref.~\srepref. In part (c), $B(\phi)$ and $B_{stress}$ are replaced by $B(\phi,t)$ and $B_{stress}(t)$, respectively.)}
\end{figure*}

\textit{Switching failure.} As magnetization leaves from $\theta=90^\circ$ towards $\theta \simeq 0^\circ$ and stress is ramped down, the torque due to stress tries to rotate the azimuthal angle $\phi$ of magnetization \emph{clockwise} rather than anticlockwise. Mathematically, note that $cos\theta$ is positive for $90^\circ \geq \theta \geq 0^\circ$ and $B(\phi,t)$ is still negative when stress has not been brought down sufficiently, i.e., still $|B_{stress}(t)| > |B_{shape}(\phi)|$ (see Fig.~\ref{fig:dynamics_illustration}(c) and corresponding discussions in text). For a \emph{slow ramp-rate} this \emph{clockwise} rotation may be considerable and magnetization can stray into a ``bad'' quadrant. Moreover, thermal fluctuations can aggravate the scenario by possibly deflecting the magnetization into the ``bad'' quadrant further. Switching may impede and magnetization vector may backtrack towards $\theta \simeq 180^\circ$ rather than traversing towards $\theta \simeq 0^\circ$ causing a switching failure. Therefore, switching failure may happen even after the magnetization has crossed the hard axis ($\theta = 90^\circ$) towards its destination $\theta \simeq 0^\circ$ (see Fig.~\ref{fig:theta_phi_motion_bad_quad_fails_illustration}). This is why it does require a sufficiently fast ramp rate during the ramp-down phase of stress. Such switching failure is an intriguing phenomenon, which cannot be conceived if magnetization is always \emph{assumed} to be on magnet's plane ($\phi=\pm90^\circ$) and only can be understood if we analyze the magnetization dynamics in a complete three dimensional potential landscape.

\textit{Fluctuation of magnetization around the easy axis (stable orientation) due to thermal noise.} From Equation (\ref{eq:stress_torque}), it can be noted that the torque on the magnetization due to shape and stress anisotropy vanishes when $sin \theta = 0$, i.e. when the magnetization vector is \emph{exactly} aligned along the easy axis ($\theta = 0^{\circ}$, 180$^{\circ}$).\cite{roy11_6,roy13_2} However, thermal fluctuations can deflect the magnetization vector from the easy axis. Considering the case when $\theta=180^\circ$, we get the following:
\begin{equation}
\phi(t) = tan^{-1} \left( \frac{\alpha h_y(t) + h_x(t)}{h_y(t) - \alpha h_x(t)} \right),
\label{eq:phi_t_thermal}
\end{equation}
\begin{equation}
\frac{d\theta}{dt} = -|\gamma| \frac{h_x^2(t) + h_y^2(t)}{\sqrt{(h_y(t)-\alpha h_x(t))^2 + (\alpha h_y(t) + h_x(t))^2}}.
\end{equation}
\noindent
From the above equation we can clearly follow that thermal torque can deflect the magnetization from the easy axis since the time rate of change of $\theta(t)$ is non-zero in the presence of thermal agitations. The initial deflection from the easy axis due to the thermal torque does not depend on the component of the random thermal field along the $z$-axis, i.e., $h_z(t)$. This is a consequence of having $\pm$$z$-axis as the easy axes of the nanomagnet. However, once the magnetization direction is even slightly deflected from the easy axis, all the three components of the random thermal field come into play. 

\textit{Thermal distribution of the initial orientation of the magnetization.} We can determine the thermal distributions of $\theta$ and $\phi$ when no stress is applied on the magnetostrictive nanomagnet by solving the Equations~(\ref{eq:theta_dynamics}) and~(\ref{eq:phi_dynamics}) with $B_{stress}$ = 0 (Refs.~\refcite{roy11_6,roy13_2}). This will yield the distribution of the magnetization's initial orientation when stress is turned on. The $\theta_{initial}$-distribution is Boltzmann peaked at $\theta$ = 0$^{\circ}$ or 180$^{\circ}$, while the $\phi_{initial}$-distribution is Gaussian peaked at $\phi = \pm 90^{\circ}$ (see Fig.~\ref{fig:thermal_theta_phi_distribution_terfenolD_1000ns}). According to the Boltzmann distribution of $\theta_{initial}$, the most probable value of $\theta$ is either 0$^{\circ}$ or 180$^{\circ}$, where stress is ineffective. This will lead to a long tail in the switching delay distribution, which is due to the fact that when magnetization starts out from $\theta = 0^{\circ}, 180^{\circ}$, it needs to wait a while before random thermal fluctuations can set the switching in motion. Thus, switching trajectories starting from an easy axis are very slow and it causes the long tail in the distribution of switching delay.\cite{roy11_6,roy13_2}

\textit{Application of a bias field to shift the peak of the initial distribution of magnetization from an easy axis.} We can eliminate the long tail in the switching delay distribution by applying a small static bias magnetic field that shifts the peak of $\theta_{initial}$ distribution away from the easy axis, so that the most probable starting orientation will no longer be an easy axis.\cite{roy11_6} This field is applied along the out-of-plane hard axis (+$x$-direction) of the nanomagnet and thus the potential energy due to the applied magnetic field becomes 
\begin{equation}
E_{mag}(t) = - M_V H\, sin\theta(t)\,cos\phi(t),
\end{equation}
where $H$ is the magnitude of magnetic field. A torque is generated due to this field, which is 
\begin{equation}
\mathbf{T_M} (t) = - \mathbf{n_m}(t) \times \nabla E_{mag}(\theta,\phi). 
\end{equation}
The presence of this torque will modify Equations~(\ref{eq:theta_dynamics}) and~(\ref{eq:phi_dynamics})  to 
\begin{multline}
\left(1+\alpha^2 \right) \cfrac{d\theta}{dt} = \frac{|\gamma|}{M_V} \lbrack 
 B_{shape,\phi}(t) sin\theta \\ - 2\alpha B(\phi,t) sin\theta cos\theta 
 \\  + \alpha M_V H\, cos\theta\,cos\phi - M_V H\, sin\phi 
 \\ + \left(\alpha P_\theta(\theta,\phi,t) + P_\phi(\theta,\phi,t) \right) \rbrack,
 \label{eq:theta_dynamics_mag}
\end{multline}
\begin{multline}
\left(1+\alpha^2 \right) \cfrac{d \phi}{dt} = \frac{|\gamma|}{M_V} 
\lbrack \alpha B_{shape,\phi}(t) + 2 B(\phi,t) cos\theta
\\ - {[sin\theta]^{-1}} \left(M_V H\, cos\theta\,cos\phi + \alpha M_V H\, sin\phi  \right)
\\ - {[sin\theta]^{-1}} \left(P_\theta(\theta,\phi,t) - \alpha P_\phi(\theta,\phi,t) \right) \rbrack \quad	(sin\theta \neq 0).
\label{eq:phi_dynamics_mag}
\end{multline}
\noindent

Note that the bias field also makes the potential energy profile of the magnet \emph{asymmetric} in $\phi$-space and the energy minimum is shifted from $\phi_{min}=\pm90^\circ$ (the plane of the magnet) to
\begin{equation}
\phi_{min} = cos^{-1}\left\lbrack \frac{H}{M_s (N_{d-xx} - N_{d-yy})} \right\rbrack.
\end{equation}
\noindent
However, the potential profile will remain \emph{symmetric} in $\theta$-space, with $\theta=0^\circ$ and $\theta=180^\circ$ remaining as the minimum energy positions. With  the material parameters and the dimensions of the nanomagnet (see Section~\ref{sec:results} later), a bias magnetic field of flux density 40 mT would make $\phi_{min} \simeq \pm87^\circ$, i.e. deflect the magnetization $\sim$3$^{\circ}$ from the magnet's plane. Application of the bias magnetic field also reduces the in-plane shape anisotropy energy barrier from 44 $kT$ to 36 $kT$ at room temperature. We assume that a permanent magnet will be employed to produce the bias magnetic field and \emph{not} by a current-carrying coil on-chip. Therefore no additional energy dissipation needs to be considered for this reason.

\textit{Energy Dissipation.} The energy dissipated during magnetization switching has two components: (1) the energy dissipated in the switching circuit for generating stress in the magnetostrictive nanomagnet with the application of a voltage, and (2) the energy dissipated internally in the nanomagnet due to Gilbert damping. We will term the first component as `$CV^2$' dissipation, where $C$ and $V$ denote the capacitance of the piezoelectric layer and the applied voltage, respectively. If the ramp rate is infinite, i.e., the voltage is turned on or off abruptly, the energy dissipated during either turn on or turn off is $(1/2)CV^2$. However, if the ramp rate is finite, this energy dissipation is reduced and the exact value will depend on the ramp rate. We calculate it following the procedure described in Ref.~\refcite{roy11_2}. 

The second component, which is the internal energy dissipation $E_d$ due to Gilbert damping, is given by the expression $\int_0^{\tau}P_d(t) dt$, where $\tau$ is the switching delay and $P_d(t)$ is the power dissipated during switching
\begin{equation}
P_d(t) = \frac{\alpha \, |\gamma|}{(1+\alpha^2) M_V} \, \left| \mathbf{T_E}(t) + \mathbf{T_M}(t)\right|^2 .
\label{eq:power_dissipation}
\end{equation}
\noindent
We sum up the two energy dissipations `$CV^2$' and $E_d$ to get the total dissipation $E_{total}$. The average power dissipation is $E_{total}/\tau$. There is no net dissipation due to random thermal torque, however, thermal fluctuations do affect both $E_d$ and `$CV^2$' dissipations. It affects $E_d$ since it raises the critical stress needed to switch with $\sim$100\% probability and `$CV^2$' dissipation is also raised since the applied voltage is proportional to stress. 

\vspace*{2mm}
So far the model presented deals with the switching of magnetization (i.e., writing a bit of information), however, the magnetization state needs to be read too, which is usually performed by using a magnetic tunnel junction (MTJ).\cite{RefWorks:577,RefWorks:555,RefWorks:572,RefWorks:76,RefWorks:74,RefWorks:33,RefWorks:300} While reading the bit of information, a material selection issue crops up since magnetostrictive materials in general cannot be utilized as the free layer of the MTJ. This is an important issue since we need to achieve a high magnetoresistance for successful read operation as required by technological applications. In Ref.~\refcite{roy15_1}, it is shown that magnetically coupling the magnetostrictive nanomagnet and the free layer can circumvent this issue. Stochastic Landau-Lifshitz-Gilbert equation of magnetization dynamics in the presence of room-temperature thermal fluctuations is solved and it is shown that such design can eventually lead to a superior energy-delay product.\cite{roy15_1}


\subsection{Interface and exchange coupled multiferroic heterostructures}
We have earlier shown in the Figure~\ref{fig:schematic_interface_coupled} the schematic diagram of the interface and exchange coupled multiferroic heterostructure devices and the axis assignment for the orientation of magnetization $M_1$. The standard spherical coordinate system with $\theta$ as polar angle and $\phi$ as azimuthal angle is chosen. The magnetization $M_1$ orients along $\theta=180^\circ$ if polarization points downward ($P_\downarrow$), while $M_1$ is along $\theta=0^\circ$ if polarization points upward ($P_\uparrow$). The lateral elliptical cross-section of $M_1$ lies on the $y$-$z$ plane ($\phi=\pm90^\circ$) with its major axis pointing to $z$-direction and minor axis in $y$-direction. The dimensions of the nanomagnet $M_1$ along the $z$-, $y$-, and $x$-axis are $a$, $b$, and $l$, respectively. So the magnetic easy axis becomes along the $\pm z$-direction and the nanomagnet's volume $\Omega=(\pi/4)abl$. When magnetization switches between its two stable states ($\theta=0^\circ, 180^\circ$), magnetization being a rotational body deflects out of magnet's plane and any deflection from the magnet's plane ($\phi=\pm90^\circ$) is termed as out-of-plane excursion.

The interface anisotropy energy in the nanomagnet $M_1$ is modeled as\cite{roy14_2}
\begin{equation}
E_{I}(\theta,t) = - M_V H_{I}(t)\,cos\theta,
\label{eq:anisotropy_interface}
\end{equation}
where $M_V=\mu_0 M_s \Omega$, $M_s$ is the saturation magnetization, and $H_I$ is the interfacial anisotropy field. If $H_I=-H_{I,max}$, the magnetization $M_1$ points along $\theta=180^\circ$ and if we vary $H_I$ from $-H_{I,max}$ to $H_{I,max}$, the magnetization orients along $\theta=0^\circ$. The total anisotropy of the magnet is the sum of the interface anisotropy alongwith the other anisotropies like magnetocrystalline anisotropy and shape anisotropy,\cite{roy11_news,roy11_6}. However, since the interfacial anisotropy is strong compared to the other anisotropies, we consider only the interfacial anisotropy (i.e., $E_{total} \simeq E_{I}$) for brevity. 

The magnetization $\mathbf{M}$ of the \emph{single-domain} nanomagnet $M_1$ has a constant magnitude of magnetization but a variable direction. Therefore, the magnetization vector can be represented by the unit vector in the radial direction $\mathbf{\hat{e}_r}$ in spherical coordinate system ($r$,$\theta$,$\phi$), i.e., $\mathbf{n_m} =\mathbf{M}/|\mathbf{M}| = \mathbf{\hat{e}_r}$. The other two unit vectors in the spherical coordinate system are $\mathbf{\hat{e}_\theta}$ and $\mathbf{\hat{e}_\phi}$ for $\theta$- and $\phi$-rotations, respectively. The torque $\mathbf{T_I}$ acting on the magnetization due to interface anisotropy can be derived from the gradient of the energy [see Equation~(\ref{eq:anisotropy_interface})] and is given by
\begin{equation}
\mathbf{T_{I}}(\theta,t) = - \mathbf{n_m} \times \nabla E_I (\theta,t) = - M_V H_{I}(t)\,sin\theta \, \mathbf{\hat{e}_\phi}. 
\label{eq:torque_interface}
\end{equation}
Note that the torque $\mathbf{T_{I}}$ acts along the out-of-plane direction, so that the magnetization can deflect out of magnet's plane ($\phi=\pm90^\circ$). 

The effect of random thermal fluctuations is incorporated via a random magnetic field and the corresponding torque $\mathbf{T_{TH}}$ is given by the Equation~(\ref{eq:T_TH})  (also see the Equations~(\ref{eq:P_theta}) and~(\ref{eq:P_phi}) for the expressions of $P_\theta$ and $P_\phi$).

The magnetization dynamics of of the nanomagnet $M_1$ under the action of the torques $\mathbf{T_{I}}$ and $\mathbf{T_{TH}}$ is described by the stochastic Landau-Lifshitz-Gilbert (LLG) equation as follows.
\begin{equation}
\frac{d\mathbf{n_m}}{dt} - \alpha \left(\mathbf{n_m} \times \frac{d\mathbf{n_m}}{dt} \right)
 = -\frac{|\gamma|}{M_V} \left\lbrack \mathbf{T_I} +  \mathbf{T_{TH}}\right\rbrack.
\end{equation}

Solving the above equation analytically, we get the following coupled equations of magnetization dynamics for $\theta$ and $\phi$:
\begin{multline}
\left(1+\alpha^2 \right) \frac{d\theta}{dt} = \frac{|\gamma|}{M_V} \, \lbrack  - \alpha M_V H_{I}(t)\, sin\theta 
				\\+ \left(\alpha P_\theta(\theta,\phi,t) + P_\phi (\theta,\phi,t) \right) \rbrack,
 \label{eq:theta_dynamics_ic}
\end{multline}
\begin{multline}
\left(1+\alpha^2 \right) \frac{d \phi}{dt} = \frac{|\gamma|}{M_V} \, \lbrack M_V H_{I}(t) - {[sin\theta]^{-1}} 
				\\ \times	\left(P_\theta (\theta,\phi,t) - \alpha P_\phi (\theta,\phi,t) \right) \rbrack 
	\quad (sin\theta \neq 0).
\label{eq:phi_dynamics_ic}
\end{multline}
We solve the above two coupled equations numerically to track the trajectory of magnetization over time, in the presence of room-temperature thermal fluctuations.\cite{roy14_2}

From Equations~(\ref{eq:theta_dynamics_ic}) and~(\ref{eq:phi_dynamics_ic}), we see that the torque acting in the $\phi$-direction is much more stronger than the torque exerted in the $\theta$-direction since the damping parameter $\alpha \ll 1$. Although, the nanomagnet has a small thickness (i.e., $l \ll b < a$ and $N_{d-xx} \gg N_{d-yy} > N_{d-zz}$, where $N_{d-zz}$, $N_{d-yy}$, and $N_{d-xx}$ are the components of the demagnetization factor along the $z$-axis, $y$-axis, and $x$-axis, respectively), magnetization cannot remain on the magnet's plane ($y$-$z$ plane, $\phi=\pm90^\circ$) due to the fact that the interface coupling energy is a few orders of magnitude higher than the shape anisotropy energy. Thus, the magnetization keeps rotating in the $\phi$-direction, but it also traverses towards the anti-parallel direction in $\theta$-space ($\theta \simeq 180^\circ$ to $\theta \simeq 0^\circ$ or vice-versa) due to damping [see Equation~(\ref{eq:theta_dynamics_ic})].

Note that exactly at $\theta=180^\circ$ or $0^\circ$, the torque acting on the magnetization due to interface anisotropy [Equation~(\ref{eq:torque_interface})] is \emph{exactly} zero, however, as described earlier in the Subsection~\ref{sec:model_strain}, thermal fluctuations can scuttle the magnetization from these points to initiate switching. At the very start of switching, the initial orientation of magnetization is not a fixed value rather a distribution due to thermal agitations (as described in the Subsection~\ref{sec:model_strain} too). Such distribution is considered during simulations. Thermal fluctuations affect the magnetization switching during the course of switching too.

\textit{Energy Dissipation.} Due to the application of voltage, we have `$CV^2$' energy dissipation, where $C$ and $V$ denote the capacitance of the ferroelectric layer and the applied voltage, respectively. We calculate it following the procedure described in Ref.~\refcite{roy11_2}. The energy dissipated in the nanomagnet due to Gilbert damping can be expressed as  $E_d = \int_0^{\tau}P_d(t) dt$, where $\tau$ is the switching delay and $P_d(t)$ is the power dissipated at time $t$ given by
\begin{equation}
P_d(t) = \frac{\alpha \, |\gamma|}{(1+\alpha^2) M_V} \, |\mathbf{T_I} (\theta(t), t)|^2 .
\label{eq:power_dissipation_ic}
\end{equation}
Thermal field with mean zero does not cause any net energy dissipation but it causes variability in the energy dissipation $E_d$ by scuttling the trajectory of magnetization.

\section{\label{sec:results}Results and Discussions} 

Here, we will review the simulation results for strain-mediated multiferroic composites,\cite{roy11_6,roy13_2} and interface and exchange coupled multiferroic heterostructures\cite{roy14_2} in subsequent subsections. The coupled sets of equations derived in the previous section are numerically solved and the performance metrics e.g., switching delay, energy dissipation during switching are reported. 

\subsection{Strain-mediated multiferroic composites}
Here we present the simulation results for strain mediated multiferroic composites.\cite{roy13_2,roy11_6} The magnetostrictive layer is  made of polycrystalline Terfenol-D and it has the following material properties -- Young's modulus (Y): 8$\times$10$^{10}$ Pa, magnetostrictive coefficient ($(3/2)\lambda_s$): +90$\times$10$^{-5}$, saturation magnetization ($M_s$):  8$\times$10$^5$ A/m, and Gilbert's damping constant ($\alpha$): 0.1 (Refs.~\refcite{RefWorks:179,RefWorks:176,RefWorks:178,materials}). We choose the dimension of the magnetostrictive layer as 100 nm $\times$ 90 nm $\times$ 6 nm, which ensures that the magnet has a single ferromagnetic domain.\cite{roy11_6,RefWorks:402} The tradeoffs between area, switching delay, and energy dissipation have been comprehensively studied for different dimensions of the nanomagnet in Ref.~\refcite{roy14_5}.

 For the piezoelectric layer, we consider lead-zirconate-titanate (PZT), which has a dielectric constant of 1000.\cite{roy11_6,roy11_news} Since we want any strain generated in the PZT layer is transferred almost completely to the magnetostrictive layer, we assume that the PZT layer is four times thicker than the magnetostrictive layer.\cite{roy11_6,roy11_news} We consider that the maximum strain of 500 ppm can be generated in the PZT layer,\cite{RefWorks:170,RefWorks:563} which would require a voltage of 66.7 mV because $d_{31}$=1.8$\times$10$^{-10}$ m/V for PZT.\cite{pzt2} Assuming this strain is transferred completely to the magnetostrictive layer, the corresponding stress in Terfenol-D is the product of the generated strain ($500\times10^{-6}$) and the Young's modulus (8$\times$10$^{10}$ Pa). Therefore, a maximum stress of 40 MPa can be generated in the Terfenol-D nanomagnet. While avoiding considerable degradation of the piezoelectric constants at low-thickness (24 nm as assumed in Ref.~\refcite{roy11_6}) piezoelectric layers is under research,\cite{RefWorks:823,RefWorks:820} we can use a higher thickness for the piezoelectric layer (e.g., 100 nm) and a concomitant amount of higher voltage, since the energy dissipation due to applying voltage is miniscule. For the piezoelectric layer, if we use materials with high piezoelectric coefficients e.g., PMN-PT, which has $d_{31}$=--3000 pm/V, and $d_{32}$=1000 pm/V (note the anisotropic strain, i.e., the signs of $d_{31}$ and $d_{32}$ are different),\cite{RefWorks:790} we can lower the operating voltage further to a few millivolts.


\begin{figure*}
\centering
\includegraphics[width=\textwidth]{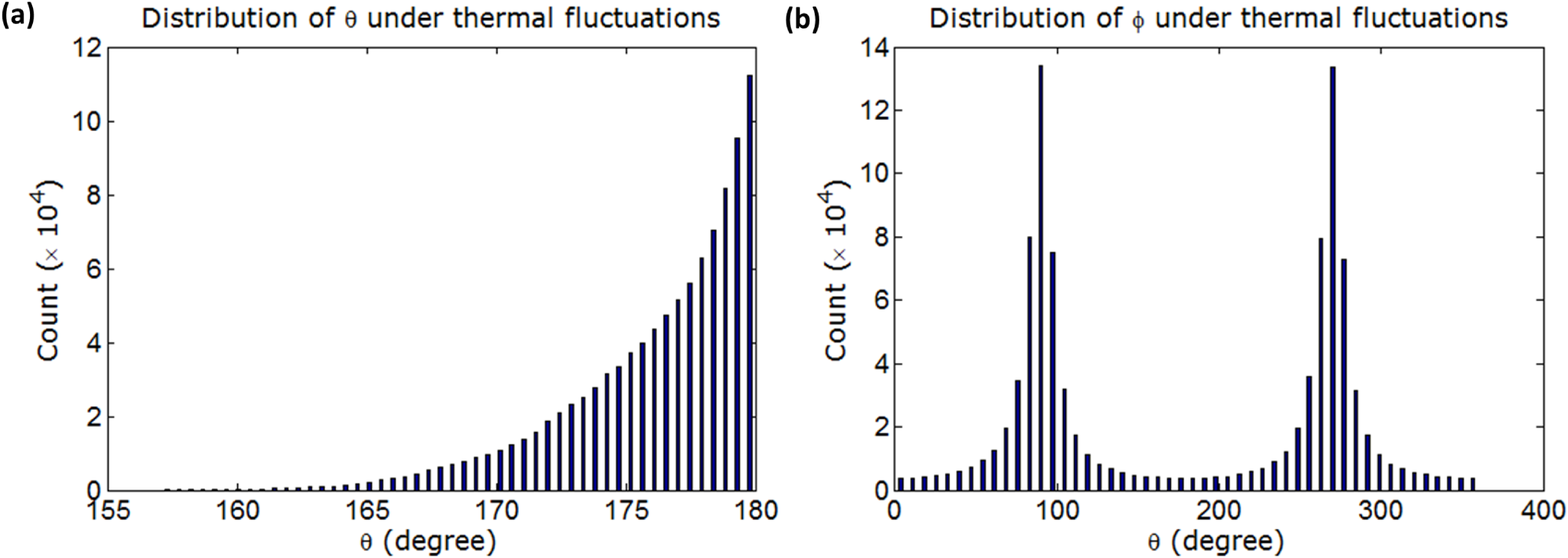}
\caption{\label{fig:thermal_theta_phi_distribution_terfenolD_1000ns} Distribution of polar angle $\theta_{initial}$ and azimuthal angle $\phi_{initial}$ due to thermal fluctuations at room temperature (300 K). 
(a) Distribution of the polar angle $\theta_{initial}$. The mean of the distribution is $\sim$175$^\circ$, while the most likely value is 180$^{\circ}$. This is a nearly exponential distribution (Boltzmann-like). 
(b) Distribution of the azimuthal angle $\phi_{initial}$. These are two Gaussian distributions with peaks centered at $90^\circ$ and $270^\circ$ (or --90$^{\circ}$), which means that the most likely location of the magnetization vector is in the plane of the nanomagnet. 
(Reprinted from Ref.~\srepref.)}
\end{figure*}

It should be noted that for the magnetostrictive layer, we need to choose a material that maximizes the product $(3/2)\lambda_s\, Y$. Terfenol-D (TbDyFe), which has 30 times higher magnetostriction coefficient in magnitude than the common ferromagnetic materials (e.g., Fe, Co, Ni), has the highest $(3/2)\lambda_s\, Y$ (Ref.~\refcite{roy11_2}). If it needs to avoid the rare-earth materials (e.g., Tb and Dy in Terfenol-D), we can also utilize Galfenol (FeGa),\cite{RefWorks:167,RefWorks:801} which has 6 times less $(3/2)\lambda_s$, but twice high $Y$ than that of Terfenol-D.

We consider that magnetization initially situates around $\theta \simeq 180^\circ$ and it fluctuates due to random thermal agitations. When a compressive stress is applied to initiate switching, magnetization starts with a certain ($\theta_{initial}$,$\phi_{initial}$) picked from the initial angle distributions. The voltage generated stress is assumed to be ramped up linearly and the stress is kept constant until the magnetization reaches the plane defined by the in-plane and the out-of-plane hard axis (i.e., the $x$-$y$ plane, $\theta = 90^\circ$). Note that when magnetization reaches at $\theta = 90^\circ$, the azimuthal angle $\phi$ may not correspond to the magnet's plane ($y$-$z$ plane, $\phi=\pm90^\circ$) and this has important consequences in switching the magnetization in the correct direction.\cite{roy13_2} In any case, the $x$-$y$ plane ($\theta = 90^\circ$) is always reached sooner or later with thermal fluctuations generating a distribution of time that magnetization takes to reach at $\theta = 90^\circ$.

After the magnetization reaches the $x$-$y$ plane, the stress is ramped down at the same rate at which it was ramped up, and reversed in magnitude to aid switching. The magnetization dynamics ensures that $\theta$ continues to rotate from $\theta = 90^\circ$ towards $0^{\circ}$.\cite{roy13_2} When $\theta$ becomes $ \leq 5^\circ$, switching is deemed to have completed. We perform a moderately large number (10,000) of simulations, with their corresponding ($\theta_{initial}$,$\phi_{initial}$) picked from the initial angle distributions, for each value of stress and ramp duration to generate the simulation results. Note that thermal fluctuations affect the magnetization dynamics during the course of switching as well. 

Figure~\ref{fig:thermal_theta_phi_distribution_terfenolD_1000ns} shows the distributions of initial angles $\theta_{initial}$ and $\phi_{initial}$ in the presence of thermal fluctuations. No bias magnetic field is applied along the out-of-plane direction (+$x$-axis) here. Note that the peak of the distribution of $\theta_{initial}$ is exactly at $\theta=180^\circ$. This means that magnetization is most likely to start from the easy axis $\theta=180^\circ$. If magnetization starts near from the easy axis, the torque acting on the magnetization would be vanishingly small. Therefore, only random thermal fluctuations can help magnetization going away from the easy axis and then magnetization can start switching. We have analyzed earlier that  a bias field can shift the peak of the distribution $\theta_{initial}$  away from the easy axis and facilitate switching, which we will present next.

\begin{figure*}
\centering
\includegraphics[width=\textwidth]{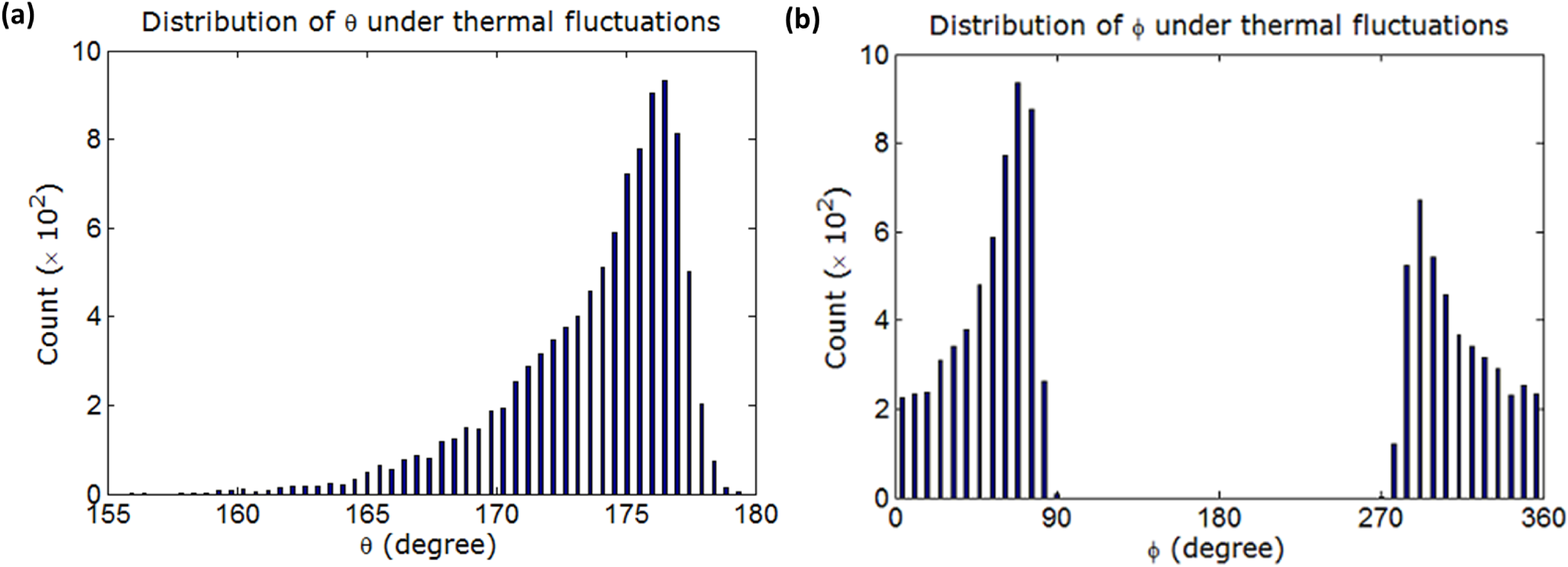}
\caption{\label{fig:thermal_theta_phi_distribution_terfenolD_mag_field} Distribution of polar angle $\theta_{initial}$ and azimuthal 
angle $\phi_{initial}$ due to thermal fluctuations at room temperature (300 K) when a magnetic field of flux density 40 mT is applied along the out-of-plane hard axis (+$x$-direction).
(a) Distribution of polar angle $\theta_{initial}$ at room temperature (300 K). The mean of the distribution is $173.7^\circ$, and
 the most likely value is 175.8$^{\circ}$.
(b) Distribution of the azimuthal angle $\phi_{initial}$ due to thermal fluctuations at room temperature (300 K). There are two distributions with  peaks centered at $\sim$65$^\circ$ and $\sim$295$^\circ$. 
(Reprinted with permission from Ref.~\japref. Copyright 2012, AIP Publishing LLC.)}
\end{figure*}

Figure~\ref{fig:thermal_theta_phi_distribution_terfenolD_mag_field} shows the distributions of initial angles $\theta_{initial}$ and $\phi_{initial}$ in the presence of thermal fluctuations when a bias magnetic field applied along the out-of-plane direction (+$x$-axis). In  Fig.~\ref{fig:thermal_theta_phi_distribution_terfenolD_mag_field}(a), note that the bias field has shifted the peak of $\theta_{initial}$ from the easy axis ($\theta = 180^{\circ}$) as we desire. The $\phi_{initial}$ distribution in  Fig.~\ref{fig:thermal_theta_phi_distribution_terfenolD_mag_field}(b) spans mostly within the interval [--90$^\circ$,+90$^\circ$] since the bias magnetic field is applied in the +$x$-direction. However, the $\phi_{initial}$ distribution is \emph{asymmetric} in the quadrants (0$^\circ$,90$^\circ$) and (270$^\circ$,360$^\circ$), which can be explained as follows. The magnetization is fluctuating around $\theta \simeq 180^\circ$ and the precessional motion of the magnetization $ [-|\gamma|/\left ( 1 + \alpha^2 \right ) \mathbf{M} \times \mathbf{H}$, where $\mathbf{M}$ is the magnetization and $\mathbf{H}$ is the effective field] due to the +$x$-directed magnetic field is such that the magnetization prefers the $\phi$-quadrant (0$^\circ$,90$^\circ$) over the $\phi$-quadrant (270$^\circ$,360$^\circ$). Hence, when the magnetization starts from $\theta \simeq 180^{\circ}$, the initial azimuthal angle $\phi_{initial}$ is more likely to be in the quadrant (0$^\circ$,90$^\circ$) than the other quadrant (270$^\circ$,360$^\circ$) in the +$x$-direction.

\begin{figure*}
\centering
\includegraphics[width=6.8in]{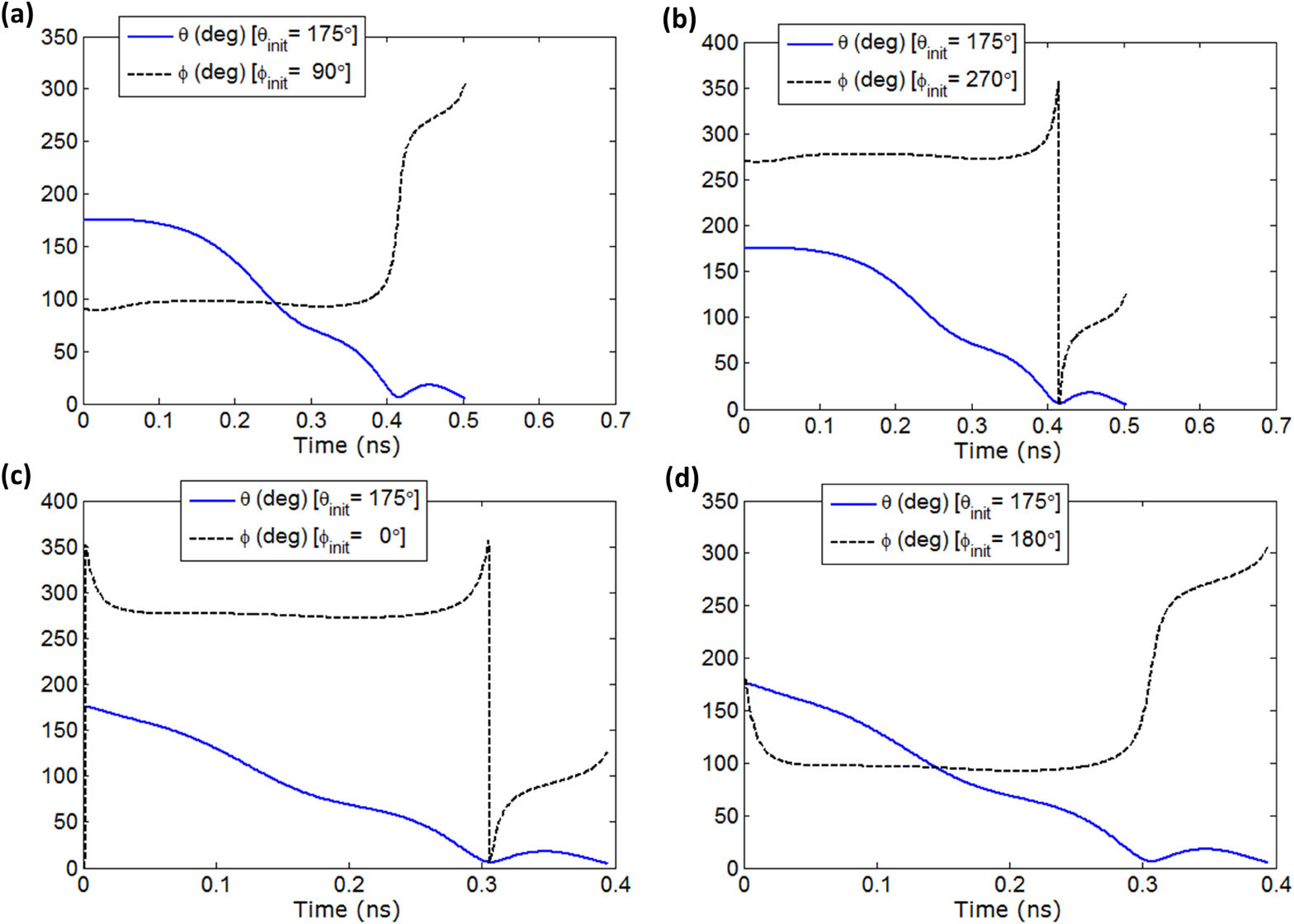}
\caption{\label{fig:dynamics_60ps_15MPa_theta_5deg} 
Temporal evolution of polar angle $\theta$ and azimuthal angle $\phi$ for a fixed $\theta_{initial}=175^\circ$ and four different values of $\phi_{initial}=\lbrace 90^\circ, 270^\circ, 0^\circ, 180^\circ \rbrace$. The applied stress is 15 MPa and the ramp duration is 60 ps. Thermal fluctuations have been ignored.
(a) $\phi_{initial}=90^\circ$. 
(b) $\phi_{initial}=270^\circ$. 
(c) $\phi_{initial}=0^\circ$. 
(d) $\phi_{initial}=180^\circ$. 
Note that when $\theta$ reaches $90^\circ$ or even earlier, $\phi$ always resides in a ``good'' quadrant [($90^\circ, 180^\circ$) or ($270^\circ, 360^\circ$)], which makes the switching successful. No bias field in the out-of-plane direction is applied to generate these simulation results. 
(Reprinted from Ref.~\srepref.)}
\end{figure*}

Figure~\ref{fig:dynamics_60ps_15MPa_theta_5deg} plots the magnetization dynamics for different values of $\phi_{initial}$ (while keeping fixed values of $\theta_{initial}$, applied stress, and ramp rate) to signify the role of out-of-plane excursion of magnetization. For Figs.~\ref{fig:dynamics_60ps_15MPa_theta_5deg}(a) and~\ref{fig:dynamics_60ps_15MPa_theta_5deg}(b), the magnetization initially lies on the plane of the magnet ($\phi_{initial}=\pm90^\circ$) and the precessional motion of magnetization due to applied stress is in the +$\hat{\mathbf{e}}_\phi$ direction, which increases $\phi$ with time. So the magnetization starts out in the ``good'' quadrants (90$^\circ$,180$^\circ$) and (270$^\circ$,360$^\circ$) for Figs.~\ref{fig:dynamics_60ps_15MPa_theta_5deg}(a) and~\ref{fig:dynamics_60ps_15MPa_theta_5deg}(b), respectively. Therefore, both the motions of magnetization [the damped motion due to applied stress and the motion due to out-of-plane excursion shown as $-2\alpha B(\phi,t)sin\theta cos\theta\,\hat{\mathbf{e}}_\theta$ and $-|B_{shape,\phi}(\phi)|sin\theta\,\hat{\mathbf{e}}_\theta$, respectively in the Fig.~\ref{fig:dynamics_illustration}(c)] are in the $-\hat{\mathbf{e}}_\theta$ direction so that $\theta$ decreases with time and the magnetization rotates in the correct direction towards $\theta = 90^{\circ}$. The increasing out-of-plane excursion of the magnetization due to $\phi$ with time is eventually opposed by the damped motion due to out-of-plane excursion [depicted as $-\alpha |B_{shape,\phi}(\phi)|\,\hat{\mathbf{e}}_\phi$ in Fig.~\ref{fig:dynamics_illustration}(c)], which attempts to bring the magnetization back to the magnet's plane ($\phi_{initial}=\pm90^\circ$). These two effects balance each other and $\phi$ assumes a stable value in the ``good'' quadrant which can be clearly observed in the plots (the flat regions in the $\phi$-plots). When $\theta$ reaches $90^{\circ}$, the torque due to stress and shape anisotropy vanishes, however, $\phi$ remains in the respective ``good'' quadrant for the cases in Figs.~\ref{fig:dynamics_60ps_15MPa_theta_5deg}(a) and~\ref{fig:dynamics_60ps_15MPa_theta_5deg}(b). At this point, stress is reversed with the same ramp rate and the damped motion due to stress and shape anisotropy eventually becomes again in the $-\hat{\mathbf{e}}_\theta$ direction. In this way, magnetization continues to rotate in the right direction towards $\theta = 0^{\circ}$. Slightly past 0.4 ns, the precessional motion due to stress and shape anisotropy continues to rotate $\phi$ and pushes it into a neighboring ``bad'' quadrant, but eventually it escapes into the other ``good'' quadrant. This brief excursion into a ``bad'' quadrant causes the ripple in $\theta$-plots. However, magnetization ends up switching successfully.

\begin{figure*}
\centering
\includegraphics[width=6.8in]{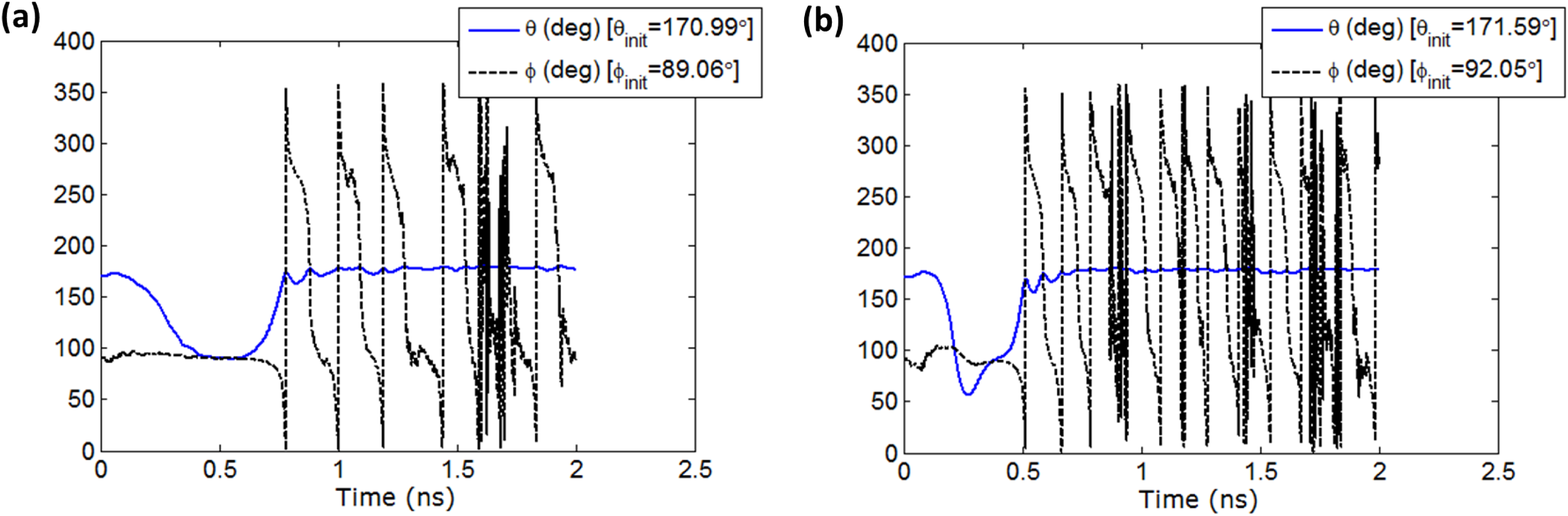}
\caption{\label{fig:dynamics_failed} 
Temporal evolution of the polar angle $\theta$ and azimuthal angle $\phi$ when magnetization fails to switch and backtracks to the initial state. Simulations are carried out at room-temperature (300 K).
(a) The applied stress is 10 MPa and the ramp duration is 60 ps. 
(b) The applied stress is 30 MPa and the ramp duration is 120 ps.
The ringing in the $\phi$-plots at the end is just due to thermal fluctuations that causes magnetization to roam around $\theta=180^\circ$. No bias field in the out-of-plane direction is applied to generate these simulation results. 
(Reprinted from Ref.~\srepref.)}
\end{figure*}

\begin{figure*} 
\centering
\includegraphics[width=80mm]{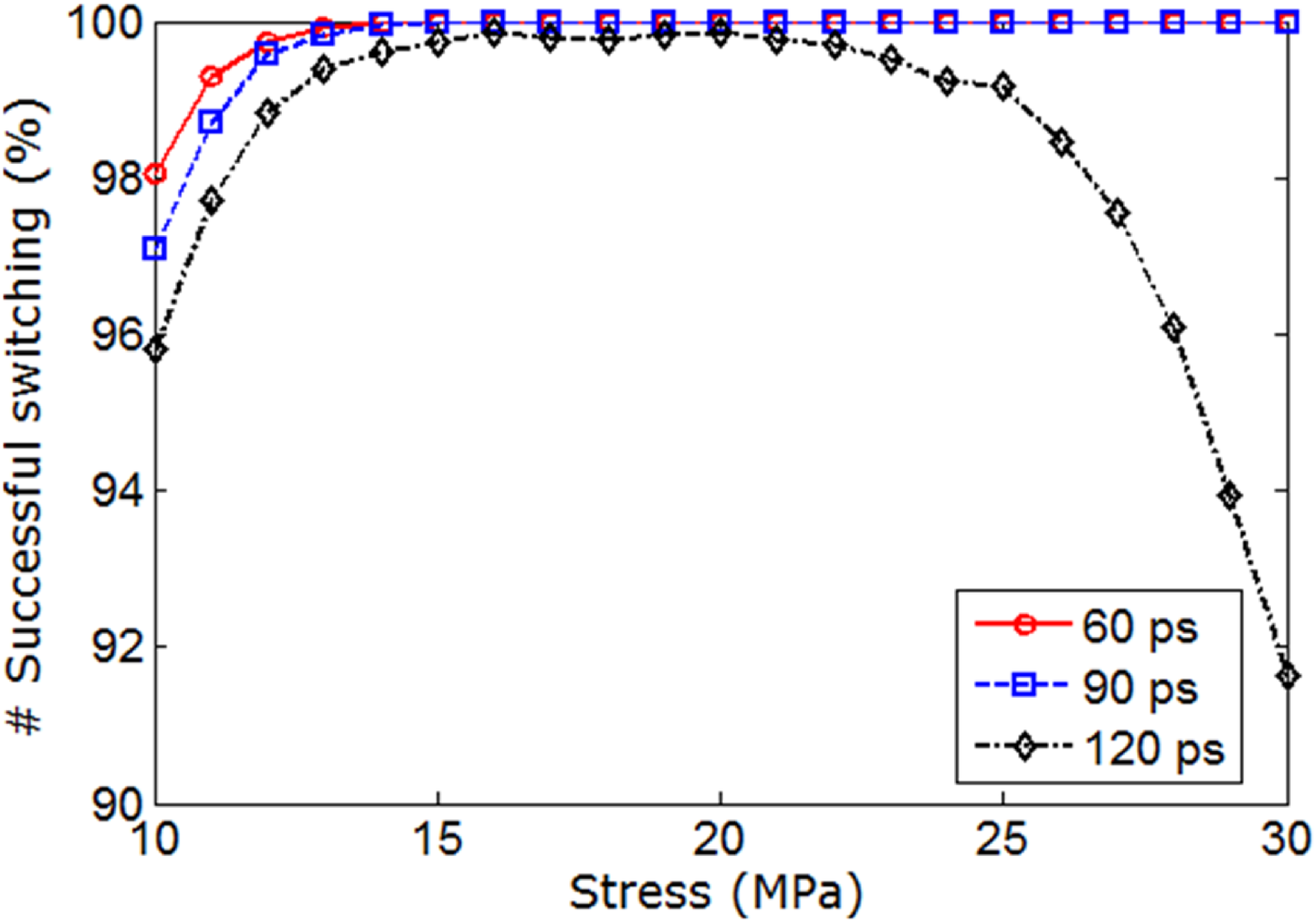}
\caption{\label{fig:thermal_stress_success_ramp_time_mag} Percentage of successful switching events at room-temperature (300 K) when  the magnetostrictive nanomagnet is subjected to stress between 10 MPa and 30 MPa with ramp duration (60 ps, 90 ps, and 120 ps) as a parameter. The critical stress at which switching becomes $\sim$100\% successful increases with ramp duration. However, at high ramp duration (e.g., 120 ps), we may not achieve $\sim$100\% switching probability for any values of stress and stress-dependence of the success probability becomes non-monotonic.
(Reprinted with permission from Ref.~\japref. Copyright 2012, AIP Publishing LLC.)}
\end{figure*}

In Figs.~\ref{fig:dynamics_60ps_15MPa_theta_5deg}(c) and~\ref{fig:dynamics_60ps_15MPa_theta_5deg}(d), the magnetization is initially lifted far out of the magnet's plane  ($\phi_{initial}=0^\circ,180^\circ$, respectively), where the huge out-of-plane shape anisotropy energy barrier cannot be overcome by the stress anisotropy and $B(\phi,t)$ remains positive, i.e., $|B_{stress}(t)| < |B_{shape}(\phi)|$. This forces magnetization to precess in the clockwise direction ($-\hat{\mathbf{e}}_\phi$) rather than in the anticlockwise direction ($+\hat{\mathbf{e}}_\phi$). Therefore $\phi$ decreases with time, which takes magnetization inside the neighboring ``good'' quadrant and eventually $|B_{stress}(t)|$ becomes greater than $|B_{shape}(\phi)|$. Then $\phi$ assumes a stable value due to the counteraction between the damped motion and the motion due to out-of-plane excursion [depicted as $-\alpha |B_{shape,\phi}(\phi)|\,\hat{\mathbf{e}}_\phi$ and $2 B(\phi,t) cos\theta\,\hat{\mathbf{e}}_\phi$ in Fig.~\ref{fig:dynamics_illustration}(c), respectively]. Thereafter, switching occurs similarly for the cases as in Figs.~\ref{fig:dynamics_60ps_15MPa_theta_5deg}(a) and~\ref{fig:dynamics_60ps_15MPa_theta_5deg}(b). Slightly past 0.3 ns, continuing $\phi$ rotation pushes $\phi$ into a neighboring ``bad'' quadrant, but eventually it escapes into the other ``good'' quadrant. Once again, this brief excursion into the ``bad'' quadrant causes the ripple in $\theta$-plots and eventually successful switching takes place in the end.

So straying into a ``bad'' quadrant for azimuthal angle $\phi$ during the ramp-down phase does not mean at all that magnetization would fail to switch. Magnetization can rotate in the other ``good'' quadrant for $\phi$ and complete the switching. Thus, there may be ripples appearing in the magnetization dynamics at the end of switching increasing the switching time, which is because of the transition of azimuthal angle $\phi$ between two ``good'' quadrants through one ``bad'' quadrant. This happens particularly near the end of switching, when the precessional motion is strong.

An interesting comparison between the switching delays in the Figs.~\ref{fig:dynamics_60ps_15MPa_theta_5deg}(a) or~\ref{fig:dynamics_60ps_15MPa_theta_5deg}(b), and the  switching delays in the Figs.~\ref{fig:dynamics_60ps_15MPa_theta_5deg}(c) or~\ref{fig:dynamics_60ps_15MPa_theta_5deg}(d) reveals that the switching delay decreases by 0.1 ns when magnetization starts from out-of-plane ($\phi_{initial}=0^\circ, 180^\circ$) compared to when magnetization starts from in-plane ($\phi_{initial}=90^\circ, 270^\circ$). This is a very consequence of the reasoning that out-of-plane excursion of magnetization in the ``good'' quadrants aids magnetization to move faster in $\theta$-space. When magnetization starts out-of-plane, magnetization spends more time deep inside a ``good'' quadrant; hence, switching gets faster than that of the case when magnetization starts in-plane of the magnet.

Figure~\ref{fig:dynamics_failed} demonstrates a couple of examples when switching fails to occur, i.e., magnetization backtracks to its original position. In Fig.~\ref{fig:dynamics_failed}(a), when the polar angle $\theta$ reaches $90^\circ$, the azimuthal angle $\phi$ has ventured into the ``bad'' quadrant ($0^\circ$,90$^\circ$) due to thermal fluctuations. Thus, switching eventually fails. In Fig.~\ref{fig:dynamics_failed}(b), when the polar angle $\theta$ reaches $90^\circ$, the azimuthal angle $\phi$ is greater than $90^\circ$ and in the ``good'' quadrant ($90^\circ$,180$^\circ$). However, after reaching around $\theta\simeq50^\circ$, the magnetization backtracks to the initial state and therefore switching fails to occur. This happens because of the long ramp duration which forces $\phi$ to decrease over time and eventually $\phi$ enters into the neighboring ``bad'' quadrant ($0^\circ$,90$^\circ$). Also during the passage of long duration of ramp, thermal fluctuations have ample opportunity to scuttle the switching. Such switching failure is depicted intuitively in the Fig.~\ref{fig:theta_phi_motion_bad_quad_fails_illustration} earlier.

\begin{figure*}
\centering
\includegraphics[width=\textwidth]{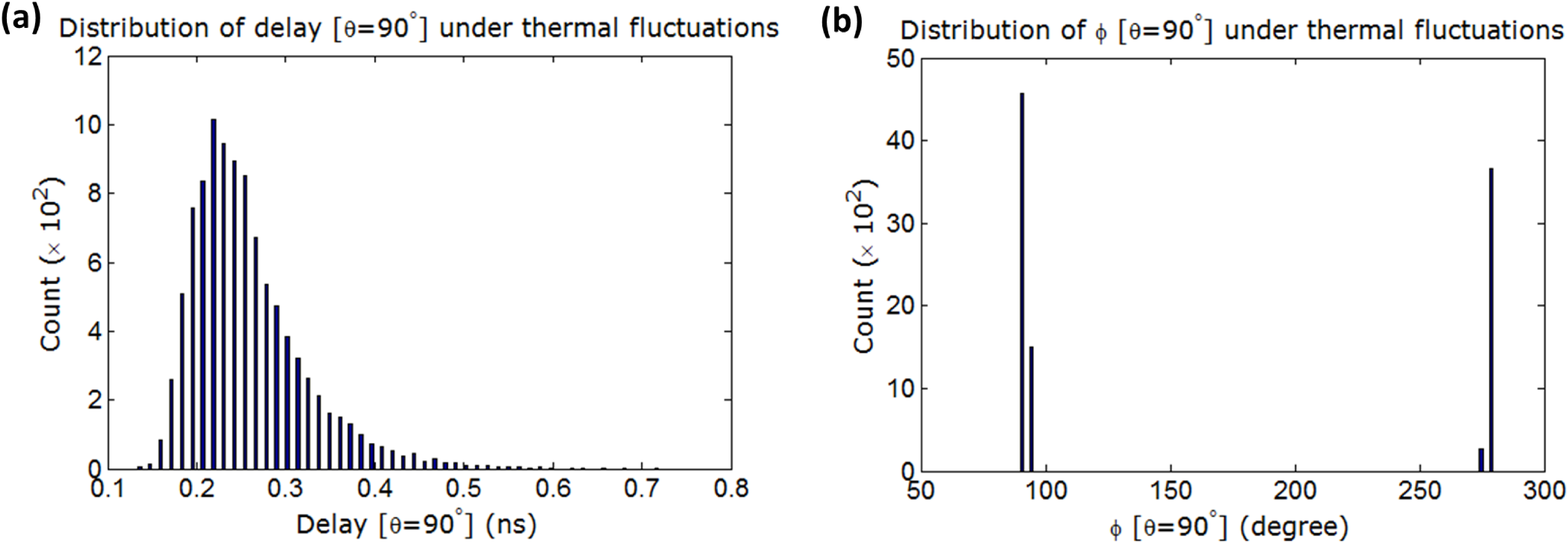}
\caption{\label{fig:thermal_distribution_delay90_60ps_15MPa_mag} Statistical distributions of different quantities when 15 MPa stress with 60 ps ramp duration is applied on the nanomagnet at room temperature (300 K). 
(a) Distribution of time taken for $\theta$ to reach $90^\circ$ starting from ($\theta_{initial}$,$\phi_{initial}$) where the latter are picked from the distributions in the presence of thermal fluctuations (see Fig.~\ref{fig:thermal_theta_phi_distribution_terfenolD_mag_field}). 
(b) Distribution of azimuthal angle $\phi$ when $\theta$ reaches $90^\circ$. Note that this figure is similar to the Fig.~S4 in Ref.~\srepref, but an out-of-plane bias field in the +$x$-direction is applied here following Ref.~\japref\; to generate these simulation results.}
\end{figure*}


Figure~\ref{fig:thermal_stress_success_ramp_time_mag} plots the percentage successful switching rates at room-temperature (300 K) when the magnetostrictive nanomagnet is subjected to stress between 10 MPa and 30 MPa with ramp duration (60 ps, 90 ps, and 120 ps) as a parameter. For each value of stress and ramp duration, a moderately large number (10,000) of simulations were performed to generate these results. Initial angle distributions at 300 K for both $\theta$ and $\phi$ are taken into account during simulations (see Fig.~\ref{fig:thermal_theta_phi_distribution_terfenolD_mag_field}). The minimum stress needed to switch the magnetization \emph{without} considering thermal fluctuations is $\sim$5 MPa, but at room temperature (300 K) this minimum stress is increased to $\sim$14 MPa for 60 ps ramp duration and to $\sim$17 MPa for 90 ps ramp duration. The minimum stress of 5 MPa \emph{without} considering thermal fluctuations is attributed to the removal of in-plane shape anisotropy energy barrier by the stress anisotropy, while at 300 K an increased magnitude of stress is required since magnetization is scuttled in ``bad'' quadrants due to thermal fluctuations. When longer ramp duration, a higher stress is required to prevent magnetization traversing into ``bad'' quadrants. Therefore, it is beneficial to reduce the ramp duration (i.e., having a faster ramp rate) to increase the success rate of switching at a lower stress level. Simulation results show that with 1 ps ramp duration, the critical stress can be reduced by $\sim$2 MPa compared to the case of 60 ps ramp duration.

For 120 ps ramp duration, $\sim$100\% success probability is unattainable for any value of stress since thermal agitations have higher latitude to scuttle the magnetization into ``bad'' quadrants while stress is ramped down. At higher stresses accompanied by a long ramp duration, there occurs higher out-of-plane excursion pushing the magnetization into ``bad'' quadrants, which further aggravates the error probability. At very long ramp duration, the success (and error) probability becomes 50\%, since the magnetization would stay in-plane of the magnet and during the ramp-down phase, random thermal fluctuations may equally scuttle the magnetization either in the ``good'' quadrants or in the `bad'' quadrants.

Figure~\ref{fig:thermal_distribution_delay90_60ps_15MPa_mag}(a) shows the distribution of time taken for magnetization polar angle $\theta$ to reach $90^\circ$ ($x$-$y$ plane). This wide distribution is caused by: (1) the initial angle distributions in Fig.~\ref{fig:thermal_theta_phi_distribution_terfenolD_mag_field}, and (2) thermal fluctuations during the course of transition from some $\theta = \theta_{initial}$  to $90^\circ$. We do need to tackle such distribution by keeping the magnetization out-of-plane far enough so that magnetization does not collapse on magnet's plane. We can use a sensing element to detect when $\theta$ reaches around 90$^{\circ}$, so that we can ramp down the stress thereafter. The sensing element can be implemented by measuring the magnetoresistance in a magnetic tunnel junction (MTJ)\cite{RefWorks:577,RefWorks:555,RefWorks:572,RefWorks:76,RefWorks:74,RefWorks:33,RefWorks:300,roy13_2,roy13,roy14}. We need to get calibrated on the magnetoresistance of the MTJ when magnetization resides at $\theta=90^\circ$ ($x$-$y$ plane). Comparing this known signal with the sensed signal of the MTJ, the stress can be ramped down. Such comparator can be implemented with these energy-efficient multiferroic devices, i.e., energy-inefficient charge-based transistors do not need to be utilized.\cite{roy_aps_2014x,roy_mrs_2014x,roy_spie_2014x} The energy dissipation in this sensing element is not considered here. It should be mentioned that usually it requires several peripheral circuitry in conjunction with the basic device itself, however, energy dissipation considering in the other required circuitry does not change the order of dissipation.\cite{rabae03,pedra02}

\begin{figure*}
\centering
\includegraphics[width=\textwidth]{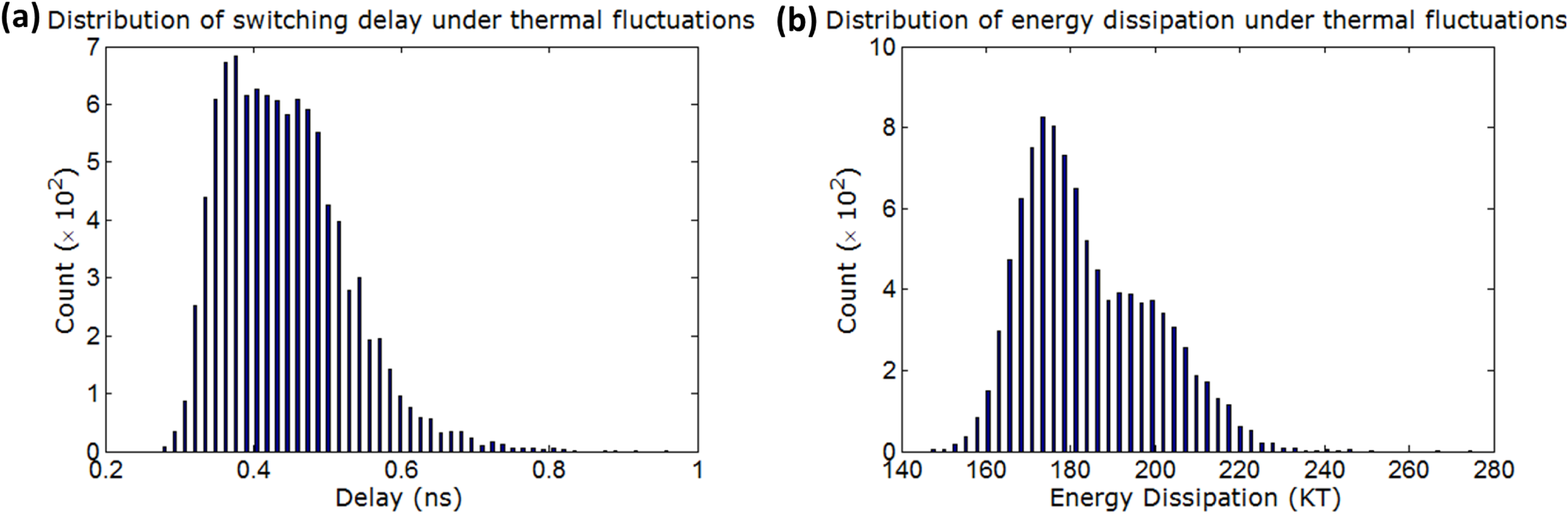}
\caption{\label{fig:thermal_distribution_delay_energy_60ps_15MPa_mag} Delay and energy distributions for 15 MPa applied stress and 60 ps ramp duration at room temperature (300 K). 
(a) Distribution of the switching delay. The mean and standard deviation of the distribution are 0.44 ns and 83 ps, respectively. 
(b) Distribution of energy dissipation. The mean and standard deviation of the distribution are 184 $kT$ and 15.5 $kT$ at room temperature, respectively.
(Reprinted with permission from Ref.~\japref. Copyright 2012, AIP Publishing LLC.)}
\end{figure*}

Some tolerance is nonetheless required since the sensing element cannot be perfect. Simulation results show that the internal dynamics works correctly as long as the stress is ramped down when $\theta$ is in the interval [85$^{\circ}$, 110$^{\circ}$], i.e. it does not have to be exactly 90$^{\circ}$. This tolerance is due to the motion arising from the out-of-plane excursion of magnetization in a ``good'' quadrant. If magnetization reaches at $\theta=90^\circ$ (even past it towards $\theta=0^\circ$) and stress is not withdrawn soon enough, then magnetization can fall on the magnet's plane ($\phi=\pm 90^\circ$, potential energy minima), upon which the success probability would be 50\% since thermal fluctuations can put magnetization in either direction of the potential landscape. 


Figure~\ref{fig:thermal_distribution_delay90_60ps_15MPa_mag}(b) shows the distribution of azimuthal angle $\phi$ when $\theta$ reaches $90^\circ$. Note that  $\phi$ always resides in the ``good'' quadrant [($90^\circ, 180^\circ$) or ($270^\circ, 360^\circ$)] and has a fairly narrow distribution. As depicted in the Fig.~\ref{fig:thermal_stress_success_ramp_time_mag}, a  high stress and  fast ramp rate are required to ensure that $\phi$ is in the ``good'' quadrants, which is conducive to successful switching.

Figure~\ref{fig:thermal_distribution_delay_energy_60ps_15MPa_mag} shows the delay and energy distributions for 15 MPa stress and 60 ps ramp duration in the presence of room-temperature (300 K) thermal fluctuations. The high-delay tail in  Fig.~\ref{fig:thermal_distribution_delay_energy_60ps_15MPa_mag}(a) is particularly associated with the switching trajectories that start very close to $\theta = 180^\circ$. In such trajectories, the torque acting on the magnetization is vanishingly small, which makes the switching sluggish at the beginning.  During this time, switching may also becomes susceptible to backtracking due to random thermal fluctuations, which may increas the delay further. Nonetheless, out of 10,000 simulations of switching trajectories peformed, there was not a single one in Fig.~\ref{fig:thermal_distribution_delay_energy_60ps_15MPa_mag}(a) where the delay exceeded 1 ns meaning that the probability of such happening is less than 0.01\%. Since the energy dissipation is the product of the power dissipation and the switching delay, similar behavior is found in Fig.~\ref{fig:thermal_distribution_delay_energy_60ps_15MPa_mag}(b). Note that the variation in the distribution of energy dissipation happens due to the internal energy dissipation (caused by Gilbert damping) in the presence of random thermal fluctuations since all the trajectories correspond to the same 15 MPa stress and 60 ps ramp duration.

\vspace*{2mm}

\begin{figure*}
\includegraphics[width=\textwidth]{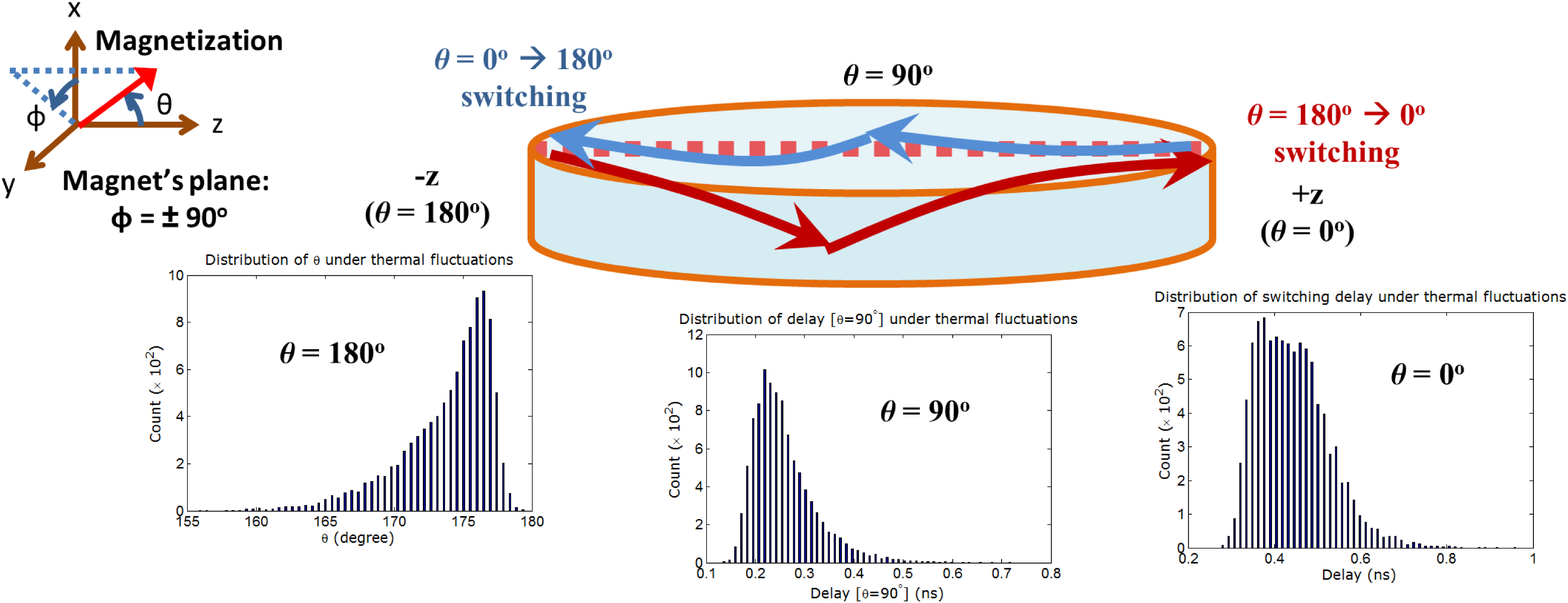}
\caption{\label{fig:magnetization_switching} 
Illustration of magnetization switching between two anti-parallel states ($\pm z$-axis). Stress rotates magnetization out of magnet's plane and when magnetization reaches the hard plane ($\theta=90^\circ$), intrinsic dynamics dictates that magnetization lifts out-of-plane in a certain direction so that a complete $180^\circ$ switching of magnetization is possible.$^{\srepref,\japref}$ While switching along the --$y$-axis rather than +$y$-axis (shown by arrows), the directions of the out-of-plane excursions would be exactly opposite.$^{\srepref}$ Three distributions are shown for $\theta=180^\circ$ to $0^\circ$ switching: the one at $\theta=180^\circ$ depicts the initial distribution of magnetization when no stress is active, the other two distributions around $\theta=90^\circ$ and $\theta=0^\circ$ correspond to 60 ps ramp period and 15 MPa stress,$^{\japref}$ which are Figs.~\ref{fig:thermal_theta_phi_distribution_terfenolD_mag_field}(a),~\ref{fig:thermal_distribution_delay90_60ps_15MPa_mag}(a), and~\ref{fig:thermal_distribution_delay_energy_60ps_15MPa_mag}(a), respectively.}
\end{figure*}

\textit{Discussion on magnetization switching.} 
We will now analyze the magnetization switching between the $180^\circ$ symmetry equivalent states based on the simulation results as depicted in the Fig.~\ref{fig:magnetization_switching}. The usual perception is that stress can rotate magnetization of a magnetostrictive nanomagnet only by $90^\circ$ from $\pm z$-axis to $\pm y$-axis (see Fig.~\ref{fig:magnetization_switching}). However, if we determine the expression of torque due to stress from the stress anisotropy energy [see Eq.~(\ref{eq:e_stress})] as
\begin{equation}
\mathbf{T_{E,stress}} = - \mathbf{\hat{e}_r} \times \nabla\,E_{stress}  = - (3/2) \, \lambda_s \sigma sin(2\theta)\, \mathbf{\hat{e}_\phi},
\label{eq:T_stress}
\end{equation}
we see that the torque due to stress acts along the out of magnet's plane ($\mathbf{\hat{e}_\phi}$ direction) and therefore magnetization lifts out-of-plane (although the out-of-plane demagnetization field due to small thickness of the nanomagnet attempts to keep the magnetization in-plane). This out-of-plane excursion of magnetization generates an \emph{intrinsic} asymmetry, which can completely switch the magnetization by $180^\circ$ (Ref.~\refcite{roy13_2}). Full $180^\circ$ switching is desirable since it facilitates having the full tunneling magnetoresistance (TMR) while electrically reading the magnetization state using a magnetic tunnel junction (MTJ).\cite{RefWorks:577,RefWorks:555,RefWorks:572,RefWorks:76,RefWorks:74,RefWorks:33,RefWorks:300}

Such full $180^\circ$ switching of magnetization (i.e., \emph{memory} operation) in piezoelectric-magnetostrictive heterostructures was first addressed by Roy in Ref.~\refcite{roy11_news}. Note that Ref.~\refcite{RefWorks:154} shows the Bennett clocking operation\cite{RefWorks:144} for \emph{logic} design, which is not memory operation since using a neighboring nanomagnet to switch another nanomagnet is tantamount to using a magnetic field,\cite{fasha13_2} the direction of which has to be reversed for switching the memory bit in either direction. Also, Ref.~\refcite{RefWorks:154} performed an incorrect analysis by \emph{assuming} that magnetization always resides in-plane, which leads to a very high switching delay and high energy-delay product compared to the traditional transistors.\cite{arXiv:1501.05941v2} This was addressed and explained by Roy.\cite{roy13_spin,fasha11,nano_edi}

Figure~\ref{fig:magnetization_switching} illustrates the magnetization switching alongwith the simulation results of different distributions in the presence of room-temperature thermal fluctuations. When magnetization starts switching from $\theta=180^\circ$, the initial orientation of magnetization is a distribution due to random thermal fluctuations. When magnetization reaches $\theta=90^\circ$, the time taken to reach there is also a distribution due to thermal fluctuations. After magnetization reaches \emph{around} $\theta=90^\circ$, stress needs to be released to switch the magnetization. However, there can be two states magnetization can end up: (1) $\theta=180^\circ$ causing a switching failure, and (2)  $\theta=0^\circ$ leading to successful switching. As shown in the Fig.~\ref{fig:magnetization_switching}, magnetization's excursion out of magnet's plane in a particular direction would lead to successful switching. As presented in the Fig.~\ref{fig:thermal_stress_success_ramp_time_mag}, a sufficiently fast ramp rate of stress accompanied by a sufficiently high stress is conducive to successful switching, i.e., the out-of-plane excursion of magnetization in the specified directions as shown in the Fig.~\ref{fig:magnetization_switching} would be maintained. Therefore, the out-of-plane excursion of magnetization generated an equivalent \emph{intrinsic} asymmetry for magnetization switching in the correct direction. If magnetization starts switching from $\theta=0^\circ$ instead of $\theta=180^\circ$, the direction of out-of-plane excursions of magnetization as shown in the Fig.~\ref{fig:magnetization_switching} would be exactly opposite.  

Note that the aforesaid magnetization switching as presented in the Fig.~\ref{fig:magnetization_switching} requires to read the magnetization state using spin valve/MTJ to sense when magnetization reaches \emph{around} $\theta=90^\circ$ (since room temperature thermal fluctuations make the traversal time a wide 	distribution as shown) so that stress can be brought down around that time. This has been discussed in the context of explaining the results in the Fig.~\ref{fig:thermal_distribution_delay90_60ps_15MPa_mag}. The sensing element reads the magnetization state and a comparator can compare the read signal with a pre-calibrated value for the MTJ resistance when magnetization's orientation is around $\theta=90^\circ$ (as explained in Ref.~\refcite{roy13_2}). It is important to note that there is tolerance around $\theta=90^\circ$, i.e., stress does not need to be withdrawn \emph{exactly} at $\theta=90^\circ$ since the sensing procedure cannot be perfect.\cite{roy13_2} It is shown that the internal magnetization dynamics provides such tolerance.\cite{roy13_2} Any additional element for comparison can be built using these energy-efficient multiferroic devices in general.\cite{roy_aps_2014x,roy_mrs_2014x,roy_spie_2014x} Researchers are trying to replace the traditional \emph{switch} based on charge-based transistors by a new possible ``ultra-low-energy'' \emph{switch} (e.g., using multiferroic composites). Therefore, any circuitry can be built with the energy-efficient switch itself rather than the conventional transistors.\cite{roy_aps_2014x,roy_mrs_2014x,roy_spie_2014x} Usually, it requires several peripheral circuitry in conjunction with the basic switch in a system.\cite{rabae03,pedra02}  While researchers report on the performance metrics of the basic switch itself, the total energy dissipation considering the other required circuitry does not change the order of energy dissipation, utilizing the respective devices.\cite{rabae03,pedra02} Therefore energy efficient technologies can be envisaged using such magnetization switching mechanism\cite{roy13_spin,roy11_news,roy11_2,roy13_2,roy11_6} and computing methodologies\cite{roy13_spin,roy13,roy14} based on such switching mechanism. 

Also such aforesaid switching is confirmed by others,\cite{RefWorks:872,RefWorks:873,RefWorks:874,RefWorks:887,RefWorks:889,RefWorks:886} but they do not perform a detailed analysis using \emph{stochastic} Landau-Lifshitz-Gilbert equation of magnetization dynamics and therefore do not conceive or mention the requirement of sensing circuitry due to the wide distribution of traversal times in the presence of room temperature thermal fluctuations as explained earlier. While it is desirable that an additional sensing procedure can be avoided, the methodology\cite{roy13_2} tackles the variation in the traversal time of magnetization particularly the crucial initial distribution of magnetization \emph{effectively} for successful switching and it also aids in decreasing the switching delay given a certain error probability. An out-of-plane field (see Ref.~\refcite{roy11_6}) can help breaking the symmetry while switching the magnetization but the variation in traversal times towards the hard axis due to the initial Boltzmann distribution of magnetization is large enough to cause switching failures (note that it requires very low switching error probability e.g., $<10^{-4}$ to be technologically viable). While further research may lower the error probability with a symmetry breaking field or possibly by other means, the critical understandings behind the switching dynamics of magnetization as explained by Roy in Ref.~\refcite{roy13_2} is an important step forward to the technological applications for building memory and logic using multiferroics.

In Ref.~\refcite{roy14}, information propagation in a chain of nanomagnets using Bennett clocking mechanism\cite{RefWorks:144} in the presence of room-temperature thermal fluctuations is analyzed. It is found that the inherent magnetization dynamics caused by the out-of-plane excursion of magnetization can lead to switching failure even if there is dipole coupling between neighboring nanomagnets that can introduce asymmetry in switching. But such asymmetry is not sufficient and therefore the asymmetry caused by the out-of-plane excursion\cite{roy13_2} can be utilized for successful switching of magnetization. Once again, here a sensing element is required to detect when magnetization reaches \emph{around} $\theta=90^\circ$ and  multiferroic devices can be utilized to build such functionality, which would not change the order of energy dissipation.

But, one contradictory viewpoint exists that any other circuitry other than the switching device itself needs to be built with traditional transistors,\cite{roy13_comment} which however undermines the research behind finding a switch \emph{replacing} the transistor itself\cite{nri,arXiv:1501.05941v2} and therefore incorrect lacking the understandings over the development of transistor based circuitry.\cite{rabae03,pedra02} Such viewpoint does not exist anywhere else in literature. A few contradicting facts vis-a-vis the comment made by Bandyopadhyay and Atulasimha (referred as BA onwards) in Refs.~\refcite{roy13_comment,biswa14_3,munir15} are as follows:\cite{arXiv:1501.05941v2} 
(1) BA are coauthors of Roy in Refs.~\refcite{roy11_news,roy11_2,roy13_2,roy11_5,roy11_6}, where energy-efficiency is claimed, requiring the sensing procedure too. In particular, energy efficiency is claimed in the presence of thermal fluctuations in Ref.~\refcite{roy11_6}, which requires the sensing element, explained in details by Roy in Ref.~\refcite{roy13_2}. Also, there is a patent\cite{roy12_patentx} following the research conceived by Roy\cite{roy11_news,roy11_2,roy11_5,roy11_6,roy13_2,roy13_spin,fasha11} that claims energy-efficiency, requiring the sensing element therein as well. Therefore, the comment made by BA in Ref.~\refcite{roy13_comment} has no technical basis.
(2) Note that Ref.~\refcite{munir15}, coauthored by BA, uses precisely shaped pulses. Such pulses need to be generated too using some circuitry. According to BA, transistors need to be utilized to build such circuitry and the system would dissipate too much energy, invalidating the claim of energy efficiency in Ref.~\refcite{munir15}. Note that one additional hardware cannot be shared between many devices distributed on a chip due to interconnect delay and loading effect. Also, note that pulse shaping is an ineffective countermeasure since it is not helping much in reducing the error probability, therefore building and using such circuitry do not make sense either. 
(3) Ref.~\refcite{biswa14_2}, in which BA are coauthors, proposed a ``toggle'' switch (as stated that ``a write cycle must be preceded by a read cycle to determine the stored bit''), which would require a similar use of spin-valve or MTJ for reading the known bit, storing it, and then using it for \emph{comparison}.  According to BA, such \emph{additional} circuitry needs to be constructed with energy-inefficient transistors, invalidating the claim of energy efficiency in Ref.~\refcite{biswa14_2}. 

According to the unsubstantiated and incorrect viewpoint as stated by BA,\cite{roy13_comment} the switching methodology confirmed by others\cite{RefWorks:872,RefWorks:873,RefWorks:874,RefWorks:887,RefWorks:889,RefWorks:886} will require the sensing circuitry too (as already explained by Roy in Ref.~\refcite{roy13_2}) and hence will be energy-inefficient. However, as explained earlier, any circuitry can be built with the energy-efficient switch itself rather than the conventional transistors\cite{roy_aps_2014x,roy_mrs_2014x,roy_spie_2014x} and furthermore BA contradict themselves as explained above. Therefore, there is no technical reasoning behind such point raised by BA in Ref.~\refcite{roy13_comment}. The other unsubstantiated points raised by BA in Ref.~\refcite{roy13_comment} on Ref.~\refcite{roy13} are also technically incorrect as explained.\cite{arXiv:1501.05941v2} Another paper\cite{RefWorks:888} also incorrectly stated that stress has to ``to drive the magnetization switch out-of-plane at first'', however, as pointed out by the Eq.~(\ref{eq:T_stress}) that the out-of-plane excursion is inherent to the magnetization dynamics.\cite{roy13_2} Such out-of-plane motion also increases the switching speed to the order of GHz.\cite{roy11_5,roy13_2} Furthermore, Ref.~\refcite{RefWorks:888} did not study the consequence of room-temperature thermal fluctuations, which is critical to the magnetization dynamics and error probability of switching. 

One of the critical points to conceive while proposing energy-efficient systems is that it must be \emph{area-efficient} as well to compete with the traditional transistors e.g., our laptops cannot be 10 times bigger. However, such area-inefficient devices have been proposed in Ref.~\refcite{biswa14}, which claims a superior design of magnetoelastic memory, compared to an earlier idea.\cite{RefWorks:850,RefWorks:559,RefWorks:848,RefWorks:849} Comparatively, the lateral area chosen by Ref.~\refcite{biswa14} is an order higher than that of chosen by Ref.~\refcite{RefWorks:850}. Also, Ref.~\refcite{biswa14} chooses to use two pairs of \emph{lateral} electrodes, used to apply stress at angles, understanding behind which is otherwise known, e.g., Ref.~\refcite{dsouza11}, which has however issues with room-temperature thermal fluctuations. The additional \emph{lateral} pads cannot be dispensed and particularly they need to be large for application of stress, thereby consuming \emph{additional large area} for \emph{each} nanomagnet apart from the area consumption by the nanomagnet itself. The area consumed by the proposal in Ref.~\refcite{biswa14} (also Refs.~\refcite{biswa14_2,biswa14_3}) is so high that the devices become of micro-scale size, therefore, the proposals are untenable for building nanoelectronics.\cite{itrs,nri} There is an ongoing drive to reduce the area-consumption,\cite{RefWorks:774,roy14_2} but the proposals in Refs.~\refcite{biswa14,biswa14_2,biswa14_3} do not bode at all on such crucial front. There is also similar issue regarding \emph{area inefficiency} in the proposal by Ref.~\refcite{RefWorks:888} due to consuming large area, which is a crucial issue and therefore untenable for meeting practical standard requirement of area density 1 Tb/in$^2$. 

Furthermore, Ref.~\refcite{biswa14} performs an incorrect analysis while comparing error probability and switching delay with that of the Refs.~\refcite{RefWorks:850,RefWorks:559,RefWorks:848,RefWorks:849}, and underestimates the MTJ resistance ratio with an incorrect statement ``The maximum value of this ratio (assuming $\eta_1=\eta_2=1$) is 2:1'' (for details see Ref.~\refcite{arXiv:1501.05941v2}). Also, there is a very basic issue behind the \emph{multi-step} switching methodology proposed in Refs.~\refcite{biswa14,biswa14_2}, which increases the switching delay exponentially while meeting a lowered error probability.\cite{arXiv:1501.05941v2} It should be noted too that Refs.~\refcite{biswa14,biswa14_2} have \emph{assumed} \emph{instantaneous} ramp for stress, which is \emph{unreasonable} and crucial, since ramp that is not fast enough causes switching failures.\cite{roy13_2,roy11_6}

Also, switching delay is a major performance metric while proposing energy-efficient systems, since devices with lower switching speed would take more time given a computation task to be performed, therefore making the energy-delay product a tenable performance metric. If we calculate the switching delay of magnetization according to Ref.~\refcite{RefWorks:154}, it will come out $\sim$1000 ns, which is clearly {untenable for building nanoelectronics},\cite{itrs,nri} While comparing with spin-transfer-torque (STT) switching mechanism, if one compares an usual performance metric i.e., switching delay-energy, clearly Ref.~\refcite{RefWorks:154} performs inferior to STT switching. Also consider the issue that 1 hour of execution would take 100 hours or more using the operation presented in Ref.~\refcite{RefWorks:154}, which however performed an \emph{incorrect} analysis.\cite{arXiv:1501.05941v2} If charge based transistors were to operate \emph{slow}, the energy dissipation would not have been an issue.\cite{rabae03,itrs,nri}

In Ref.~\refcite{munir15}, the authors proposed to reduce error probability of switching in a system of \emph{two} dipole-coupled magnetostrictive nanomagnets in strain-mediated multiferroic heterostructures using voltage (stress) pulse shaping. The authors conclude that high success probability of switching \emph{cannot} be achieved at high switching speed ($\sim$1 ns), and therefore their proposed system is only applicable for niche applications (but therein too it is inferior to the traditional transistors). However, such conclusion lacked the critical understandings behind such high error probability for general-purpose logic applications. Fortunately, such analysis and a possible solution along the line of the analysis is present in literature\cite{roy14} using Bennett clocking\cite{RefWorks:144} for logic design, on which Ref.~\refcite{munir15} (and arXiv version Ref.~\refcite{munir14}) has made some \emph{incorrect} statements. Earlier, a subset of the authors of Ref.~\refcite{munir15} also published a paper\cite{fasha13} on a \emph{four} magnet system using the Bennett clocking mechanism, where the authors also came up with a similar conclusion of high error probability and the demise of multiferroic nanomagnetic logic, however, without relevant analysis similar to the case as in the Ref.~\refcite{munir15}. There are several technical issues in Ref.~\refcite{munir15} as well.\cite{arXiv:1501.05941v2} Due to the incorrect analysis performed in the Refs.~\refcite{fasha13,munir15}, in Ref.~\refcite{RefWorks:894}, the authors cast doubt on that front, which is incorrect too.

\begin{figure*}
\centering
\includegraphics[width=80mm]{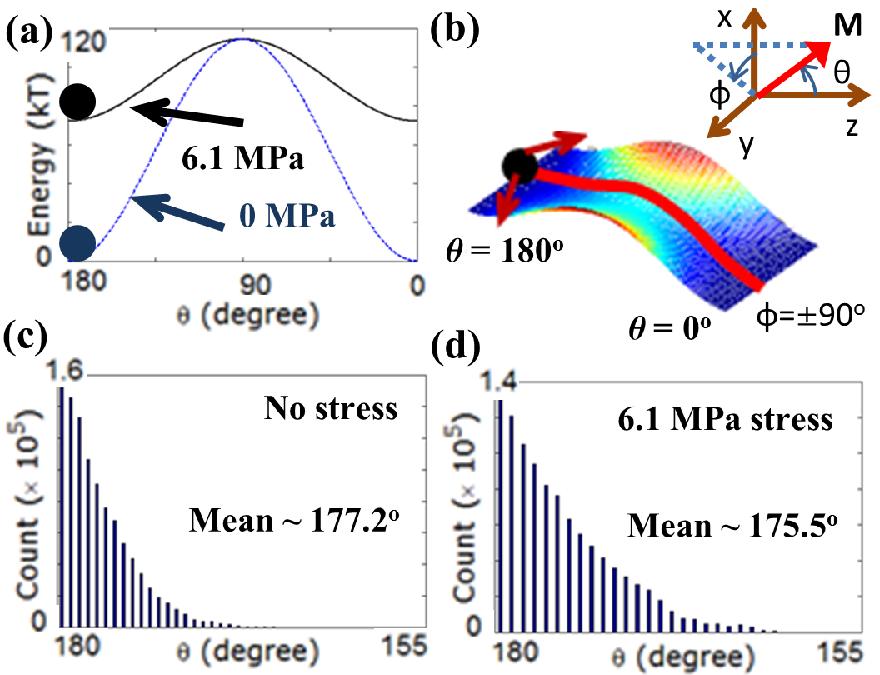}
\caption{\label{fig:APL_103_232401_2013}
(a) Potential landscapes of the nanomagnet with no stress and 6.1 MPa stress, as in Ref.~\aplSAW.
(b) Three-dimensional potential landscape showing the deflection of magnetization out of magnet's plane ($y$-$z$ plane, $\phi=\pm90^\circ$).
(c) Room-temperature (300 K) thermal distribution of $\theta$ when no stress is active.
(d) Room-temperature (300 K) thermal distribution of $\theta$ when 6.1 MPa stress is applied.
}
\end{figure*}

Rather than using Bennett clocking mechanism\cite{RefWorks:144} for logic design, Ref.~\refcite{roy13} proposed universal logic gates (NAND/NOR) utilizing a \emph{single} device with multiple contacts on the device to add up the strains generated in the piezoelectric layer,\cite{comsol} and a \emph{Set} input to preset the non-volatile magnetization state and facilitate concatenation. Ref.~\refcite{roy13} uses the switching methodology presented in Ref.~\refcite{roy13_2} and it is clearly mentioned while referring the Ref.~\refcite{roy13_2} that ``Computing methodologies utilizing such 180$^\circ$ switching mechanism between the two stable states of a shape-anisotropic magnetostrictive nanomagnet have not been proposed so far''. However, BA in Ref.~\refcite{roy13_comment} casts doubt on that, which has no technical basis as explained earlier. Moreover, BA in Ref.~\refcite{biswa14_3} follow the same idea presented in Ref.~\refcite{roy13} by Roy and uses a $\sim90^\circ$ (precisely $86.4^\circ$) switching mechanism following an earlier idea by others.\cite{RefWorks:850,RefWorks:559,RefWorks:848,RefWorks:849} Therefore, \emph{magnetization does not switch a complete $180^\circ$} in Ref.~\refcite{biswa14_3} and this leads to a low	tunneling magnetoresistance (TMR) while reading the magnetization state and it has serious consequence on \emph{read} error probability. Ref.~\refcite{biswa14_3} states that it requires a sensing circuitry for operation of the proposal in the Ref.~\refcite{roy13} and indicates as if it is an issue with the computing proposal in Ref.~\refcite{roy13}. First, Ref.~\refcite{biswa14_3} does not point out at all that the sensing element is required for complete $180^\circ$ switching only and it does not require so if $90^\circ$ switching mechanism is used. 

Ref.~\refcite{biswa14_3} just uses resistors and potential divider (see Fig.~1 in Ref.~\refcite{biswa14_3}) to accommodate multi-inputs (rather than using \emph{intrinsic} strain-addition property of piezoelectrics, while following the central idea in Ref.~\refcite{roy13}) and the authors want to use traditional transistors for any other circuit element, which makes the proposal in Ref.~\refcite{biswa14_3} energy-inefficient.\cite{arXiv:1501.05941v2} (Note that a circuit element cannot be shared between many devices distributed on a chip due to interconnect delay and loading effect.) It is stated in Ref.~\refcite{biswa14_3} (in the supplementary material) that ``The dissipation in the resistance $R$ can be negligible as we can make this resistance arbitrarily high.'' This is incorrect (any standard electrical engineering undergraduate textbook can be consulted) since $RC$ delay will be too high. Therefore, the design proposed in Ref.~\refcite{biswa14_3} is untenable. There are several technical issues in Ref.~\refcite{biswa14_3} as well.\cite{arXiv:1501.05941v2}

It should be noted that the switching as depicted in the Fig.~\ref{fig:magnetization_switching} just toggles the magnetization state upon application and removal of stress. Therefore, it requires to read the magnetization state before writing a bit. If we want to switch the magnetization along a specified direction, it is possible to use spin-transfer-torque with spin-polarized current which was first proposed in Ref.~\refcite{roy10} by Roy, however, it incurs much higher energy dissipation due to utilizing current-induced spin-torque mechanism. Unless new strategies to lower the energy dissipation in spin-transfer-torque switching can be achieved, a target of 1 aJ energy dissipation per switching a bit of information cannot be reached.

Ref.~\refcite{biswa13} follows the idea presented in the Ref.~\refcite{roy10} with the addition of surface acoustic wave (SAW) and also follows the formulation from Refs.~\refcite{roy11_6,roy11_3}. The basic idea was to switch the magnetization by STT along the desired direction when magnetization comes at the hard-plane ($\theta=90^\circ$) upon application of stress.\cite{roy10} However, thermal fluctuations cause a wide distribution for the time taken by magnetization to reach at $\theta=90^\circ$ as explained in Ref.~\refcite{roy13_2}. This is why STT current needs be kept active for almost half of the duration of switching.\cite{biswa13} Figure~3 in Ref.~\refcite{biswa13} shows that the energy dissipation with a stress of 6.1 MPa is $5\times10^9$ kT at room-temperature (300 K), which is $\sim$20 pJ. This is 4-5 orders of magnitude higher than that of traditional transistors and therefore {untenable for building nanoelectronics}.\cite{itrs,nri} Also, Fig.~3 in Ref.~\refcite{biswa13} shows the energy dissipation when no stress is present (only switching in STTRAM), which is $\sim40\times10^9$ kT = 160 pJ. Note that Ref.~\refcite{biswa13} has considered the material Terfenol-D as free layer in STT switching to calculate the  energy dissipation. However, the material that is commonly used for the free layer is CoFeB,\cite{RefWorks:786,RefWorks:774,RefWorks:33,RefWorks:564} which has Gilbert damping parameter $\alpha$ an order lower\cite{RefWorks:786,RefWorks:774} than Terfenol-D ($\alpha$ of Terfenol-D is 0.1 used by Ref.~\refcite{biswa13}). The critical current of switching is proportional to the damping parameter\cite{RefWorks:786,RefWorks:774} and the energy dissipation is proportional to the square of the switching current. Therefore, Ref.~\refcite{biswa13} calculated \emph{incorrectly} the switching current an order higher (23 mA) and energy dissipation about two orders higher ($\sim$160 pJ) for STT switching (Ref.~\refcite{RefWorks:786} \emph{correctly} determined switching current $\sim$1 mA and energy dissipation $\sim$1 pJ experimentally). Clearly, the comparison with STT switching performed in Ref.~\refcite{biswa13} is \emph{incorrect} and actually the energy dissipation in STT switching is 12.5 times lower (in stead of 8 times higher as \emph{incorrectly} claimed by Ref.~\refcite{biswa13}) than the hybrid scheme proposed in Ref.~\refcite{biswa13}.

Also, the energy dissipation calculated in Ref.~\refcite{biswa13} due to SAW is missing a crucial point that will make the energy dissipation exceedingly high. Ref.~\refcite{biswa13} says that ``SAW is \emph{global} and affects every memory cell.'' This is represented in Fig.~\ref{fig:APL_103_232401_2013}(a) that with the application of 6.1 MPa stress the barrier height decreases but it is not sufficient enough to make the potential landscape monostable and cause switching. However, when stress is applied the magnetization \emph{does} rotate. The distribution in Fig.~\ref{fig:APL_103_232401_2013}(d) upon application of stress is wider than the distribution in Fig.~\ref{fig:APL_103_232401_2013}(c) when no stress is active. Therefore, the energy dissipation upon application of stress and removal of stress \emph{must be calculated}, which is \emph{missed} by Ref.~\refcite{biswa13}. This energy dissipation turns out to be $\sim$40 kT for one cell. Considering just 1 MB memory, clearly such strategy would dissipate an energy which is quite worse than that of transistors.\cite{itrs,nri}

Ref.~\refcite{RefWorks:847} embraces the same concept of using STT (as proposed in Ref.~\refcite{roy10}) and does not consider the sensing procedure (explained in Ref.~\refcite{roy13_2}), which is required for successful switching in strain-mediated multiferroic composites in the presence of thermal fluctuations without STT, as explained earlier. This is why the Ref.~\refcite{RefWorks:847} \emph{incorrectly} terms Refs.~\refcite{roy11_2,roy13_2} as stochastically unstable. Since, using STT leads to energy-inefficiency\cite{roy10,biswa13,RefWorks:847}, symmetry breaking by other means rather than using STT is desirable.

In Ref.~\refcite{roy14_2}, it is shown that interface and exchange coupling in multiferroic heterostructures can facilitate switching of magnetization in a particular direction without using spin-torque mechanism, while incurring miniscule energy dissipation and there is no necessity of a sensing element as required for strain-mediated multiferroic composites.\cite{roy13_2} The switching methodology presented in the Ref.~\refcite{roy14_2} can be harnessed for logic design and computing purposes too.\cite{roy13,roy14} In the next subsection, we will present the simulation results for magnetization switching in such interface and exchange coupled multiferroic heterostructures.

\vspace*{2mm}

\begin{figure*}
\centering
\includegraphics[width=\textwidth]{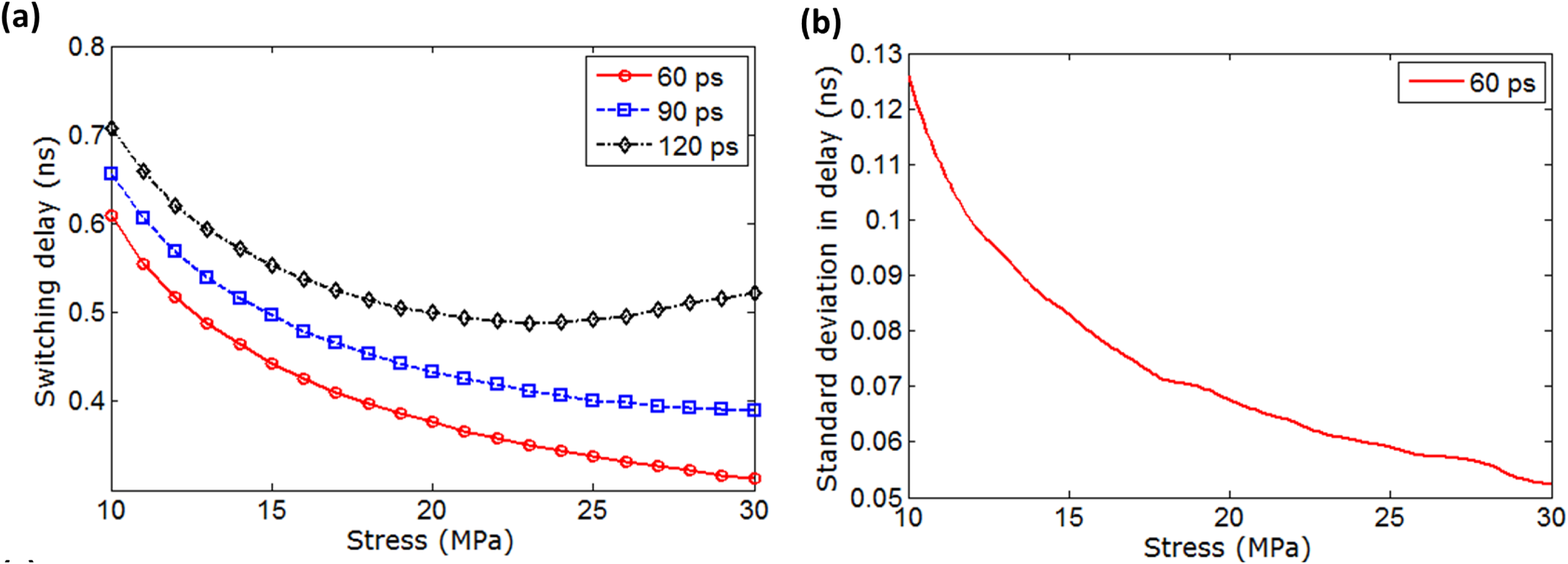}
\caption{\label{fig:thermal_mean_stress_delay_and_std_mag} 
(a) The thermal mean of the switching delay (at 300 K) versus stress (10--30 MPa) for different ramp durations (60 ps, 90 ps, 120 ps). Switching may fail at low stress levels and also at high stress levels for long ramp durations. Failed attempts are excluded when computing the mean.
(b) The standard deviations in switching delay versus stress (10--30 MPa) for 60 ps ramp duration at 300 K. We consider only the successful switching events in determining the standard deviations. The standard deviations in switching delay  for other ramp durations are of similar magnitudes and show similar trends.
(Reprinted with permission from Ref.~\japref. Copyright 2012, AIP Publishing LLC.)}
\end{figure*}

\vspace*{2mm}
Figure~\ref{fig:thermal_mean_stress_delay_and_std_mag}(a) shows the thermally averaged switching delay versus stress with different ramp duration (60 ps, 90 ps, 120 ps) as a parameter. Only the successful switching events are considered here since the switching delay metric does not make sense for an unsuccessful event. For a certain stress, if the ramp duration is decreased (i.e., the ramp rate is increased), the stress reaches its maximum value quicker and switches the magnetization faster, thereby decreasing the switching delay. For ramp durations of 60 ps and 90 ps, the switching delay decreases with increasing stress since an increasing stress anisotropy rotates the magnetization faster. However, for 120 ps ramp duration, the dependence of switching delay decreases with stress is non-monotonic, due to the exactly the same reasoning that caused the non-monotonicity in Fig.~\ref{fig:thermal_stress_success_ramp_time_mag}. A high stress accompanied by a long ramp duration is harmful since during the ramp-down phase it rotates the magnetization in ``bad'' quadrants leading to increased switching delay and even switching failures.

Figure~\ref{fig:thermal_mean_stress_delay_and_std_mag}(b) shows the standard deviation of switching delay distribution with stress for 60 ps ramp duration. At higher values of stress, the increased stress anisotropy energy makes the potential landscape more deep and resist the thermal fluctuations more effectively. Therefore, the spread in switching delay diminishes with a higher stress.

\begin{figure*}
\centering
\includegraphics[width=\textwidth]{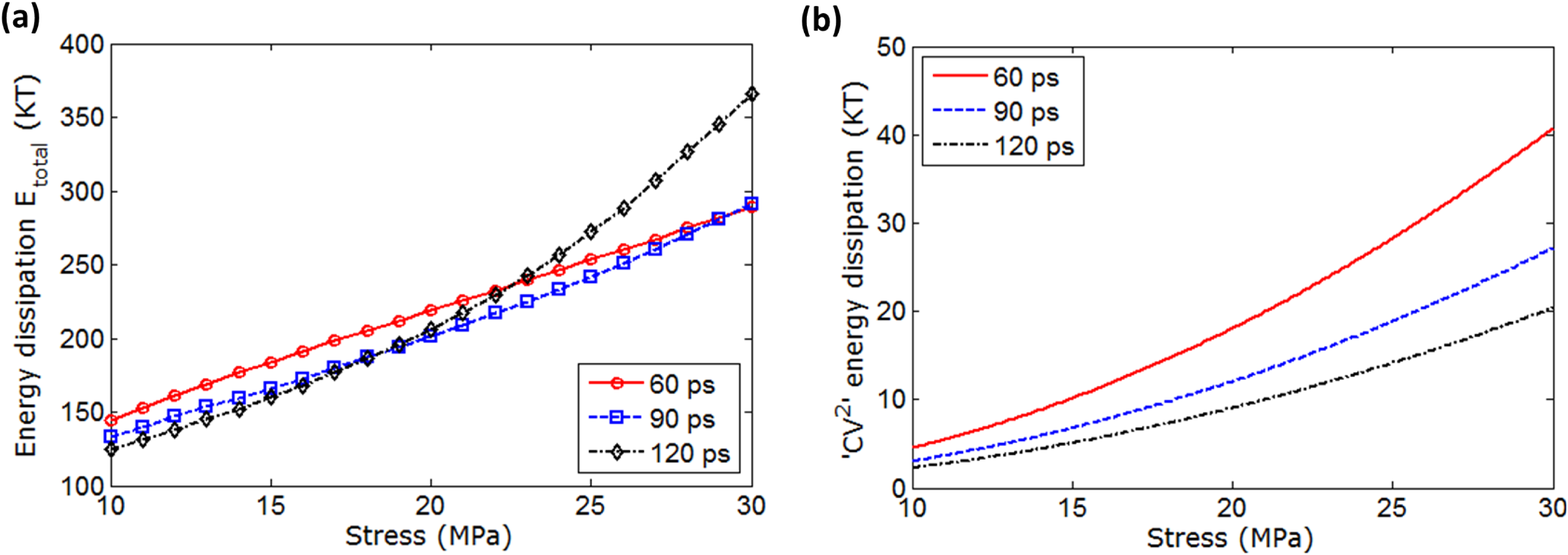}
\caption{\label{fig:thermal_mean_stress_energy_CV2_mag} 
(a) Thermal mean of the total energy dissipation versus stress (10--30 MPa) for different ramp durations (60 ps, 90 ps, 120 ps). Once again, failed switching attempts are excluded when computing the mean.
(b) The `$CV^2$' energy dissipation in the external circuit as a function of stress for different ramp durations. The dependence on voltage is 
not exactly quadratic since the voltage is not applied abruptly, but instead ramped up gradually and linearly in time.
(Reprinted with permission from Ref.~\japref. Copyright 2012, AIP Publishing LLC.)}
\end{figure*}

Figure~\ref{fig:thermal_mean_stress_energy_CV2_mag}(a) shows the thermal mean of the total energy dissipated to switch the magnetization as a function of stress for different ramp durations. For a certain ramp duration, the average power dissipation ($E_{total}/\tau$) increases with stress and for a certain stress it decreases with increasing ramp duration. A higher stress leads to more `$CV^2$' dissipation (see Fig.~\ref{fig:thermal_mean_stress_energy_CV2_mag}(b)) and also more internal dissipation because an increasing stress anisotropy results in a higher torque. A slower switching (i.e. more adiabatic) decreases the power dissipation. But the switching delay curves in the Fig.~\ref{fig:thermal_mean_stress_delay_and_std_mag}(a) show the opposite trend. At a slower ramp rate (higher ramp duration), the average power dissipation $E_{total}/\tau$ is always smaller than that of a higher ramp rate. However, the switching delay does not decrease as fast as with higher values of stress (in fact switching delay may increase for higher ramp duration, see Fig.~\ref{fig:thermal_mean_stress_delay_and_std_mag}(a)), which is why the energy dissipation curves in  Fig.~\ref{fig:thermal_mean_stress_energy_CV2_mag}(a) exhibit the cross-overs. 

\begin{figure*}
\centering
\includegraphics[width=80mm]{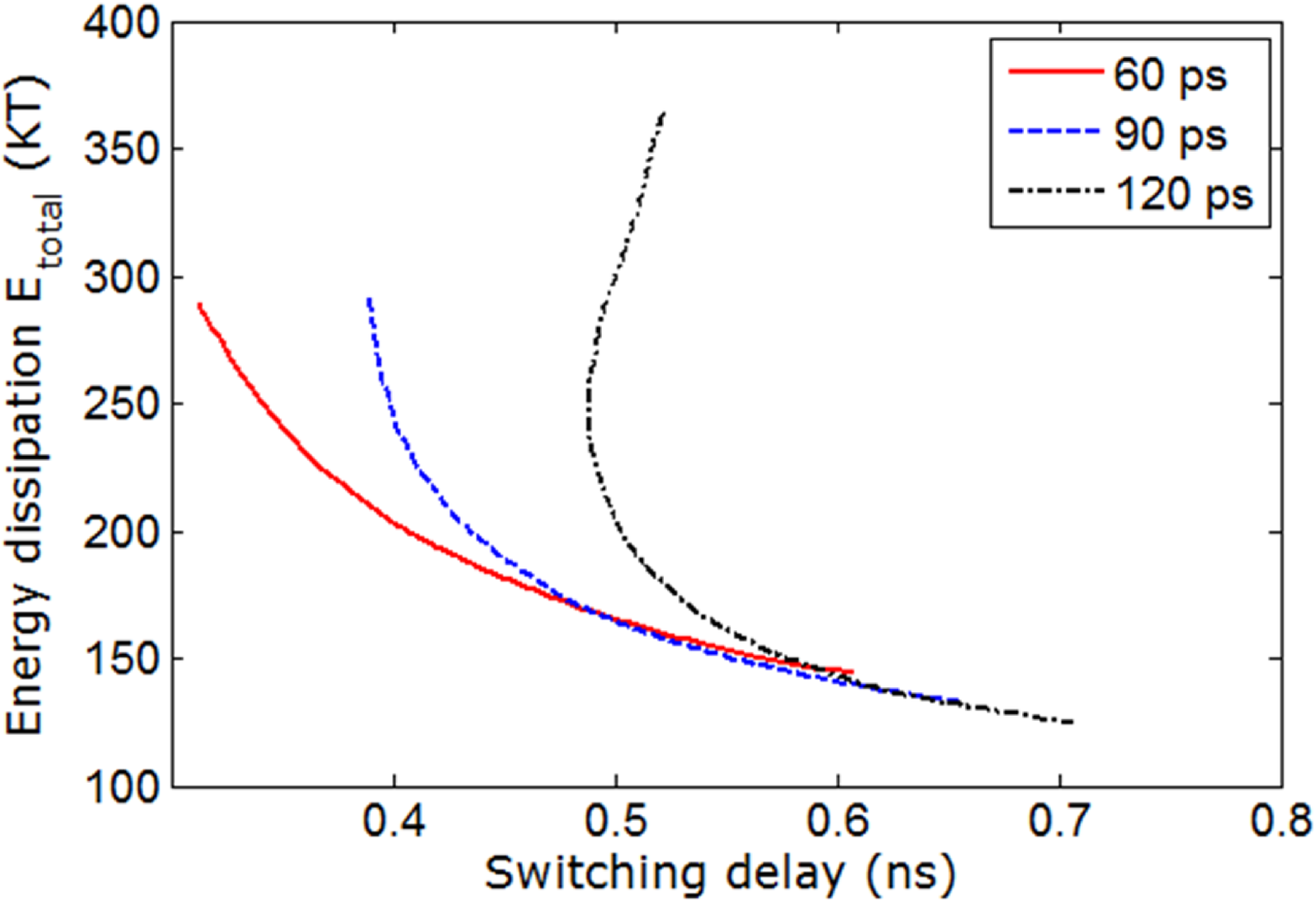}
\caption{\label{fig:thermal_mean_delay_energy_mag} 
Thermal mean of the switching delay versus thermal mean of the total energy dissipation for different stress levels (10--30 MPa) and different ramp durations (60 ps, 90 ps, 120 ps). Once again, failed switching attempts are excluded when computing the mean.}
\end{figure*}

Figure~\ref{fig:thermal_mean_stress_energy_CV2_mag}(b) plots the `$CV^2$' energy dissipation due to the application of voltage-induced stress for different ramp durations. Stress is proportional to the applied voltage $V$, and therefore the `$CV^2$' energy dissipation increases with stress for a certain ramp duration. This `$CV^2$' energy dissipation however is a small fraction of the total energy dissipation ($<$ 15\%) particularly because a miniscule voltage-generated stress is required to switch the magnetization of a magnetostrictive nanomagnet in a piezoelectric-magnetostrictive multiferroic. The `$CV^2$' dissipation decreases with an increasing ramp duration (i.e., slower ramp rate) for a certain stress since the switching becomes more \emph{adiabatic}. This component of the energy dissipation would come as several orders of magnitude higher if we switch the magnetization with an external magnetic field\cite{RefWorks:142} or spin-transfer-torque.\cite{RefWorks:7}

Note that we would require an adiabatic circuit to take advantage of the energy saving due to using an adiabatic pulse. In any case, using a piezoelectric layer of higher piezoelectric coefficients e.g., utilizing lead magnesium niobate-lead titanate (PMN-PT) instead of lead-zirconate-titanate (PZT), the energy dissipation can be reduced further. Also, PMN-PT can generate anisotropic strain, which allows us to work with lower voltage for a required strain, thereby reducing the energy dissipation even further.  PMN-PT layer has a dielectric constant of 1000, $d_{31}$=--3000 pm/V, and $d_{32}$=1000 pm/V (Ref.~\refcite{RefWorks:790}). With the piezoelectric layer's thickness $t_{piezo}$=24 nm (Ref.~\refcite{roy11_6}), $V=1.9$ mVs (2.9 mVs) of voltages would generate 20 MPa (30 MPa) compressive stress [$\sigma=Y\,d_{eff}\,(V/t_{piezo})$, where $d_{eff}=(d_{31}-d_{32})/(1+\nu)$] in the magnetostrictive Terfenol-D layer, which has $Y=80$ GPa,\cite{roy11_6} and Poisson's ratio $\nu=0.3$ (Ref.~\refcite{RefWorks:821}). Modeling the piezoelectric layer as a parallel plate capacitor ($\sim$100 nm lateral dimensions), the capacitance C=2.6 fF and thus $CV^2$ energy dissipation turns out to be $<$ 0.1 aJ. This is without considering any energy saving with the help of an adiabatic circuit. Such miniscule energy dissipation is the basis of ultra-low-energy computing using these multiferroic devices.\cite{roy11_news,roy13_spin,roy13,roy14}

Figure~\ref{fig:thermal_mean_delay_energy_mag} plots the switching delay versus energy dissipation for different ramp durations. This plot can be extracted from the Fig.~\ref{fig:thermal_mean_stress_delay_and_std_mag}(a) (stress versus switching delay) and Fig.~\ref{fig:thermal_mean_stress_energy_CV2_mag}(a) (stress versus energy dissipation). This plot signifies that as the switching delay increases, energy dissipation deceases. This points out the usual delay-energy trade-off. For 120 ps ramp duration, there is a opposite trend at higher stress values, the reasoning behind which has been already described while explaining the results in the Fig.~\ref{fig:thermal_stress_success_ramp_time_mag}.

\subsection{Interface and exchange coupled multiferroic heterostructures}
Here we present the simulation results for interface and exchange coupled multiferroic heterostructures.\cite{roy14_2} Fig.~\ref{fig:schematic_interface_coupled} shows that the nanomagnets ($M_1$ and $M_2$ layers) and the ferroelectric $P$-layer are made of Fe and $PbTiO_3$, respectively, while the spacer layer is made of $Au$ and the thicknesses of the trilayer $M_1$/spacer/$M_2$ are 1/4/1 monolayers.\cite{RefWorks:649,RefWorks:676,RefWorks:677,RefWorks:672} The Fe layer has a unit cell length of 0.287 nm and it possess the following material parameters: saturation magnetization ($M_s$) = 1e6 A/m, and damping parameter ($\alpha$) = 0.01.\cite{RefWorks:650,RefWorks:674,RefWorks:675} The elliptical lateral cross-section ($y$-$z$ plane, Fig.~\ref{fig:schematic_interface_coupled}) of the vertical stack has a dimension of $15\,nm \times 7\,nm$. The $P$-layer has a unit cell length of 0.388 nm and it has 5 layers in the vertical direction ($x$-direction, Fig.~\ref{fig:schematic_interface_coupled})\cite{RefWorks:649}. The energy difference between the P-alignment and AP-alignment is 10 meV/atom,\cite{RefWorks:649} and the absolute value of energy is calculated to be about 10 eV or 385 kT at room-temperature. This huge interface coupling makes potential landscape of $M_1$ \emph{monostable} at $\theta=180^\circ$ or at $\theta=0^\circ$ depending on the $P$-alignment or the $AP$-alignment in the trilayer, respectively. Due to the \emph{monostable} energy landscape and huge energy barrier, spontaneous switching of magnetization between $\theta=180^\circ$ and $\theta=0^\circ$ cannot take place. The interface coupling energy is a few orders of magnitude higher than the shape anisotropic energy and hence consideration of shape anisotropy does not make any significant difference, however, it is included during the simulations.

If we consider the $P$-layer as a parallel plate capacitor and use a relative dielectric constant of 1000,\cite{RefWorks:673} the capacitance $C$ of the layer becomes $\sim$0.4 fF. If the $P$-layer is accessed with a 10 $\mu m$ long silver wire with resistivity $\sim$2.6 $\mu \Omega$-$cm$,\cite{interconnect} the resistance $R$ can be calculated as $\sim$3 $k\Omega$. Therefore, the $RC$ time constant turns out to be the order of 1 ps. We assume that the ferroelectric $PbTiO_3$ has a coercive voltage of 20 MV/m\cite{RefWorks:670} and hence a voltage $V\simeq40$ mv is required to switch its polarization. Note that polarization switching is possible in less than 100 ps\cite{RefWorks:430} and a voltage ramp with period $T = 100$ ps or more is considered to enforce the quasistatic (adiabatic) assumption ($T \gg RC$). Without any adiabatic assumption, the metric $CV^2$ can be calculated as $0.5$ aJ and hence the energy dissipation due to the application of the voltage is miniscule. With 100 ps ramp period, the ``$CV^2$" dissipation is determined as a negligible value of 0.01 aJ.\cite{roy11_2,roy11_6} Note that we do not calculate any \emph{standby} leakage current through the thin ferroelectric since the device operation is non-volatile, i.e., it is possible to turn off the voltage without loosing the information in long term. However, during the active mode of operation, the leakage needs to be considered, however, the tunneling current is small ($<$ 1 nA, Ref.~\refcite{RefWorks:792}) leading to negligible energy dissipation. Another issue that needs to be considered is ferroelectric fatigue, which may make the coercive field higher over time,\cite{RefWorks:800,RefWorks:799} i.e., it would require a higher voltage to switch the polarization. However, the energy dissipation due to applied voltage is miniscule, and therefore it does not appear to be a bottleneck provided the polarization switches and the interface coupling between the $P$-layer and the trilayer persists. In any case, further progress in the experimental front can handle such issues better.

\begin{figure*}
\centering
\includegraphics[width=80mm]{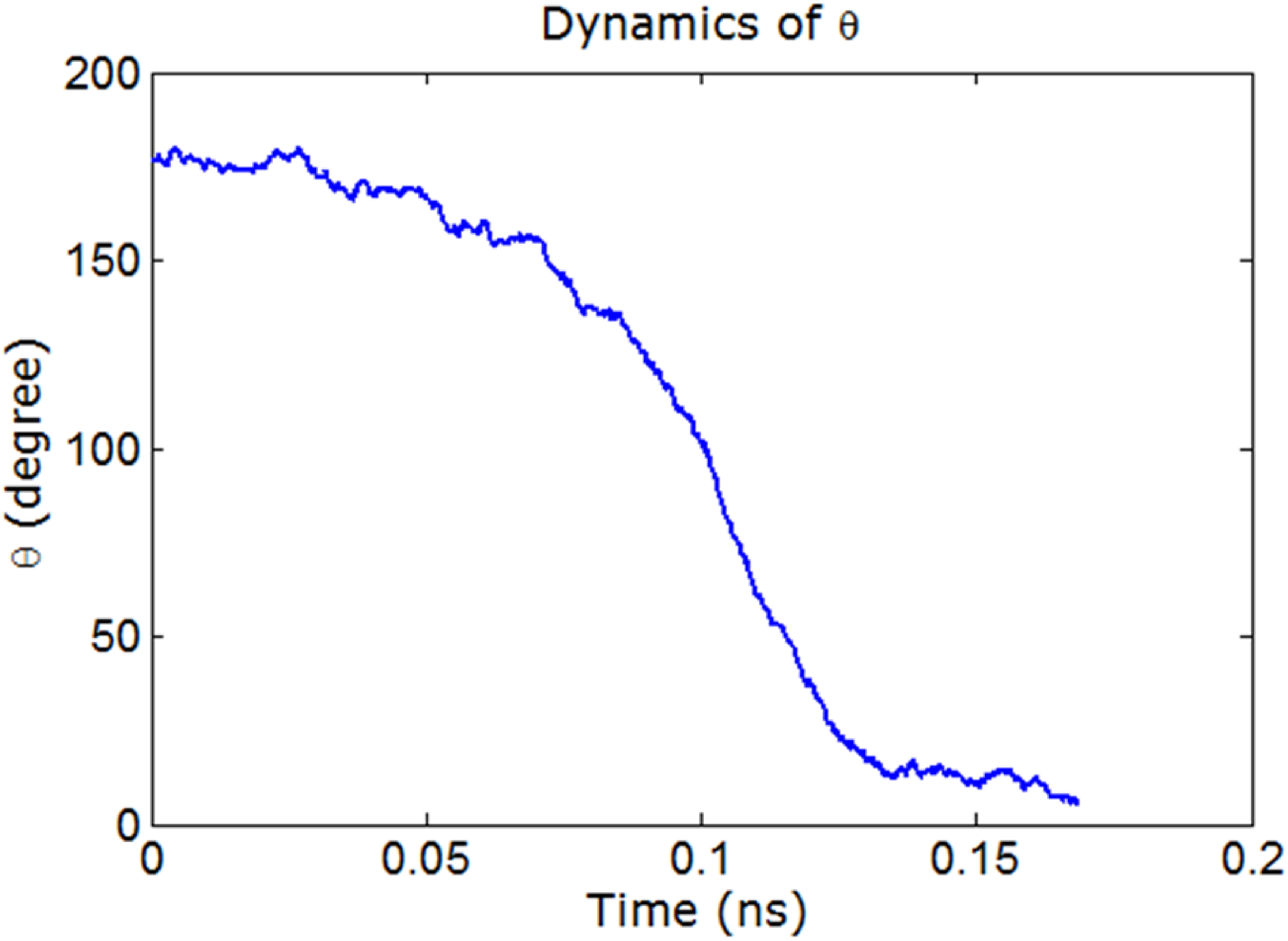}
\caption{\label{fig:thermal_single_run} A sample dynamics of magnetization while switching from $\theta \simeq 180^\circ$ to $\theta \simeq 0^\circ$ in the presence of room-temperature (300 K) thermal fluctuations. The ramp period is 100 ps and the switching delay is 168.5 ps. The energy dissipation due to Gilbert damping is 1.42 aJ.
(\textcopyright IOP Publishing.  Reproduced by permission of IOP Publishing from Ref.~\jphysdref. All rights reserved.)
} 
\end{figure*}

Figure~\ref{fig:thermal_single_run} presents a sample magnetization dynamics in the presence of room-temperature (300 K) thermal fluctuations. The ramp period is considered to be 100 ps and the it turns out from the simulation that switching has completed in less than 175 ps. Note that during the course of switching, random thermal kicks have forced magnetization to backtrack temporarily, however, the strong interface anisotropy has eventually enforced magnetization to switch from $\theta \simeq 180^\circ$ to $\theta \simeq 0^\circ$.

\begin{figure*}
\centering
\includegraphics[width=80mm]{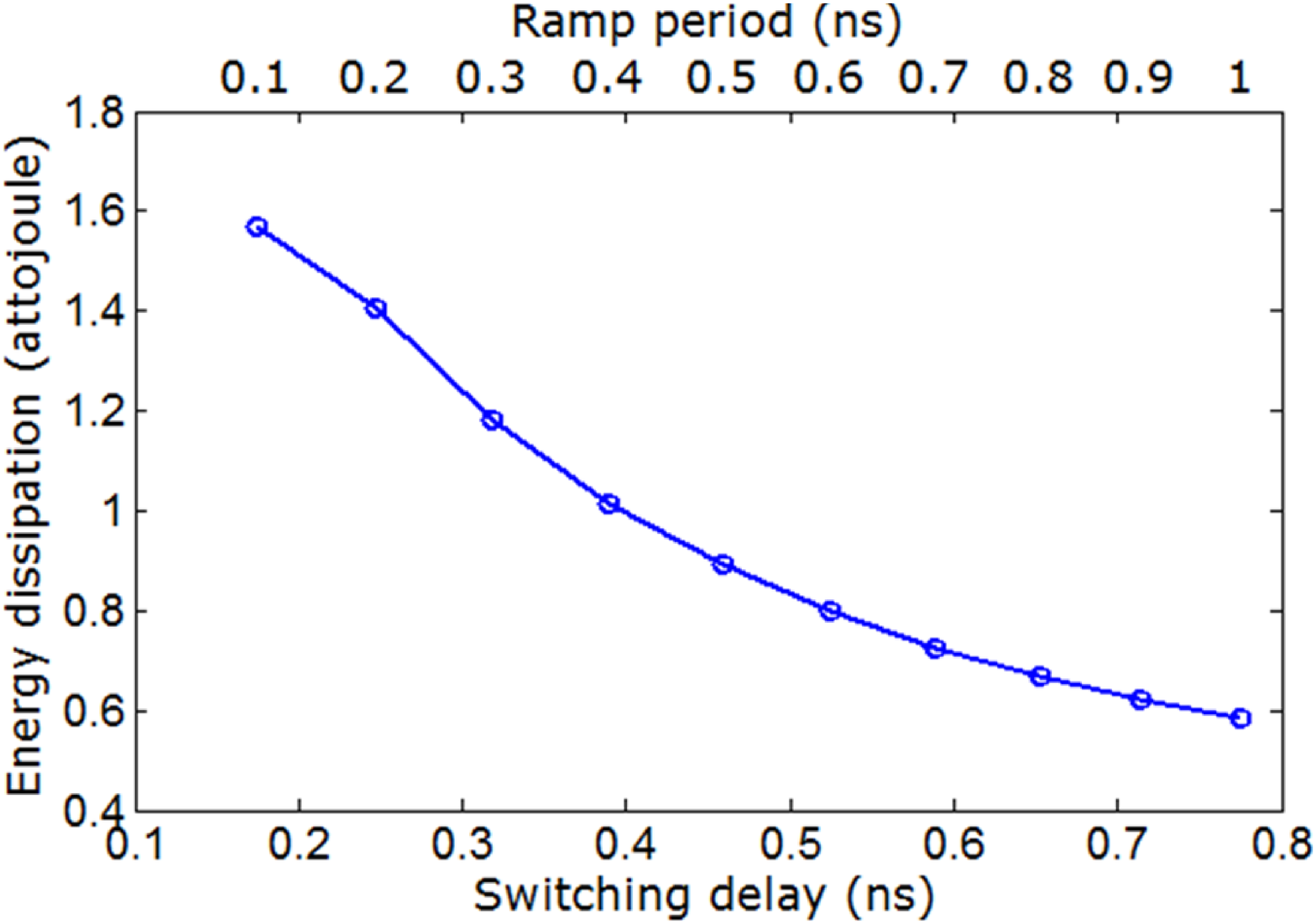}
\caption{\label{fig:delay_energy_ramp} Switching delay-energy trade-off as a function of ramp period (upper axis). For a faster ramp, the switching becomes faster but the energy dissipation goes higher. Each point is generated from 10000 simulations in the presence of room-temperature (300 K) thermal fluctuations and the average values of switching delays and energy dissipations are plotted. For 0.1 ns ramp period, the average (max) switching delay is 175.3 ps (330.8 ps), while the mean energy dissipation is 1.56 aJ. For a slower ramp with period 1 ns, the average (max) switching delay is 775.2 ps (1003.5 ps), while the mean energy dissipation is 0.58 aJ.
(\textcopyright IOP Publishing.  Reproduced by permission of IOP Publishing from Ref.~\jphysdref. All rights reserved.)
} 
\end{figure*}

Figure~\ref{fig:delay_energy_ramp} plots the average switching delay versus average energy dissipation for a range of  ramp periods (0.1 ns -- 1 ns). A moderately large number of simulations (10,000) in the presence of room-temperature (300 K) thermal fluctuations are performed and they are averaged to generate each point in the curve. For each trajectory of the 10,000 simulations, when the magnetization reaches $\theta \leq 5^\circ$, the switching is deemed to have completed. Note that as we increase the ramp period of applied voltage across the heterostructure, the switching delay also increases and less energy is dissipated in the switching process, elucidating the well-established delay-energy trade-off for a device in general. The results clearly demonstrate that that switching in sub-nanosecond delay is possible while dissipating a miniscule amount of energy of $\sim$1 aJ. Note that the ``$CV^2$'' energy dissipation is a couple of orders of magnitude lower than the energy dissipation due to Gilbert damping and it decreases with the increase of ramp period since the switching becomes more adiabatic.\cite{roy11_2,roy11_6} While the Figure~\ref{fig:delay_energy_ramp} provides the mean of the switching delay distribution, the standard deviation in switching delay is also an important performance metric. the standard deviation for ramp period of 0.1 ns is about 22 ps and it increases about twice when the ramp period is increased to 1 ns. At higher ramp period, thermal fluctuations have more time to scuttle the magnetization and cause variability in switching time, increasing the standard deviation.  

To understand the effect of temperature on the performance metrics, simulations have been performed at an elevated temperature (400 K) and the metrics switching delay and energy dissipation turn out to be similar (within 5\%) compared to that of room-temperature (300 K) case.\cite{roy14_2} Interestingly, the mean switching delay at a higher temperature T = 400 K decreases compared to the case at T = 300 K, which can be traced out from the reasoning that the initial deflection of magnetization due to thermal fluctuations increases at a higher temperature. Hence, magnetization is likely to start far away from the easy axis at a higher temperature for different trajectories, leading to the decrease in the mean switching delay. It turns out that this decrease in mean switching delay at T = 400 K is very small (less than 2\%) compared to that of T = 300 K. However, the trend of standard deviation in switching delay with the increase in temperature shows an opposite trend to that of the mean. It can be understood by considering that the standard deviation of random thermal field at higher temperature increases with temperature [see Eq.~(\ref{eq:ht})]. The mean energy dissipation decreases with increasing temperature and this decrease at T = 400 K is quite small (less than 3\%) compared to the case of T = 300 K. This once again signifies the switching delay-energy trade-off for this device.

So far we have considered the writing of a bit of information by switching the magnetization from one state to another, however, the magnetization state needs to be read too. In this interface and exchange coupled structure, the giant magnetoresistance (GMR)\cite{RefWorks:433,RefWorks:434} of the trilayer is calculated to be of the order of 30\%,\cite{RefWorks:649,RefWorks:688} which provides a way to read the magnetization states (P-alignment or AP-alignment). Although this GMR is not that high compared to tunneling magnetoresistance (TMR),\cite{RefWorks:577,RefWorks:555,RefWorks:572,RefWorks:76,RefWorks:74,RefWorks:33,RefWorks:300} suitable design strategies can be possibly be devised to work with this moderate value of GMR and also it may be possible to increase the magnetoresistance by suitable material choice and design. It is also argued in Ref.~\refcite{RefWorks:649} that even with the variance in the smaller thicknesses of the layers, it is still possible to interface-couple the polarization and magnetization in the proposed structures. It should be noted that the modeling of interface anisotropy is not limited to the way that is performed here, however, any \emph{strong} interface-coupled system would facilitate switching of magnetization from one state to the another. 

The important consequence of having such a strong interface anisotropy is that we can achieve devices of very small lateral area, and it is therefore possible to cram an enormous amount of devices on a single chip. If we use an area density of $10^{-12}$ cm$^{-2}$, the dissipated power would be 10 mW/cm$^2$ considering 1 aJ energy dissipation in a single nanomagnet with 1 ns switching delay and 10\% switching activity (i.e., 10\% of the magnets switch at a given time). Such extremely dense and ultra-low-energy non-volatile computing systems can be powered by energy harvesting systems without the need of an external battery.\cite{roundyfx,anton,lu,jeon}

\section{\label{sec:conclusions}Summary and outlook}

We have reviewed the dynamical systems study for electrical field-induced magnetization switching in strain-mediated multiferroic composites, and interface and exchange coupled multiferroic heterostructures. The magnetization switching dynamics in strain-mediated multiferroic composites using Landau-Lifshitz-Gilbert equation revealed intriguing phenomena in binary switching mechanism. It is shown that binary switching in a `symmetric' potential landscape can be successful in the presence of room-temperature thermal fluctuations. To achieve such symmetry-breaking, we require the following two criteria: (1) a sufficiently high stress that keeps the magnetization more out of magnet's plane inside the so-called ``good'' quadrants; and (2) a sufficiently fast ramp rate that reduces the possibility of backtracking causing switching failure. A high stress and a fast ramp rate also increase the switching speed and counters the detrimental effects of thermal fluctuations. This can potentially open up a new methodology of binary switching since tilting the potential landscape would not be necessary and such findings would stimulate experimental research to establish the proposed methodology of binary switching. As stated, it requires to sense when magnetization reaches around hard-plane so that stress can be brought down thereafter to achieve successful switching by breaking the symmetry of having equal probability of successful switching and switching failure in a`symmetric' potential landscape. Other ways to break this symmetry may be harnessed and is subject to further research. Note that the aforesaid switching in strain-mediated multiferroic composites just toggles the magnetization state without being able to maintain the direction of switching. It is shown that magnetization switching in interface and exchange coupled multiferroic heterostructures can maintain the direction of switching.

The calculated performance metrics from LLG simulations e.g., switching delay and energy dissipation show a profound promise for technological applications. The results show that switching can take place in sub-nanosecond delay while expending a miniscule amount of energy of $\sim$1 attojoule. This energy dissipation is at least 2-3 orders of magnitude lower than that of the other emerging devices. So multiferroic magnetoelectrics are intriguing in respect to both basic physics of binary switching and applied physics. Also, a strong interface and exchange coupling in multiferroic heterostructures enforces error-resiliency during the switching process and facilitates to scale down the lateral dimensions to unprecedented dimensions of $\sim$10 nm even in the presence of room-temperature thermal fluctuations. This is very crucial since it will help competing with the traditional charge-based electronics and consequently for building future nanoelectronics. Due to these superior performance characteristics of multiferroic magnetoelectrics as described here, currently it is of immense interest to analyze different possible theoretical designs followed by experimental demonstrations. Successful experimental implementations must tackle the issue of process variation at low dimensions, which traditional transistors face too. Processors built on such devices can lead to unprecedented applications that can work by harvesting energy from the environment without the need of an external battery e.g., medically implanted device to warn an impending epileptic seizure by monitoring the brain signals, wearable computers powered by body movements etc. Moreover, the basic building blocks are non-volatile permitting instant turn-on computer, facilitating computational designs, and avoiding energy dissipation for bit storage elements in a system. This can potentially perpetuate Moore's law beyond the year 2020 and turn out to be an unprecedented opportunity for ultra-low-energy computing in our future information processing systems. Experimental efforts are emerging and successful experimental implementation can potentially pave the way for our future nanoelectronics.


\end{multicols}
\end{document}